\documentclass[useAMS,referee,usenatbib]{biom}
\usepackage{graphicx}
\usepackage{graphics}
\usepackage{amsmath}
\usepackage{amssymb}

\newcommand\independent{\protect\mathpalette{\protect\independenT}{\perp}}
\def\independenT#1#2{\mathrel{\rlap{$#1#2$}\mkern2mu{#1#2}}}

\newcommand{\expit}{\mbox{expit}}
\newcommand{\calA}{\mathcal{A}}
\newcommand{\calD}{\mathcal{D}}
\newcommand{\calW}{\mathcal{W}}

\newcommand{\calT}{\mathcal{T}}
\newcommand{\calS}{\mathcal{S}}
\newcommand{\Ns}{N^*}
\newcommand{\Ys}{Y^*}
\newcommand{\Ts}{T^*}

\newcommand{\hatLam}{\widehat{\Lambda}}

\newcommand{\bgamma}{\mbox{\boldmath $\gamma$}}
\newcommand{\bSigma}{\mbox{\boldmath $\Sigma$}}
\newcommand{\hatgamma}{\widehat{\bgamma}}
\newcommand{\hatSig}{\widehat{\bSigma}}
\newcommand{\sumin}{\sum^n_{i=1}}
\newcommand{\sumjD}{\sum^{D}_{j=1}}
\newcommand{\calN}{\mathcal{N}}

\newcommand{\bx}{\mbox{\boldmath $x$}}
\newcommand{\bX}{\mbox{\boldmath $X$}}
\newcommand{\xbar}{\overline{\bx}}
\newcommand{\Xbar}{\overline{\bX}}
\newcommand{\abar}{\overline{a}}
\newcommand{\Abar}{\overline{A}}
\newcommand{\taubar}{\overline{\tau}}
\newcommand{\Tbar}{\overline{\calT}}
\newcommand{\Tbb}{\boldsymbol{\mathfrak{T}}}
\newcommand{\bh}{\mbox{\boldmath $h$}}
\newcommand{\bH}{\mbox{\boldmath $H$}}
\newcommand{\Nbar}{\overline{N}}
\newcommand{\Ybar}{\overline{Y}}
\newcommand{\hatq}{\widehat{q}}
\newcommand{\Tfrak}{\mathfrak{T}}
\newcommand{\Zbb}{\mathbb{Z}}
\newcommand{\calO}{\mathcal{O}}

\newcommand{\prodmM}{\prod^{M}_{m=0}}
\newcommand{\calZ}{\mathcal{Z}}
\newcommand{\Hm}{H^{-}}
\newcommand{\tilh}{\widetilde{\bh}}
\newcommand{\tilH}{\widetilde{\bH}}

\newcommand{\bPsi}{\mbox{\boldmath $\Psi$}}

\newcommand{\sumln}{\sum^n_{\ell=1}}
\newcommand{\sumjpD}{\sum^{D}_{j'=1}}

\newcommand{\bS}{\mbox{\boldmath $S$}}
\newcommand{\bA}{\mbox{\boldmath $A$}}
\newcommand{\bG}{\mbox{\boldmath $G$}}

\newcommand{\btheta}{\mbox{\boldmath $\theta$}}
\newcommand{\balpha}{\mbox{\boldmath $\alpha$}}
\newcommand{\bdelta}{\mbox{\boldmath $\delta$}}

\title[Generalized Logrank-type Test for SMARTs]{A Generalized Logrank-type Test for Comparison of Treatment Regimes in
  Sequential Multiple Assignment Randomized Trials}

\author{Anastasios A. Tsiatis$^{*}$\email{tsiatis@ncsu.edu} and
Marie Davidian$^{**}$\email{davidian@ncsu.edu} \\
Department of Statistics, North Carolina State University, Raleigh, NC, USA.}

\begin{document}






\begin{abstract}
  The sequential multiple assignment randomized trial (SMART) is the
  ideal study design for the evaluation of multistage treatment
  regimes, which comprise sequential decision rules that recommend
  treatments for a patient at each of a series of decision points
  based on their evolving characteristics.  A common goal is to
  compare the set of so-called embedded regimes represented in the
  design on the basis of a primary outcome of interest.  In the study
  of chronic diseases and disorders, this outcome is often a time to
  an event, and a goal is to compare the distributions of the
  time-to-event outcome associated with each regime in the set.  We
  present a general statistical framework in which we develop a
  logrank-type test for comparison of the survival distributions
  associated with regimes within a specified set based on the data
  from a SMART with an arbitrary number of stages that allows
  incorporation of covariate information to enhance efficiency and can
  also be used with data from an observational study.  The framework
  provides clarification of the assumptions required to yield a
  principled test procedure, and the proposed test subsumes or offers
  an improved alternative to existing methods.  We demonstrate
  performance of the methods in a suite of simulation
  studies.  The methods are applied to a SMART in patients with acute
  promyelocytic leukemia. \vspace*{0.3in}
  \end{abstract}

  \begin{keywords}
    Augmented inverse probability weighting; Dynamic treatment regime;
   Estimating function; Potential outcomes
  \end{keywords}

  \maketitle

\section{Introduction}\label{s:intro}

Clinicians caring for patients with chronic diseases and disorders
typically make sequential treatment decisions at key points in a
patient's disease or disorder progression.  For example, treatment of
cancer patients involves a series of decisions at the times of
diagnosis, remission or failure to achieve remission, recurrence, and
so on.  Clinicians make these decisions based on accrued information
on the patient, including how the patient responded to previous
treatments, with the goal of optimizing a long-term health outcome.

A treatment regime is a sequence of decision rules, each rule
corresponding to a key decision point and mapping accrued information
on the patient to a recommended treatment option chosen from among
those feasible for the patient.  Thus, treatment regimes formalize the
clinical decision-making process and provide a strategy for treating
an individual patient.  Sequential multiple assignment randomized
trials (SMARTs) \citep{Lavori,Murphy2005}, which involve multiple
stages of randomization of participants, where each stage corresponds
to a key decision point, are ideally suited to the study of treatment
regimes, and SMARTs have been conducted in a range of disease and
disorder areas
\citep{almirall2014introduction,KidwellCancer,lorenzoni2023use,bigirumurame2022sequential}.
In a SMART, the treatment options to which a subject can be assigned
at any stage can depend on the subject's past history, including
previous treatments and responses to them, and the design of the SMART
defines a set of regimes, known as the SMART's embedded regimes.
Primary or secondary analyses in a SMART often focus on comparison of
all or a subset of the embedded regimes on the basis of a health
outcome of interest.

There is an extensive literature on methods for inference on treatment
regimes, including well-established methods for inference on a
specified set of given regimes, such as the embedded regimes in a
SMART \citep{nahum2019introduction,TsiatisBook}.  The majority assume
a continuous or discrete outcome that is observed for all subjects,
and interest focuses on comparison of the mean outcomes that would be
achieved if the entire patient population were to receive treatments
according to the rules in each regime in the set.  In many chronic
diseases and disorders, the relevant outcome is a time to an event,
e.g., in cancer, disease-free or overall survival time, and, as in
conventional (single-stage) clinical trials, interest is instead in
comparison of the event-time (survival) distributions if the patient
population were to receive treatments according to the rules in each
regime.  A further complication that arises in the context of a SMART
or other study giving rise to data to be used for this purpose is that
the outcome may be censored for some subjects.

\begin{figure}
   \centering
\includegraphics[width=12cm]{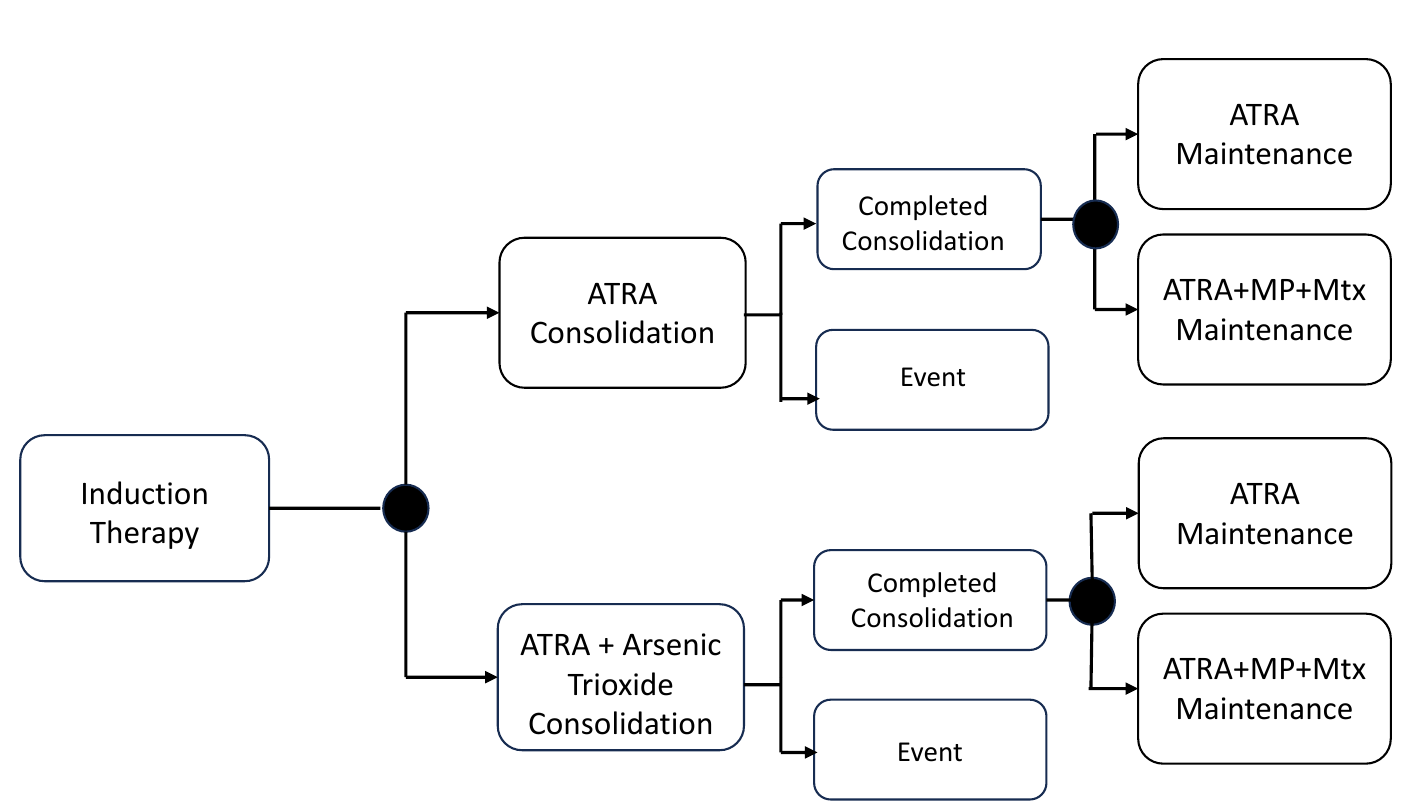}
\caption{\label{f:c9710} Schematic depicting the design of
  Study C9710.  Solid circles represent points of randomization. The
  four embedded regimes implied by this design are of the form ``Give
  consolidation therapy $a$ initially; if consolidation is completed
  before the event occurs, give maintenance therapy $b$,'' where $a =$
ATRA or ATRA+Arsenic Trioxide and $b = $ ATRA or ATRA+MP+Mtx.}
\end{figure}

This situation is exemplified by North American Leukemia Intergroup
Study C9710, coordinated by the Cancer and Leukemia Group B, now part
of the Alliance for Clinical Trials in Oncology, in patients with
acute promyelocytic leukemia (APL) \citep{Powell2010}.  The primary
outcome was a composite of time to failure to achieve complete
remission (CR), relapse after achieving CR, or death, whichever came
first, referred to as event-free survival (EFS) time.  Study C9710 was
a SMART with two stages, as depicted in Figure~\ref{f:c9710}: All
subjects were to receive the same induction chemotherapy and were then
randomized at stage 1 to one of two consolidation therapies, two
courses of all-trans-retinoic acid (ATRA) or two courses of ATRA
combined with arsenic trioxide; thus, the first decision point
corresponds to choice of consolidation therapy.  Those not
experiencing the event by completion of consolidation therapy were to
be randomized at the second stage to receive maintenance therapy of
ATRA alone or ATRA plus oral methotrexate (Mtx) and mercaptopurine
(MP); thus, the second decision point corresponds to choice of maintenance
therapy for patients who complete consolidation. Study C9710 involves
four embedded regimes; e.g., one of these administers ATRA
consolidation therapy at the first decision and, if the patient does
not experience an event by the end of consolidation, administers
ATRA+Mtx+MP maintenance therapy at the second decision.  For some
subjects, EFS time was not observed because it did not occur during
the study period, i.e., was administratively censored, or was censored
due to loss to follow up.  Of necessity, subjects for whom the outcome
occurred or was censored during consolidation were not re-randomized.
Although the original goal was not to compare the embedded regimes,
the trial presents the opportunity for further insight on how best to
sequence these therapies in the treatment of APL patients.

A number of methods for comparing regimes in a specified set of
regimes on the basis of a time-to-event outcome have been proposed.
Some authors focus on comparing survival probabilities only at a
single time point \citep*{Lunceford}, while others propose
logrank-type tests for comparing survival distributions
\citep{GuoTsiatis,FengWahed,LiMurphy,KidwellLogrank,Li2014}.  Early
works limit comparisons to two regimes that assign different stage 1
treatments.  \citet{KidwellLogrank} and \citet{Li2014} propose
tests for comparing $\geq 2$ regimes, including ``shared
path'' regimes for which the stage 1 treatment is the same.  The
former authors restrict to comparison of the embedded regimes in a
particular two-stage SMART with four embedded regimes.  The latter
consider comparison of regimes involving, in principle, any number of
decision points based on data from a SMART or observational study,
although the approach is demonstrated only in the case of the same
two-stage four-regime setting as \citet{KidwellLogrank}.  Both methods have
potential shortcomings, discussed in the sequel.

In this article, we propose a general framework from which to derive a
logrank-type test for comparing an arbitrary set of regimes for any
number of decision points; e.g., in a SMART, all or a subset of the
embedded regimes.  In addition to providing a test statistic for the
null hypothesis of no difference in survival distributions for a set
of regimes, unlike existing methods, the framework also yields a
covariate-adjusted statistic that uses baseline and intermediate
covariate information to increase efficiency, resulting in more
powerful tests.  Although we focus on SMARTs, the methodology also can
be used with observational data and includes the approach of
\citet{Li2014}.  In Section~\ref{s:framework}, we present the
framework and the proposed test procedures, and we constrast our
approach to existing methods.  We present results of a suite of
simulation studies demonstrating performance of the methods in
Section~\ref{s:sims}, and we apply the methods to Study C9710 in
Section~\ref{s:example}.

\section{Statistical framework and methodology}
\label{s:framework}

\subsection{Problem formulation}
\label{ss:basic}

We first characterize a treatment regime when the outcome is a time to
an event.  Consider $K$ decision points, where at Decision $k$,
$k=1,\ldots,K$, $\calA_k$ is the finite set of all treatment options.
Following \citet[Section~8.3.1]{TsiatisBook}, let $\bx_1$ be the
information available on an individual at Decision 1 at time
$\tau_1=0$ (baseline), such as demographics, medical history
variables, clinical and physiologic measures, and so on; and let
$a_1 \in \calA_1$ be the treatment option given at Decision 1.  For
$k=2,\ldots,K$, if an individual does not experience the event between
Decisions $k-1$ and $k$, let $\tau_k$ be the time Decision $k$ is
reached; $\bx_k$ be additional information obtained between Decisions
$k-1$ and $k$; e.g., updated clinical and physiologic measures, timing
and severity of adverse reactions to and response to prior treatments, 
etc.; and $a_k \in \calA_k$ be the option given at Decision $k$.
Then, at Decision 1, the history of information available on an
individual is $\bh_1=(\tau_1,\bx_1)$, and, defining
$\abar_k = (a_1,\ldots,a_k)$, $k=1,\ldots,K$, and $\taubar_k$ and
$\xbar_k$ similarly, the accrued history of information available at
Decision $k$ for an individual who reaches Decision $k$ without having
experienced the event is $\bh_k = (\taubar_k,\xbar_k,\abar_{k-1})$,
$k=2,\ldots,K$.  For an individual who experiences the event at time
$t$ between Decisions $k-1$ and $k$, the history for $j = k,\ldots,K$
is $\bh_j = (\taubar_k,\xbar_k,\abar_{k-1},t)$.  A treatment regime
$d$ is a set of decision rules
$d = \{ d_1(\bh_1),\ldots, d_K(\bh_K)\} = (d_1,\ldots,d_K)$, where,
for an individual for whom $\bh_k$ indicates the event has not yet
occurred, the $k$th rule $d_k(\bh_k)$ returns a treatment option in
the subset $\Psi_k(\bh_k) \subseteq \calA_k$ that is feasible for an
individual with history $\bh_k$.  If $\bh_k$ indicates that the event
has occurred, $d_k(\bh_k)$ makes no selection, and $\Psi_k(\bh_k)$ is
the empty set.  As in \citet[Section~6.2.2]{TsiatisBook}, at Decision
$k$ there may be $\ell_k$ subsets of $\calA_k$ that are feasible sets
for different histories.  If
$s_k(\bh_k) = 1,\ldots,\ell_k$ indicates the relevant feasible subset
for $\bh_k$, $d_k(\bh_k)$ is typically a composite of subset-specific
rules; write
$d_k(\bh_k) = \sum_{l=1}^{\ell_k} I\{ s_k(\bh_k)=l\} d_{k,l}(\bh_k)$,
where $d_{k,l}(\bh_k)$ selects options from the $l$th subset.

\begin{figure}
   \centering
\includegraphics[width=12cm]{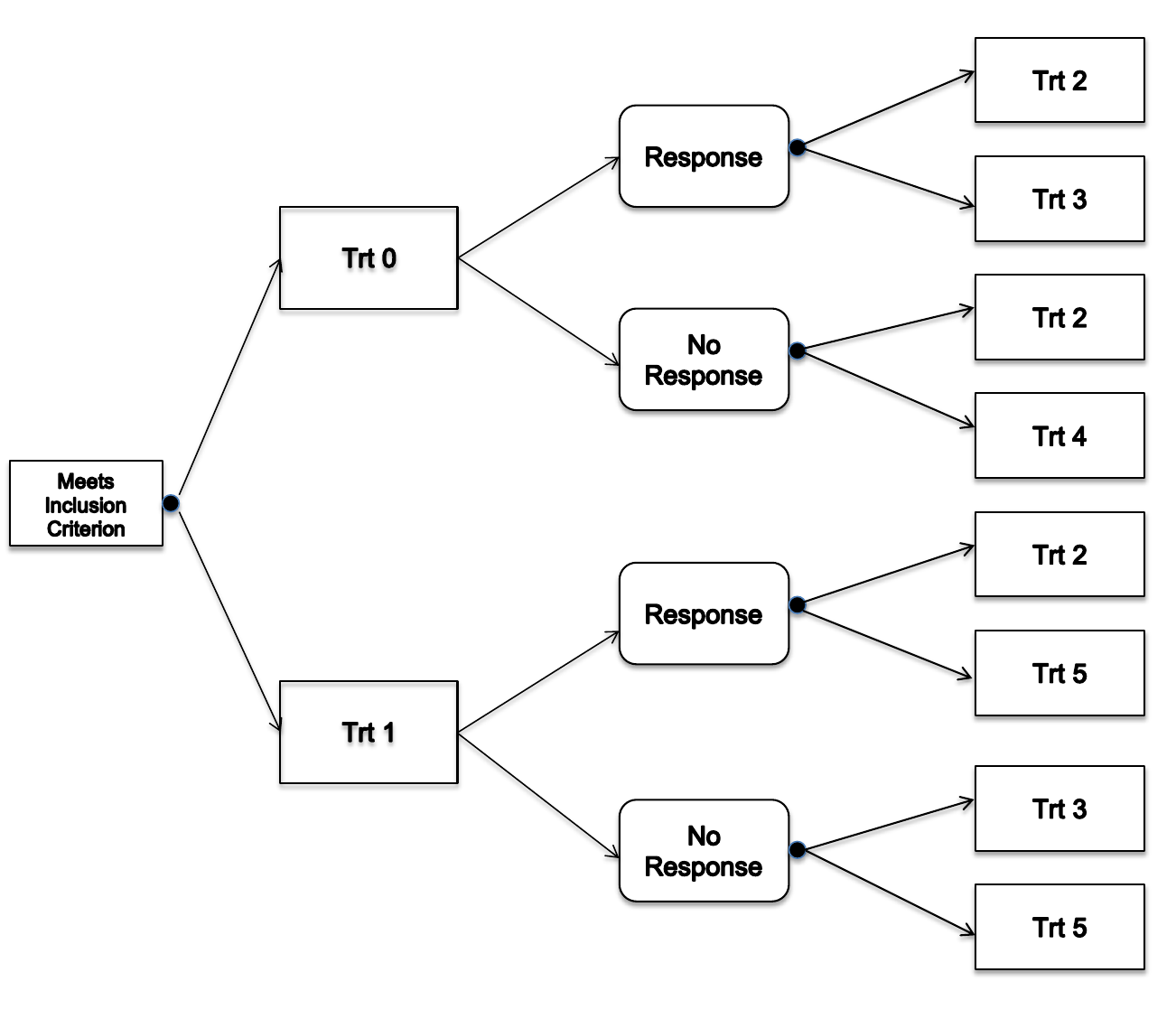}
\caption{\label{f:eight} Schematic depicting the design of a two-stage
  SMART with $\calA_1=\{0,1\}$, $\calA_2 = \{2,3, 4,5\}$ in which
  subjects who do not experience the event prior to Decision 2 are
  classified as responders or nonresponders to Decision 1 treatment
  and are randomized to stage 2 treatments within feasible subsets of
  $\calA_2$ depending on their stage 1 treatment and response
  status; here, $\ell_2=4$. Solid circles represent points of randomization.  The eight
  embedded regimes implied by this design are of the form ``Give Trt $a$
  initially; if the event does not occur before Decision 2 and
  response, give Trt $b$, otherwise if the event does not occur before
  Decision 2 and no response, give Trt $c$,'' where the eight regimes
  correspond to
  $(a,b,c) = (0,2,2), (0,2,4), (0,3,2), (0,3,4), (1,2,3), (1,2,5),
  (1,5,3) (1,5,5)$.}
\end{figure}

In a SMART, the feasible sets and regimes that can be studied are
dictated by the design.  For example, consider the SMART in
Figure~\ref{f:eight}, with $K=2$.  Here,
$\Psi_1(\bh_1) = \calA_1 = \{0, 1\}$ for any $\bh_1$, and regimes that
can be studied based on this trial have rules $d_1(\bh_1)$ that return
one of the treatments in $\calA_1$.  At Decision 2, $\Psi_2(\bh_2)$ is
the empty set for subjects who experience the event prior to Decision
2; otherwise, $\Psi_2(\bh_2)$ is a subset of $\calA_2= \{2, 3, 4, 5\}$
determined by the stage 1 treatment and response status contained in
histories $\bh_2$ for which the event has not yet taken place, so that
regimes that can be studied have rules $d_2(\bh_2)$ that are a
composite of subset-specific rules.  E.g., if $\bh_2$ indicates that
$a_1=1$ and the individual is a responder, $\Psi_2(\bh_2) = \{2, 5\}$, and
a subset-specific rule selects Decision 2
treatment from this subset. The SMART in
Figure~\ref{f:eight} has eight embedded regimes satisfying these
conditions.

We formalize the null hypothesis of interest in terms of the potential
outcomes framework in \citet[Section~8.3.2]{TsiatisBook}, who define
for any sequence of treatment options
$\abar_K \in \calA_1 \times \cdots \times \calA_K$ that could
hypothetically be given to an individual, the potential number of
decision points that would be reached and potential times at which
they would be reached, potential covariate information that would
arise, and the associated potential event time.   As in (8.50) of
\citet{TsiatisBook}, denote by $\calW^*$ the collection of these
potential variables for all possible sequences.  For any regime $d$
and associated feasible sets, it is possible to define in terms of
$\calW^*$ $\Ts(d)$, the potential time to the event if a randomly
chosen individual in the population were to receive treatments by
following the rules in $d$.  Denote the hazard rate of experiencing
the event at time $u$ for an individual following the rules in $d$ by
$\lambda(u,d) = \lim_{du \rightarrow 0} du^{-1} P\{ u \leq \Ts(d) <
u+du | \Ts(d) \geq u\}$, $u \geq 0$.  Then, given a set
$\calD = \{d^1,\ldots,d^D\}$ of $D$ regimes of interest, e.g., the
embedded regimes in a SMART or a subset thereof, we are interested in
testing the null hypothesis
\begin{equation}
  H_0\mathord{:}  \,\,\,\lambda(u,d^1) = \cdots = \lambda(u,d^D), \,\,\, u \geq 0,
  \label{eq:null}
  \end{equation}
versus the alternative hypothesis that at least one hazard rate
differs from the rest.

Analogous to the formulation of the standard logrank test in a
conventional (single-stage) multi-arm clinical trial, to motivate our proposed test of
$H_0$ in (\ref{eq:null}), for the regimes $d^j$, $j=1,\ldots,D$, of
interest, assume a proportional hazards relationship; namely
\begin{equation}
\lambda(u,d^j) = \lambda_0(u) \exp(\beta_j), \hspace{0.1in} j=1,\ldots,D-1.
  \label{eq:prophazards}
  \end{equation}
  Under (\ref{eq:prophazards}), the baseline hazard $\lambda_0(u)$ is
  $\lambda(u,d^D)$, and (\ref{eq:null}) corresponds to
  $H_0\mathord{:}\,\, \beta_1 = \cdots = \beta_{D-1} = 0.$ As with the
  standard logrank test, we do not necessarily believe that
  (\ref{eq:prophazards}) holds.  In fact, (\ref{eq:prophazards})
  generally cannot hold; e.g., intuitively, the hazards for two
  ``shared path'' regimes cannot be proportional under departures from
  $H_0$.  Rather, (\ref{eq:prophazards}) is solely a convenient
  mechanism by which to motivate a test of (\ref{eq:null}).

  Toward that goal, we consider the possibly misspecified model
  (\ref{eq:prophazards}) and develop a test statistic in the spirit of
  generalized score tests \citep{Boos}. 
  For any regime $d$, let $\Ns(u,d) = I\{ \Ts(d) \leq u\}$ and
  $\Ys(u,d) = I\{\Ts(d) \geq u\}$ be the event-time counting process
  and at-risk process, respectively, associated with $\Ts(d)$.
  Assuming (\ref{eq:prophazards}) holds and letting
  $\Lambda_0(u) = \int_0^u \lambda_0(s)\,ds$ and
  $\Lambda(u,d) =\int_0^u \lambda(s,d)\,ds$ denote the cumulative
  hazard functions associated with $\lambda_0(u)$ and $\lambda(u,d)$,
  respectively, $E\{ d\Ns(u,d) - d\Lambda(u,d) \Ys(u,d)\} = 0$.  Then
  for the set $\calD$ of regimes of interest,
  \vspace*{-0.15in}
  \begin{equation*}
  \begin{aligned}
&E\{ d\Ns(u,d^D) - d\Lambda_0(u) \Ys(u,d^D)\} = 0, \\
E\{ d\Ns(u,d^j) &- d\Lambda_0(u) \exp(\beta_j) \Ys(u,d^j)\} = 0, \,\, j=1,\ldots,D-1,
  \end{aligned}
  \end{equation*}
  which defines a dynamic regime marginal structural
  model \citep*{Orellana}.  If we were able to observe $\Ts(d^j)$ for all
  $j=1,\ldots,D$, and if (\ref{eq:prophazards}) were correct, then,
  analogous to \citet*{Yangetal}, for each regime
  $j=1,\ldots,D$, we would be led to a set of estimating functions 
  for  $\Lambda_0(u)$, $u \geq 0$, and $\beta_1,\ldots,\beta_{D-1}$.
That associated with 
$\Lambda_0(u)$, $u \geq 0$, is the infinite-dimensional expression
$$d\Ns(u,d^j) - d\Lambda_0(u) \exp\Big\{ \sum_{j'=1}^{D-1} \beta_{j'}
I(j'=j)\Big\} \Ys(u,d^j), \,\,\, u \geq 0;$$
that associated with   $\beta_{j'}$, $j'=1,\ldots,D-1$, is
$$\int^\infty_0 \, I(j'=j)  \Big[d\Ns(u,d^j) - d\Lambda_0(u) \exp\Big\{ \sum_{j'=1}^{D-1}
\beta_{j'} I(j'=j)\Big\}\Ys(u,d^j)\Big].$$ Note that this expression
is equal to zero except when $j'=j$.  Following \citet{Orellana} and
\citet{Yangetal}, we can combine these estimating functions across the
$D$ regimes in $\calD$ via a weighted sum, yielding the estimating
functions
\begin{align}
  \sum^D_{j=1}  &w(u, d^j) \Big[ d\Ns(u,d^j) - d\Lambda_0(u) 
                  \exp\Big\{ \sum_{j'=1}^{D-1} \beta_{j'}I(j'=j)\Big\}
                  \Ys(u,d^j)\Big], \,\, u \geq 0, \label{eq:estfuncLam} \\
                &\int^\infty_0 \, w(u, d^j) \{d\Ns(u,d^j) - d\Lambda_0(u)
                  \exp(\beta_j) \Ys(u,d^j)\}, \hspace{0.1in} j=1,\ldots,D-1, \label{eq:estfuncbeta}
  \end{align}
where, for any regime $d$, $w(u,d)$ is a possibly time-dependent weight
function discussed later.  

Now, under $H_0\mathord{:}\,\, \beta_1 = \cdots = \beta_{D-1} = 0$, if it were
possible to observe $\Ts(d^j)$, $j=1,\ldots,D$, for each of the $n$
participants in a SMART, indexing individuals by $i = 1,\ldots,n$,
from (\ref{eq:estfuncLam}), for any $u \geq 0$, the ``score equation''
associated with the ``nuisance parameter'' $\Lambda_0(u)$ is
\begin{equation}
 \sumin \sum^D_{j=1} w(u, d^j)\{ d\Ns_i(u,d^j) -d\Lambda_0(u)\Ys_i(u,d^j)\} =0,
\label{eq:scoreLam}
 \end{equation}
 yielding
 $d\hatLam_0(u) = \big\{\sumin \sumjD w(u,d^j)
 d\Ns_i(u,d^j)\big\}/\big\{\sumin \sumjD w(u,d^j) \Ys_i(u,d^j)\big\}$.
 From (\ref{eq:estfuncbeta}), the ``score equation'' associated
 with $\beta_j$ is  $\sumin \int^\infty_0\, w(u,d^j) \{ d\Ns_i(u,d^j) -
 d\Lambda_0(u)\Ys_i(u,d^j) \}$; substituting
$d\hatLam_0(u)$ yields the $(D-1)$-dimensional ``score vector'' $\Tbb^*$
associated with $\beta_1,\ldots,\beta_{D-1}$, with $j$th element
\vspace*{-0.15in}
\begin{equation}
\sumin \int^\infty_0\, w(u,d^j)   \{ d\Ns_i(u,d^j) - d\hatLam_0(u)\Ys_i(u,d^j) \}.
\label{eq:scorebeta}
\end{equation}
Thus, a test statistic for $H_0$ could be constructed as a quadratic
form in $\Tbb^*$.

Of course, for a given set of regimes $\calD$, we cannot observe
$\Ts(d^j)$, $j=1,\ldots,D$, for each subject in a SMART (or
observational study).  Thus, to exploit these developments, we
must relate these potential outcomes and 
(\ref{eq:scoreLam})-(\ref{eq:scorebeta}) to the data actually
available, discussed next.

  \subsection{Data and test procedure}
  \label{ss:test}

  As in Study C9710, the event time for some subjects may be censored.
  Thus, following \citet[Section~8.3.2]{TsiatisBook}, the observed
  data on a participant in a $K$-stage SMART (or observational study)
  can be represented as
\begin{equation}
\calO = \{ \kappa, \calT_1, \bX_1, A_1, \calT_2, \bX_2, A_2, \ldots, \calT_\kappa,
\bX_\kappa, A_\kappa, U, \Delta),
\label{eq:data}
\end{equation}
where $\kappa$ is the observed number of decision points reached by
the subject before experiencing the event or censoring,
$1 \leq \kappa \leq K$; $A_1 \in \calA_1$ is the treatment option
received at time $\calT_1 = 0$ (baseline); $\calT_k$,
$k=2,\ldots,\kappa$, are the observed times at which the subject
reached Decisions $2,\ldots,\kappa$, at which they received treatment
options $A_k \in \calA_k$, $k=2,\ldots\kappa$; and $\bX_1$ is baseline
covariate information and $\bX_k$, $k=2,\ldots,\kappa$, is the
additional covariate information collected on the individual between
Decisions $k-1$ and $k$.  In (\ref{eq:data}), $U$ is the observed time
to the event or censoring, whichever comes first, and $\Delta=1$ or 0 as $U$
is the event or censoring time.  Finally, writing
$\Abar_k = (A_1,\ldots,A_k)$, $k=1,\ldots,\kappa$, and $\Xbar_k$ and
$\Tbar_k$ similarly, the observed history up to Decision $k$ is
$\bH_1 = (\calT_1, \bX_1)$, $k=1$, and, for $k = 2,\ldots,K$,
$\bH_k = (\Tbar_k,\Xbar_k,\Abar_{k-1})$, $\kappa \geq k$, and
$\bH_k = (\Tbar_\kappa,\Xbar_\kappa,\Abar_\kappa,U,\Delta)$,
$\kappa < k$.  As in (8.62) of \citet[Section~8.3.2]{TsiatisBook}, the
history available up to time $u\geq 0$ can be expressed as
\begin{align*}
H(u) = \{ (\calT_1, &\bX_1, A_1), I(\kappa\geq 2, \calT_2 \leq u), 
I(\kappa\geq 2, \calT_2 \leq u) (\calT_2, \bX_2, A_2), \ldots, 
    I(\kappa\geq k, \calT_k \leq u), \\
   & I(\kappa\geq k, \calT_k \leq u) (\calT_k, \bX_k, A_k),
  k=3,\ldots,K,  I(U < u), (U,\Delta)I(U < u)\}.
\end{align*}  

It is possible to obtain analogs to (\ref{eq:scoreLam}) and
(\ref{eq:scorebeta}) in terms of the observed data (\ref{eq:data})
under standard identifiability assumptions, which are discussed in
detail in \citet[Section~8.3.2]{TsiatisBook}.  The consistency
assumption in these authors' (8.66) states that the observed data are
equal to their potential analogs under the treatments actually
received.  The sequential randomization assumption (SRA) states that
$\calW^* \independent A_k | (\bH_k, \kappa \geq k)$, $k=1,\ldots,K$, 
where ``$\independent$'' denotes statistical independence, which is
guaranteed by randomization in a SMART but must be made based on
domain considerations in an observational study.  We assume that
censoring is noninformative in the sense that the cause-specific
hazard for censoring satisfies
$\lambda_c\{ u | H(u), \calW^*\} = \lim_{du \rightarrow 0} du^{-1} P\{
u \leq U < u+du,\Delta=0 | U\geq u, H(u), \calW^*\} = \lim_{du
  \rightarrow 0} du^{-1} P\{ u \leq U < u+du,\Delta=0 | U\geq u,
H(u)\} = \lambda_c\{ u | H(u)\}$, say, so that censoring depends only
on information observed through time $u$.  The positivity assumption
states roughly that
$\omega_k(\bh_k,a_k) = P(A_k=a_k|\bH_k=\bh_k, \kappa \geq k) > 0$ for
all $\bh_k$ such that $P(\bH_k=\bh_k, \kappa\geq k) >0$ and
$a_k \in \Psi_k(\bh_k)$, $k=1,\ldots,K$, which is true by design in a
SMART; see \citet[Section~8.3.2]{TsiatisBook} for a precise
formulation.

Under these assumptions, analogs to (\ref{eq:scoreLam}) and
(\ref{eq:scorebeta}) in terms of the observed data are based on
inverse probability weighting.  For regime $d$ and $u \geq 0$,
define 
\begin{equation}
  C(u,d) = \prod^\kappa_{k=1} \left[ I\{\calT_k>u\} + I\{\calT_k \leq
    u\} I\{A_k = d_k(\bH_k)\} \right], \,\,\,\, u \geq 0,
\label{eq:Cdu}
\end{equation}
the indicator of whether or not the treatments received by an
individual through $u$ are consistent with having followed the rules
in $d$ through $u$.  Similarly, define 
\begin{equation}
\pi(u,d) = \prod^\kappa_{k=1}  \left[ I\{\calT_k>u\} + I\{\calT_k \leq
    u\} \omega_k\{ \bH_k,d_k(\bH_k) \} \right], \,\,\,\, u \geq 0.
  \label{eq:pidu}
\end{equation}
Dependence of $C(u,d)$ and $\pi(u,d)$ on $H(u)$ is
suppressed for brevity.  As an example, consider the SMART in
Figure~\ref{f:eight}, with $K=2$ and the embedded regime
$d = (d_1,d_2)$ that gives Trt $a$ at Decision 1, and, if the event
does not occur before Decision 2, gives Trt $b$ if an individual is a
responder and Trt $c$ if not.  Because in this SMART
$\Psi_1(\bh_1)=\{0,1\}$ for all $\bh_1$,
$I\{A_1 = d_1(\bH_1)\} = I(A_1=a)$ and
$\omega_1\{ \bH_1,d_1(\bH_1)\} = P(A_1=a)$, the probability of being
randomized to Trt $a$ at stage 1.  Thus, if $\kappa=1$, as
$\calT_1=0$, $C(u,d) = I(A_1=a)$ and $\pi(u,d) = P(A_1=a)$ for all
$u \geq 0$.  If $\kappa=2$, letting $R_2 = 1~(0)$ if an individual
responds (does not respond) to stage 1 treatment, where $R_2$ is a
function of $\bX_2$ and thus $\bH_2$, $d_2(\bh_2)$ is a composite of
two subset-specific rules: $d_{2,1}(\bh_2)$ selects Trt $c$ for
$\bh_2$ such that $r_2=0, a_1 = a$, and $d_{2,2}(\bh_2)$ selects Trt
$b$ for $\bh_2$ such that $r_2=1, a_1=a$.  Thus,
$I\{ A_2 = d_2(\bH_2)\} =R_2 I(A_2 = b) + (1-R_2) I(A_2= c)$, and
$\omega_2\{ \bH_2, d_2(\bH_2)\} = R_2 P(A_2 = b | A_1=a,
R_2=1,\kappa=2) + (1-R_2) P(A_2 = c | A_1=a, R_2=0, \kappa=2)$, where
$P(A_2=a_2 | A_1=a,R_2=r, \kappa=2)$ is the probability of being
randomized to Trt $a_2$ at stage 2 after being randomized to Trt $a$
at stage 1 and reaching stage 2 as a responder ($r=1$) or nonresponder
($r=0$).  Then if $\kappa=2$, for $u \geq 0$,
$C(u,d) = I(A_1=a)\big[ I\{\calT_2>u\} + I\{\calT_2 \leq u\}\{R_2
I(A_2 = b) + (1-R_2) I(A_2= c)\}\big]$ and
$\pi(u,d) = P(A_1=a) \big[ I\{\calT_2>u\} + I\{\calT_2 \leq u\}\{R_2
P(A_2 = b | A_1=a, R_2=1,\kappa=2) + (1-R_2) P(A_2 = c | A_1=a, R_2=0,
\kappa=2)\}\big]$.

Denote the survival function for censoring under the noninformative
censoring assumption as
$K_c\{u | H(u)\} = \exp[ -\int^u_0 \lambda_c\{ s|H(s)\}\,ds]$, and
define $N(u) = I(U \leq u, \Delta=1)$ and $Y(u) = I(U \geq u)$.  Then,
under the above assumptions, we argue in Section~\ref{s:webA} of the
Appendix that an observed data analog to (\ref{eq:scoreLam}) is given
by
\begin{equation}
\sumin \sum^D_{j=1} \Omega_i(u,d^j)\{ dN_i(u) -d\Lambda_0(u) Y_i(u)\} =0,\,\,\,\,\,
\Omega(u,d) = \frac{C(u,d) I(U \geq u) }{\pi(u,d) K_c\{u | H(u)\} } w(u, d),
\label{eq:Omega}
\end{equation}
so that
$d\hatLam_0(u) = \big\{\sumin \sumjD \Omega_i(u,d^j)
dN_i(u)\big\}/\big\{\sumin \sumjD \Omega_i(u,d^j)
Y_i(u)\big\}$. Similarly, under $H_0$, an analog to the ``score
equation'' for $\beta_j$ is 
$\sumin \int^\infty_0 \, \Omega_i(u, d^j) \{dN_i(u) - d\Lambda_0(u)
Y_i(u)\}$, $j=1,\ldots,D-1$,
so that, substituting $d\hatLam_0(u)$, the analog to
(\ref{eq:scorebeta}) in terms of the observed data is the
$(D-1)$-dimensional ``score vector'' $\Tbb =
(\Tfrak^1,\ldots,\Tfrak^{D-1})^T$, where 
\begin{equation}
\Tfrak^j = \sumin \int^\infty_0\, \Omega_i(u,d^j)   \{ dN_i(u) -
d\hatLam_0(u)Y_i(u) \}, \hspace{0.1in} j=1,\ldots,D-1.
\label{eq:scorebetaobs}
\end{equation}
Although we have exemplified (\ref{eq:Cdu}) and (\ref{eq:pidu}) in the
context of SMARTs, where $\omega_k(\bh_k,a_k)$, $k=1,\ldots,K$, are
the known randomization probabilities used in the design at each
stage, the formulation of (\ref{eq:scorebetaobs}) is valid if the data
(\ref{eq:data}) arise from an observational study, in which case
$\omega_k(\bh_k,a_k)$ are likely unknown; see
Section~\ref{ss:improving}.

The quantity $\Omega(u,d)$ defined in (\ref{eq:Omega}) involves the censoring
survival function $K_c\{u | H(u)\}$, a function of the history $H(u)$.
Thus, in principle, censoring can depend on covariates and treatments
received observed up to time $u$.  As in a well-conducted
conventional, single-stage clinical trial and in
\citet[Section~8.3.3]{TsiatisBook}, censoring in a SMART may be
primarily administrative, in which case it is reasonable to assume
that, in addition to being noninformative in the sense defined above,
censoring is independent of $H(u)$, and write $K_c(u)$.  This
assumption is analogous to that underlying the standard logrank test
for a conventional multi-arm clinical trial, namely, that censoring is
noninformative and independent of baseline covariates and treatment
assignment.  Under this condition, we can take
$w(u,d^j) = K_c(u)$, $j=1,\ldots,D$, which leads to
\vspace*{-0.15in}
\begin{equation}
\Omega(u,d^j) = \frac{C(u,d^j) I(U \geq u) }{\pi(u,d^j) },
  \label{eq:OmegaK}
  \end{equation}
  so that (\ref{eq:scorebetaobs}) and thus $\Tbb$ does not depend on
  the censoring distribution.  In fact, analogous to the usual logrank
  test, (\ref{eq:OmegaK}) holds if censoring depends only on stage 1
  (baseline) treatment assignment, which we write as $K_c(u|a_1)$, and
  we take $w(u,d^j) = K_c(u | a^j)$, where $a^j$ is the stage 1
  treatment assigned by regime $d^j$.  We assume henceforth that
  censoring is independent of $H(u)$ with the possible exception of
  stage 1 treatment and that $w(u,d)$ is specified so that
  (\ref{eq:OmegaK}) holds; more complex dependence on $H(u)$ is
  discussed in Section~\ref{s:webB} of the Appendix.

\begin{figure}
   \centering
\includegraphics[width=12cm]{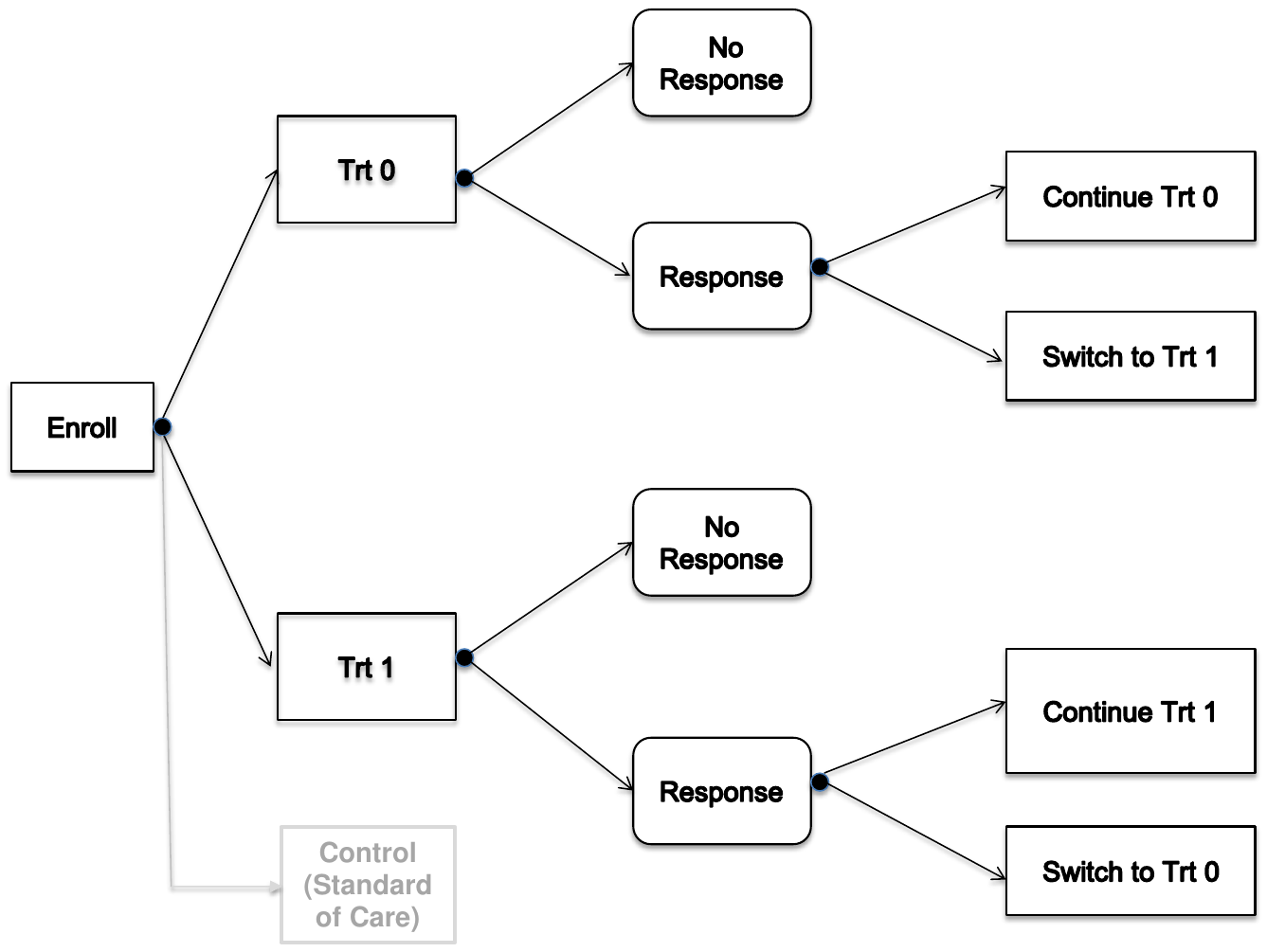}
\caption{\label{f:four} Schematic depicting the design of two possible
  two-stage SMARTs, without and with a control (``standard of care'')
  regime to which subjects can be randomized up front.  In the former
  design, with no control regime, $\calA_1=\{0,1\}$; in the latter,
  $\calA_1 = \{0, 1, $control\}.  In either design, subjects who
  receive stage 1 options Trt 0 or 1 and do not experience the event
  prior to Decision 2 are classified as nonresponders or responders to
  Decision 1 treatment, nonresponders continue stage 1 treatment, and
  responders are randomized to either continue stage 1 treatment or
  switch; thus, for such subjects $\calA_2 = \{0, 1\}$, where 0
  indicates continue and 1 indicates switch.  Solid circles represent
  points of randomization.  The four embedded regimes excluding the
  possible control regime are of the form ``Give Trt $a$ initially; if
  the event does not occur before response status is ascertained and
  nonresponse, continue; else if response, give $b$,'' where the four
  regimes correspond to $(a,b) = (0,0), (0,1), (1,0), (1, 1)$.}
\end{figure}

  In principle, a test statistic for $H_0$ can be constructed as a
  quadratic form in $\Tbb$ depending on a (large-sample) approximation
  to the covariance matrix of $\Tbb$ under $H_0$.  However, an
  obstacle to obtaining an approximate covariance matrix is that
  $\Tfrak^j$ in (\ref{eq:scorebetaobs}) for each $j = 1,\ldots,D-1$
  and thus $\Tbb$ is not a sum of independent and identically
  distributed (iid) mean-zero terms, so that standard asymptotic
  theory does not apply.  Defining
  $d\Nbar_i(u) = \sumjD \Omega_i(u,d^j) dN_i(u)$,
  $\Ybar_i(u) = \sumjD \Omega_i(u,d^j) Y_i(u)$, and
  $\hatq(u,d^j) = \big\{\sumin \Omega_i(u,d^j) Y_i(u)\big\}/\sumin
  \Ybar_i(u)$, we show in Section~\ref{s:webC} of the Appendix that $n^{-1/2}$ times
  (\ref{eq:scorebetaobs}) can be written as
\begin{equation}
  \label{eq:Tdj}
\begin{aligned}
  n^{-1/2} \Tfrak^j = n^{-1/2} \sumin \Tfrak^j_i &+ o_p(1), \hspace*{0.35in}
 \Tfrak^j_i = \left[ \vphantom{\sum^D_{j'=1}} \int^\infty_0  \Omega_i(u,d^j) \{ dN_i(u) - d\Lambda_0(u)Y_i(u)
 \} \right.\\ &- \left. \sum^D_{j'=1} \int^\infty_0  \Omega_i(u,d^{j'}) q(u,d^j) \{
  dN_i(u) - d\Lambda_0(u)Y_i(u) \} \right]
  \end{aligned}
\end{equation}
where $q(u,d)$ is the limit in probability of $\hatq(u,d)$, so that
$n^{-1/2} \Tbb = n^{-1/2} \sumin \Tbb_i + o_p(1)$,
$\Tbb_i = (\Tfrak^1_i,\ldots,\Tfrak^{D-1}_i)^T$, $i=1,\ldots,n$, and thus
$n^{-1/2} \Tbb$ is asymptotically equivalent to $n^{-1/2}$ times a sum
of mean-zero iid
terms, and the asymptotic covariance matrix of $n^{-1/2} \Tbb$ can be
approximated by $\bSigma = n^{-1} \sumin (\Tbb_i \Tbb_i^T)$.  Substituting
$\hatq(u,d^j)$ for $q(u,d^j)$ and $d\hatLam_0(u)$ for $d\Lambda_0(u)$
in (\ref{eq:Tdj}), denoting the resulting approximation to
$\Tfrak^j_i$ as $\widehat{\Tfrak}^j_i$, and letting
$\widehat{\Tbb}_i = (\widehat{\Tfrak}^1_i,\ldots,
\widehat{\Tfrak}^{D-1}_i)^T$, $\bSigma$ can be estimated by
$\hatSig = n^{-1} \sumin (\widehat{\Tbb}_i \widehat{\Tbb}_i^T)$, yielding
the test statistic $\Zbb = n^{-1} \Tbb^T \hatSig^{-} \Tbb$, where
$\hatSig^{-}$ is a generalized inverse of $\hatSig$.  As in 
\citet*{WahedChi}, we use a generalized inverse, as $\bSigma$ and thus
$\hatSig$ may be singular due to a dependency structure in $\bSigma$
induced by some SMART designs and the set of regimes of
interest.  E.g., for the SMART in Figure~\ref{f:eight}, if $\calD$ is
the set of all $D=8$ embedded regimes, the hazards for the four
regimes starting with the same stage 1 treatment involve a linear
dependency, so that $\bSigma$ $(7 \times 7)$ 
has rank 5.  In the SMART in Figure~\ref{f:four}, ignoring
the control, there is no such dependency, so if
$\calD$ is the set of $D=4$ embedded regimes, $\bSigma$ is of full
rank (= 3).  In general, under $H_0$, $\Zbb$ has an approximate
$\chi^2_\nu$ distribution with degrees of freedom
$\nu = \mbox{rank}(\bSigma)$.

In (\ref{eq:scorebetaobs}) and (\ref{eq:Tdj}), integration from 0 to
$\infty$ implies that the range of integration is from 0 to at most
the largest observed event time.  In many studies, the number of
subjects at risk at event times leading up to this time may be small;
moreover, even fewer of these subjects may have treatment experience
consistent with a particular regime.  These features may lead to small
numbers of subjects having undue influence on the test statistic.
Thus, analogous to \citet{KidwellLogrank}, we recommend truncating the
range of integration at a time $L$ equal to or less than the largest
observed event time.  As a rule of thumb, we suggest choosing $L$ so
that 1\% to 4\% of subjects remain at risk at $L$.

\subsection{Improving efficiency}
\label{ss:improving}

In a SMART, the randomization probabilities $\omega_k(\bh_k,a_k)$,
$k=1,\ldots,K$, are known and are treated as known in the inverse
probability weighting via (\ref{eq:OmegaK}) in the foregoing
developments.  However, although counterintuitive, it is well known
\citep[e.g., ][]{TsiatisSemi} that estimating these probabilities can
lead to more efficient inferences.  Accordingly, we propose positing
and fitting parametric models $\omega_k(\bh_k,a_k; \bgamma_k)$, e.g.,
logistic models fitted by maximum likelihood (ML), yielding
ML estimators $\hatgamma_k$.  In a SMART, these models can be chosen
so that the ML estimators for the randomization probabilities are the
sample proportions of subjects randomized to each feasible treatment
given their histories.  E.g., in the SMART in Figure~\ref{f:eight},
for $a_1=$ Trt 0 and response status $r_2 = 1$, the estimator for
the probability of being assigned to Trt 2 at stage 2 is the sample
proportion subjects who received Trt 0, reached stage 2 as responders,
and were assigned to Trt 2.  In an observational study, the unknown
$\omega_k(\bh_k,a_k)$ must be modeled and fitted, typically via
logistic regression/ML.  We discuss modeling and fitting of
$\omega_k(\bh_k,a_k)$ in SMARTs and observational studies in
Section~\ref{s:webB} of the Appendix.

In Section~\ref{s:webD} of the Appendix, following results in \citet{TsiatisSemi},
we present an argument demonstrating that, if the randomization
probabilities in a SMART are modeled and estimated by ML as above and
the fitted models $\omega_k(\bh_k,a_k; \hatgamma_k)$, $k=1,\ldots,K$,
are substituted in place of the known probabilities, as long as
estimation of $\bgamma = (\bgamma_1^T,\ldots,\bgamma_K^T)^T$ is taken
into appropriate account in a modified test statistic, the resulting
test of $H_0$ will be more powerful than that based on $\Zbb$.  In an
observational study, in which the $\omega_k(\bh_k,a_k)$ are unknown and
must be modeled and fitted, the argument also demonstrates how
estimation of $\bgamma$ should be taken into account to lead to a test
that is valid as long as the models are correctly specified.

The form of the modified test statistic follows from Section~\ref{s:webD} of the Appendix
and is given as follows.  Let $\Tbb(\hatgamma)$ be $\Tbb$ with the
models $\omega_k(\bh_k,a_k; \hatgamma_k)$, $k=1,\ldots,K$, fitted by
ML substituted, and let $\bSigma_\gamma$ be the asymptotic covariance
matrix of $\Tbb(\hatgamma)$.  Let $\bS_\gamma(\bgamma)$ be the score
vector associated with $\bgamma$; i.e., the vector of partial
derivatives with respect to $\bgamma$ of the loglikelihood associated
with $\bgamma$ for the posited models. See Section~\ref{s:webD} of the Appendix for examples
of the form of $\bS_\gamma(\bgamma)$.  For $j=1,\ldots,D-1$, let
$\widehat{\Tfrak}^j_i(\hatgamma)$, $i=1,\ldots,n$, denote
$\widehat{\Tfrak}^j_i$ as defined above with the fitted models
$\omega_k(\bh_k,a_k; \hatgamma_k)$, $k=1,\ldots,K$, substituted.  For
each $j=1,\ldots,D-1$, carry out a linear regression of the
$\widehat{\Tfrak}^j_i(\hatgamma)$ on $\bS_{\gamma,i}(\hatgamma)$ and
form for subjects $i = 1,\ldots,n$ the residuals from this fit, which we
denote as $\widehat{\Tfrak}^{j,R}_i(\hatgamma)$, $i=1,\ldots,n$.  Then
we show in Section~\ref{s:webD} of the Appendix that, letting
$\widehat{\Tbb}^R_i(\hatgamma) = \{
\widehat{\Tfrak}^{1,R}_i(\hatgamma), \ldots,
\widehat{\Tfrak}^{D-1,R}_i(\hatgamma)\}^T$, the estimator for
$\bSigma_\gamma$ is given by
$\hatSig_\gamma = n^{-1} \sumin \{ \widehat{\Tbb}^R_i(\hatgamma)
\widehat{\Tbb}^R_i(\hatgamma) ^T\}$, and the modified test statistic
is constructed as 
$\Zbb(\hatgamma) = n^{-1}\widehat{\Tbb}^R(\hatgamma)^T \hatSig^{-}_\gamma
\widehat{\Tbb}^R(\hatgamma)$, where $\widehat{\Tbb}^R(\hatgamma) = 
\sumin \widehat{\Tbb}^R_i(\hatgamma)$.  We also present an argument that,
asymptotically, the test of $H_0$ based on $\Zbb(\hatgamma)$ is at
least as powerful as that based on $\Zbb$.  As above, truncation of
integration at some value $L$ should also be implemented.

In fact, as we discuss in Section~\ref{s:webD} of the Appendix, if there are components of
the history at any decision point that are associated with the
outcome, it is possible to obtain an even more powerful test by
exploiting these associations.  The associated test statistic is
obtained by regressing the $\widehat{\Tfrak}^j_i(\hatgamma)$,
$i=1,\ldots,n$, not only on $\bS_{\gamma,i}(\hatgamma)$ but also on
suitably chosen functions of the histories at each decision point to
obtain the residuals $\widehat{\Tfrak}^{j,R}_i(\hatgamma)$,
$i=1,\ldots,n$. Only components of the history
thought to be strongly associated with the outcome should be
incorporated, as including components that are only weakly associated
could degrade performance if $n$ is not large.  See Section~\ref{s:webD} of the Appendix for
a demonstration and further discussion.  

When $n$ is not large, any of the proposed tests based on $\Zbb$ or
$\Zbb(\hatgamma)$ can be anticonservative; i.e., reject $H_0$ too
often, leading to inflated type I error.  This behavior is due to the
fact that the estimator $\hatSig$ or $\hatSig_\gamma$, as appropriate,
understates the uncertainty in the components of $\Tbb$ or
$\Tbb(\hatgamma)$, which is not uncommon with methods based on
semiparametric theory.  In Section~\ref{s:webE} of the Appendix, we propose a
bias correction to $\hatSig$ and $\hatSig_\gamma$ that, when the
bias-corrected covariance matrix is used to form the test statistic,
yields a test that more closely achieves the nominal level of significance.  In the
simulations in Section~\ref{s:sims}, we demonstrate that such
anticonservatism is ameliorated when the bias correction is used.

\subsection{Comparison to existing methods}
\label{ss:compare}

\citet{KidwellLogrank} and \citet{Li2014} were the first to propose
logrank-type tests for comparing regimes, including ``shared path''
regimes, on the basis of a time-to-event outcome.  The approach of
\citet{KidwellLogrank} is restricted to SMARTs with $K=2$ stages
analogous to that in Figure~\ref{f:four} without the control regime
and focuses on the four embedded regimes. The approach of
\citet{Li2014} allows the data to arise from an observational study
and in principle can be applied to settings with any number of
decision points, although it is formulated and evaluated only in a
two-stage observational setting analogous to the four-regime SMART
considered by \citet{KidwellLogrank} and for regimes of the form,
e.g., with $K=2$, ``Give $a_1$; if response give $a_2$, otherwise
continue $a_1$.''  Of necessity, modeling and fitting of
$\omega_k(\bh_k,a_k)$, $k=1,\ldots,K$, is required. Both approaches do
not support incorporation of covariate information to improve power
and assume that censoring is noninformative and independent
of $H(u)$.

Our motivation for developing the proposed methods is to provide a
unified framework for general $K$ in which comparison of the regimes
in an entirely arbitrary set of regimes of interest within the class
of ``feasible regimes'' given the available data
\citep[Section~6.2.3]{TsiatisBook} can be carried out based on data
from a SMART or observational study and that allows consideration of
censoring mechanisms beyond independent censoring, as might be
expected with observational data.  We focus primarily on SMARTs and
independent censoring in this article.  The ``score vector'' on which
the test statistic of \citet{Li2014} is based is equivalent to our
$\Tbb$ comprising elements in (\ref{eq:scorebetaobs}) with
$w(u,d^j) = K_c(u)$, $j=1,\ldots,D$, as in (\ref{eq:OmegaK}); because
the \citet{Li2014} formulation does not include a weight function
analogous to $w(u,d)$, generalization to censoring mechanisms other
than noninformative, independent censoring as in Section~\ref{s:webB}
of the Appendix is not accommodated.  Regarding the covariance matrix
used to form the test statistic of \citet{Li2014}, the authors
contend, contrary to the results in Section~\ref{ss:improving}, that,
in our notation, estimation of $\bgamma$ in models
$\omega_k(\bh_k,a_k; \bgamma_k)$, $k=1,\ldots,K$, does not have an
effect on the asymptotic distribution of the test statistic, so that
no account need be taken of estimation of $\bgamma$.  In
Section~\ref{s:webD} of the Appendix, we show that there is a
misstatement in the proof that \citet{Li2014} present to justify this
claim.  Thus, their test statistic as proposed, which uses a
covariance matrix that does not take account of this estimation, is
expected to be conservative, as seen in their simulations, where it
rejects $H_0$ at a rate lower than the nominal level for even very
large sample sizes.

The test of \citet{KidwellLogrank} is based on a vector of pairwise
comparisons of the hazards corresponding to three of the embedded
regimes against a fourth reference regime, where known randomization
probabilities are used.  Derivation of the covariance matrix of this
vector used to form the authors' proposed test statistic is based on
counting process martingale theory.  However, as we discuss in
Section~\ref{s:webD} of the Appendix, the components of the vector
need not be martingales with respect to the filtration given by
\citet{KidwellLogrank}.  Thus, although the vector has mean zero and
is asymptotically normal, the authors' covariance formula derived from
this perspective may be biased for the true sampling covariance
matrix.  In simulations, we have observed that the extent to which
this feature may impact inferences is dependent on the data generative
scenario: as shown in Section~\ref{s:sims}, depending on the scenario,
the test may be robust to this feature or may not achieve the nominal
level of significance even for large sample sizes.

\section{Simulation studies}\label{s:sims}

We present results of a suite of simulation studies, each with 5000
Monte Carlo data sets, under several data generative scenarios and
SMART designs with $K=2$.  For each scenario, we evaluate type I error
under $H_0$ in (\ref{eq:null}) and power under alternatives to $H_0$
for level of significance 0.05 using the test statistic of
\cite{KidwellLogrank}, $\Zbb_{\textrm{KW}}$, say, where applicable,
and several versions of the proposed test statistic with estimated
randomization probabilities, $\Zbb(\hatgamma)$, denoted as
$\Zbb_{\textrm{U,nocov}}$ and $\Zbb_{\textrm{C,nocov}}$ without
incorporation of covariates using the uncorrected and bias-corrected
versions of the covariance matrix $\hatSig_\gamma$, respectively; and
as $\Zbb_{\textrm{U,cov}}$ and $\Zbb_{\textrm{C,cov}}$, which do
incorporate covariates.

Scenarios 1 and 2 involve a SMART as in Figure~\ref{f:four} without
the control regime, so with four embedded regimes, and use the
generative process of \citet{KidwellLogrank}, modified to possibly
involve covariates.  We first generated baseline covariate
$X_1 \sim \calN(0,1)$ and stage 1 treatment $A_1$ as Bernoulli(0.5).
We then generated response status $R$ as Bernoulli(0.4); and, if
$R=1$, intermediate covariate $X_2$ as Bernoulli($p_{X_2}$), where,
for
$\btheta_{X_2} = (\theta_{X_2,1}, \theta_{X_2,2},\theta_{X_2,3})^T$,
$p_{X_2} = \expit(\theta_{X_2,1} + \theta_{X_2,2} X_1 + \theta_{X_2,3}
A_1)$ and $\expit(u) = e^u /(1+e^u)$, and $A_2$ as Bernoulli(0.5),
so that $\bX_2 = (R,X_2)^T$.  Then, given
$\btheta = (\theta^{NR}_1, \theta^{NR}_0, \theta^{R}_1, \theta^{R}_0,
\theta^{RE}_{11},
\theta^{RE}_{10},\theta^{RE}_{01},\theta^{RE}_{00})^T$,
$\bdelta^{NR}=(\delta^{NR}_1, \delta^{NR}_0)^T$,
$\bdelta^{R}=(\delta^{R}_1, \delta^{R}_0)^T$,
$\balpha_\ell =
(\alpha_{\ell,11},\alpha_{\ell,10}\alpha_{\ell,01}\alpha_{\ell,00})^T$,
$\ell=1,2$, with
$\lambda^{NR}_r = \theta^{NR}_r \exp(\delta^{NR}_r X_1)$,
$\lambda^R_r = \theta^R_r \exp(\delta^{R}_rX_1)$, $r=0,1$, and
$\lambda^{RE}_{rs} = \theta^{RE}_{rs} \exp\{\alpha_{1,rs} X_1 +
\alpha_{2,rs} (X_2-p_{X_2})\}$, $r,s = 0, 1$, for $R=0$, we generated
potential event times $T^{NR}_r$, $r = 0, 1$, as
exponential($\lambda^{NR}_r$); and, for $R=1$, potential times to
response $T^{R}_r$, $r = 0, 1$, as exponential($\lambda^{R}_r$) and
times from response to event $T^{RE}_{rs}$ as
exponential($\lambda^{RE}_{rs}$), $r, s = 0, 1$.  With
$T^{NR} = A_1 T^{NR}_1+(1-A_1)T^{NR}_0$ for $R=0$ and
$T^R = A_1 \{ A_2 (T^R_1+T^{RE}_{11}) + (1-A_2) (T^R_1+T^{RE}_{10}) \}
+ (1-A_1) \{ A_2 (T^R_0+T^{RE}_{01}) + (1-A_2) (T^R_0+T^{RE}_{00}) \}$
for $R=1$, the event time was $T = R T^R + (1-R) T^{NR}$.  Censoring
time $C$ was generated as uniform($0, c_{\textrm{max}})$, and 
$U = \min(T,C)$, $\Delta=I(T \leq C)$.  If the event time was
censored prior to response, we redefined $R=0$.  Thus, for $R=1$, the
time to Decision 2 $\calT_2 = A_1 T^{R}_1+(1-A_1)T^{R}_0$ and
$\kappa = 2$, and, as nonresponders never reach Decision 2,
$\kappa=1$ for $R=0$.

In Scenario 1, we took $c_{\textrm{max}} = 3.80$, resulting in 
30\% to 40\% censoring.  Under $H_0$, Scenario 1(a) takes
$\btheta = (1/0.91,1/0.91,1/0.5,1/0.5,1,1,1,1)^T$, with 
$\btheta_{X_2}$, $\bdelta^{R}$, $\bdelta^{NR}$, $\balpha_\ell$,
$\ell=1,2$, vectors of zeroes, so that time to event is not associated
with covariates, and thus duplicates the first null hypothesis
scenario of \citet{KidwellLogrank}.  Scenario 1(b) is the same but
with covariate associations induced by taking
$\btheta_{X_2} = (0, 0.15,0)^T$, $\bdelta^{R}=(0.7,0.7)^T$,
$\bdelta^{NR}=(0.3,0.3)^T$, $\balpha_\ell=(0.7,0.7,0.7,0.7)^T$,
$\ell=1,2$.  In Table~\ref{t:results}, under both 1(a) and
1(b), all tests are anti-conservative for $n=250$.  Those based on
$\Zbb_{\textrm{C,nocov}}$, $\Zbb_{\textrm{C,cov}}$, and
$\Zbb_{\textrm{KW}}$ achieve the nominal level for $n = 500, 1000$;
including unimportant covariates as in 1(a)
does not degrade performance.  The tests based on
$\Zbb_{\textrm{U,nocov}}$ and $\Zbb_{\textrm{U,cov}}$ are
anti-conservative, supporting use of the bias correction.  As 
an alternative to $H_0$ under 1(b), we took
$\btheta = (1/0.91,1/1.15,1/0.9,1/0.5,1/2,1/2.33,1/1.11,1/0.67)^T$, with all
other quantities the same.  From Table~\ref{t:results}, for tests that
achieve the nominal level under $H_0$, incorporation of covariates
associated with the event time using the proposed methods yields
increases in power.

As discussed in Section~\ref{s:webF} of the Appendix, the test of \citet{KidwellLogrank} may
be expected to be robust to departure from the martingale property
under Scenario 1.  Under $H_0$, Scenarios 2(a) and 2(b) are the same
as 1(a) and 1(b) except with $c_{\textrm{max}} = 8$,
resulting in 25\% to 30\% censoring, and
$\btheta = (1/0.91,1/0.91,1/0.5,1/0.5,1/3,1/3,1/3,1/3)^T$.  As
Table~\ref{t:results} shows, all tests are anti-conservative with
$n=250$; however, for $n=500, 1000$, the tests based on
$\Zbb_{\textrm{C,nocov}}$ and $\Zbb_{\textrm{C,cov}}$ achieve the
nominal level, while those based on $\Zbb_{\textrm{KW}}$ continue to
be anti-conservative.  In Section~\ref{s:webF} of the Appendix, we speculate that this
behavior may reflect a strong departure from the martingale property
for these Scenarios.  For an alternative to Scenario 2(b), we took
instead $\btheta = (1/0.35,1/0.9,1/0.5,1/0.5,1/3.3,1/3.3,1/3,1/3)^T$; again,
Table~\ref{t:results} shows gains in power when associated covariates are
incorporated in the proposed methods.

Scenario 3 uses a different generative strategy and mimics the design
of Study C9710.  We first generated $\bX_1 = (X_{11},X_{12})^T$, where
$X_{11} \sim \calN(0,1)$ and $X_{12} \sim$ uniform$(0,1)$, and $A_1$
as Bernoulli(0.5).  For given $\alpha_{1D}$, $\alpha_{1SS}$,
$\alpha_{2AL}$, $\psi$, and $\zeta$, we took
$\btheta_{1D} = (\alpha_{1D}, 0.5 \psi,0.5 \psi, -0.26 \zeta)^T$ and
$\btheta_{1SS} = (\alpha_{1SS}, 0.5 \psi,0.5 \psi, 0.24 \zeta)^T$,
generated potential event time $T_D$ as
exponential($\lambda_{1D}$),
$\lambda_{1D} = \exp\{\theta_{1D,1} + \theta_{1D,2} X_{11} +
\theta_{1D,3} (X_{12}-0.5) + \theta_{1D,4} (A_1-0.5)\}$ and potential
time to Decision 2 $T_{SS}$ as
exponential($\lambda_{1SS}$),
$\lambda_{1SS} = \exp\{\theta_{1SS,1} + \theta_{1SS,2} X_{11} +
\theta_{1SS,3} (X_{12}-0.5) + \theta_{1SS,4} (A_1-0.5)\}$, and took
$S = \min(T_D,T_{SS})$, $R = I(T_{SS} < T_D)$.  If $R=1$, 
with $\btheta_{X_2} = (0.2, 0.5\psi, 0.4 \psi, 0.12 \zeta)^T$ and
$\btheta_{2AL} = (\alpha_{2AL}, 0.5 \psi, -0.52 \psi, 0.6 \psi, -0.1
\zeta, -0.11 \zeta)^T$, we generated $X_2$ as
Bernoulli($p_{X_2}$),  $p_{X_2} = \expit(\theta_{X_2,1} + \theta_{X_2,2} X_{11} + \theta_{X_2,3}
X_{12} + \theta_{X_2,4} A_1)$, $A_2$ as Bernoulli(0.5),
and ``added life'' post-response $T_{AL}$ as exponential($\lambda_{2,AL}$), 
$\lambda_{2AL} = \exp\{\theta_{2AL,1} + \theta_{2AL,2} X_{11} +
\theta_{2AL,3} (X_{12}-0.5) + \theta_{2AL,4} (X_2 - p_{X2}) + \theta_{2AL,5}
(A_1-0.5) + \theta_{2AL,6}(A_2-0.5) \}$.  We then took 
$T = (1-R) T_D + R (T_{SS} + T_{AL})$ and, with $C$ 
uniform($0, c_{\textrm{max}})$,  $U = \min(T,C)$, $\Delta=I(T \leq
C)$; if $R=1$ and $C < S$, redefine $R=0$.  For $R=1$, $\calT_2
= T_{SS}$ and $\kappa=2$; else, $\kappa=1$.

Scenarios 3(a) - 3(c)  reflect different possible departures from
the martingale property through specification of $\alpha_{1D}$ and
$\alpha_{2AL}$; the property roughly holds with
$\alpha_{1D} = \alpha_{2AL}$.  In all cases, $\alpha_{1SS} = -4.2$,
$\psi=1.5$, and $\zeta=0$ corresponds to $H_0$ while $\zeta>0$
produces alternatives to $H_0$.  For Scenarios 3(a) and 3(b),
$c_{\textrm{max}} = 500$, resulting in about 40\% to 45\% censoring;
for Scenario 3(c), $c_{\textrm{max}} = 300$ for about 30\%
censoring. For Scenario 3(a), $\alpha_{1D} = \alpha_{2AL} = -5.5$;
from Table~\ref{t:results}, under $H_0$, the tests based on
$\Zbb_{\textrm{C,nocov}}$, $\Zbb_{\textrm{C,cov}}$, and
$\Zbb_{\textrm{KW}}$ all achieve the nominal level, and incorporation
of covariates yields increased power under alternatives.  Under
Scenarios 3(b) and 3(c), $\alpha_{1D} = -4.5$, $\alpha_{1AL} = -5.5$
and $\alpha_{1D} = -5.5$, $\alpha_{1AL} = -3.5$, respectively. While
the tests based on $\Zbb_{\textrm{C,nocov}}$ and $\Zbb_{\textrm{C,cov}}$ achieve the
nominal level under $H_0$ in both cases for all $n$, that based on
$\Zbb_{\textrm{KW}}$ is conservative under 3(b) and anti-conservative
under 3(c), which persists across all $n$, possibly reflecting 
departure from the martingale property; see Section~\ref{s:webF} of the Appendix.

Scenarios 4 and 5 involve data generative strategies similar to those
for Scenario 3; details are in Section~\ref{s:webF} of the Appendix.  Scenario 4 includes an
additional control arm as in Figure~\ref{f:four}, so that
$\calA_1 = \{0, 1, $control\}, with equal randomization to treatments
0, 1 and control (i.e., randomization probability 1/3 for each) at
Decision 1 and randomization probability 0.5 to each treatment option
at Decision 2.  Scenario 5 follows a design like that in
Figure~\ref{f:eight}, with eight embedded regimes and randomization
probabilities of 0.5 for each treatment option in each feasible set at
each decision point.  Thus, in both scenarios, the test of
\citet{KidwellLogrank} is not applicable.  Under $H_0$, the proposed
tests using the bias-corrected version of $\hatSig_\gamma$ achieve the
nominal level in all cases, and under alternatives to $H_0$,
incorporation of covariates associated with the time-to-event outcome
leads to gains in power.

The foregoing results focus on comparison of the hazard rates
associated with all embedded regimes represented in a SMART.  A secondary
analysis of interest in many SMARTs is comparison of specific pairs of
embedded regimes, e.g., the most and least resource intensive or burdensome or
each embedded regime against a control regime in SMARTs like that in
Figure~\ref{f:four}.  In Section~\ref{s:webF} of the Appendix, we report on simulation
studies of the performance of the proposed tests and, where
applicable, that of tests based on the statistic of
\citet{KidwellLogrank}, for comparing the hazards for pairs of
embedded 
regimes, including ``shared path'' regimes.  An additional analysis of
interest in some SMARTs is to compare more complex ``feasible
regimes'' with rules depending on covariate information; we report on
a simulation study using the proposed methods for this purpose in
Section~\ref{s:webF} of the Appendix.

\begin{table}[h]
  \caption{Simulation results under the data generative scenarios
    described in the text under both the null hypothesis $H_0$ in
    (\ref{eq:null}) and alternatives based on 5000 Monte Carlo (MC)
    data sets, with all tests conducted with level of significance
    0.05.  Entries are MC proportions of times the test based on the
    indicated test statistic rejected $H_0$.
    $\Zbb_{\textrm{U,nocov}}$ and $\Zbb_{\textrm{C,nocov}}$ denote the
    proposed generalized logrank statistics $\Zbb(\hatgamma)$ based on
    estimated randomization probabilities but without incorporation of
    covariates using the uncorrected and corrected versions of the
    covariance matrix $\hatSig_\gamma$, respectively;
    $\Zbb_{\textrm{U,cov}}$ and $\Zbb_{\textrm{C,cov}}$ denote the
    same but incorporating all covariates as described in
    Section~\ref{s:webD} of the Appendix using the uncorrected and
    corrected versions of the covariance matrix $\hatSig_\gamma$,
    respectively; and $\Zbb_{\textrm{KW}}$ denotes the test statistic
    proposed by \citet{KidwellLogrank}. Entries under $H_0$ have MC
    standard error of approximately 0.003; those under alternatives
    have MC standard error of about 0.006.  In Scenarios 3-5, $\zeta$
    dictates the alternative; see the text.}
  \label{t:results}
  \centering
  \begin{footnotesize}
    \begin{tabular}{clcccccc} \Hline
\textbf{Scenario} &  $n$ & $\zeta$ &$\Zbb_{\textrm{U,nocov}}$
      & $\Zbb_{\textrm{C,nocov}}$  & $\Zbb_{\textrm{U,cov}}$  &  $\Zbb_{\textrm{C,cov}}$
      & $\Zbb_{\textrm{KW}}$\\
      \hline 
  &  \multicolumn{7}{c}{\textbf{Null Scenarios}} \\*[0.06in]   

1(a) & 250 & -- & 0.076 & 0.063 & 0.074 & 0.062 & 0.056\\ 
       & 500 & -- & 0.060 & 0.053 & 0.058 & 0.052 & 0.053\\
       & 1000 & -- & 0.054 & 0.051 & 0.054 & 0.051 & 0.053 \\*[0.06in]   

1(b) & 250 & -- & 0.071 & 0.061 & 0.073 & 0.059 & 0.059\\
       & 500 & -- & 0.060 & 0.054 & 0.057 & 0.050 & 0.054\\ 
     & 1000 & -- & 0.056 & 0.053 & 0.053 & 0.050 & 0.055\\*[0.06in]

 2(a) & 250 & -- & 0.075  &  0.060 & 0.081 & 0.066 & 0.081 \\
        & 500 & -- & 0.060  &  0.053 & 0.061 & 0.053 & 0.081 \\
        & 1000 & -- & 0.055 & 0.051 & 0.056 & 0.053 & 0.081\\*[0.06in]

2(b) & 250 & -- & 0.071  &  0.060 & 0.074 & 0.063 & 0.070 \\
       & 500 & -- & 0.061  &  0.054 & 0.059 & 0.051 & 0.072 \\
       & 1000 &-- & 0.054 & 0.050 & 0.054 & 0.051 & 0.065\\*[0.06in]

  3(a) & 250 & 0.00 & 0.061 & 0.051 & 0.057 & 0.049 & 0.052\\
         & 500 & 0.00 & 0.054 & 0.049 & 0.053 & 0.050 & 0.049\\
         & 1000 & 0.00 & 0.051 & 0.048 & 0.050 & 0.048 & 0.050\\*[0.06in]     
      
3(b) & 250 & 0.00 & 0.057 & 0.050 & 0.060 & 0.050 & 0.042 \\
       & 500 & 0.00 & 0.053 & 0.048 & 0.052 & 0.049 & 0.040 \\
       & 1000 & 0.00 & 0.054 & 0.050 & 0.052 & 0.049 & 0.044 \\*[0.06in]
      
3(c) & 250 & 0.00 & 0.063 & 0.051 & 0.064 & 0.053 & 0.074 \\
       & 500 &  0.00 & 0.056 & 0.053 & 0.057 & 0.051 & 0.079 \\
       & 1000 & 0.00 & 0.050 & 0.048 & 0.057 & 0.053 & 0.078 \\*[0.06in]
      
4 & 375 & 0.00 & 0.058 & 0.050 & 0.062 & 0.055 & --\\
   & 750 & 0.00 & 0.055 & 0.052 & 0.056 & 0.052 & --\\
   & 1500 & 0.00 & 0.050 & 0.049 & 0.051 & 0.049 & --    \\*[0.06in]  

5  &  500 & 0.00 & 0.055 & 0.048 & 0.055 & 0.049 & -- \\
    & 1000 & 0.00 & 0.056 & 0.051 & 0.054 & 0.050 & -- \\*[0.1in] 

                  & \multicolumn{7}{c}{\textbf{Alternative Scenarios}} \\*[0.06in]

1(b) & 500 & -- & 0.534 & 0.487 & 0.573  & 0.558 & 0.468 \\    
       & 1000 & -- & 0.813 & 0.807 & 0.891 & 0.887 & 0.784\\*[0.06in]

2(b) & 500 & -- & 0.455 & 0.438 & 0.534 & 0.515 &  0.451\\    
       & 1000 & -- & 0.769 & 0.761 & 0.842 & 0.836 & 0.766\\*[0.06in]

3(a) & 250 & 3.50 & 0.726 & 0.706 & 0.827 & 0.807 & 0.720\\
       & 500 & 2.50 & 0.719 & 0.707 & 0.821 & 0.812 &  0.727 \\
      & 1000 & 1.75 & 0.699 & 0.694 & 0.808 & 0.802 & 0.709\\*[0.06in] 
      
3(b) & 250 & 2.25 & 0.731 & 0.711 & 0.813 & 0.794 & 0.709 \\
       & 500 & 1.65 & 0.749 & 0.736 & 0.834 & 0.826 & 0.730 \\
      & 1000 & 1.25 & 0.817 & 0.812 & 0.880 & 0.875 & 0.805 \\*[0.06in]

3(c) & 250 & 7.00 & 0.687 & 0.656 & 0.808 & 0.784 & 0.559 \\
      & 500 & 5.25 & 0.734 & 0.717 & 0.857 & 0.846 & 0.561 \\
      & 1000 & 3.50 & 0.664 & 0.656 & 0.812 & 0.808 & 0.466 \\*[0.06in]                                           

4 & 375 & 2.00 &  0.713 & 0.701 & 0.786 & 0.774 & --\\
   & 750 & 1.50 &  0.774 & 0.767 & 0.852 & 0.848 & --\\
  & 1500 &1.00  & 0.728 & 0.724 & 0.823 & 0.818 & --\\*[0.06in]  

 5 & 500 & 4.00 & 0.742 & 0.729 & 0.874 & 0.862 & -- \\
    & 1000 & 3.00 & 0.695 & 0.687 & 0.847 & 0.842 & -- \\
    \hline
    \end{tabular}
    \end{footnotesize}
\end{table}

\section{Application to Study C9710}\label{s:example}

As in Section~\ref{s:intro}, Study C9710 is a SMART with $K=2$
decision points.  At Decision 1, $\calA_1 = \{0, 1\}$, where 0 (1)
corresponds to ATRA (ATRA + arsenic trioxide) consolidation therapy;
at Decision 2, $\calA_2 = \{0, 1\}$, where 0 (1) corresponds to ATRA
(ATRA+Mtx+MP) maintenance therapy.  Only subjects who did not
experience the event or for whom the event was not censored before the
end of consolidation, referred to as ``responders,'' with $\kappa=2$,
were randomized to maintenance therapy.  Section~\ref{s:webG} of the Appendix provides
details of the data, which involve $n=467$ subjects, of which 310 were
responders ($\kappa=2$) and 157 experienced the event of censoring
before completing consolidation ($\kappa=1$).  Baseline covariate
$\bX_1$ comprises ten variables, and the additional information
collected between study entry and stage 2, $\bX_2$, includes 46
variables indicating adverse events arising during induction and/or
consolidation therapy; see Section~\ref{s:webG} of the Appendix.  Equal randomization was
used at both decisions, i.e., $\omega_1(\bh_1,a_1) = 1/2$ for all
$\bh_1$ and $a_1 \in \calA_1$, and $\omega_2(\bh_2,a_2) = 1/2$ for all
$\bh_2$, $a_2 \in \calA_2$, and $\kappa=2$.  As in
Section~\ref{ss:improving} and shown in Section~\ref{s:webG} of the Appendix, we specified
models $\omega_k(\bh_k,a_k; \bgamma_k)$, $k=1, 2$, and fitted these by
ML to obtain estimated randomization probabilities.  The maximum
follow-up time is 4694 days, with maximum time to EFS of 4000; in all
analyses, we took $L=4200$, resulting in about 2\% truncation.

\begin{figure}
   \centering
\includegraphics[width=12cm]{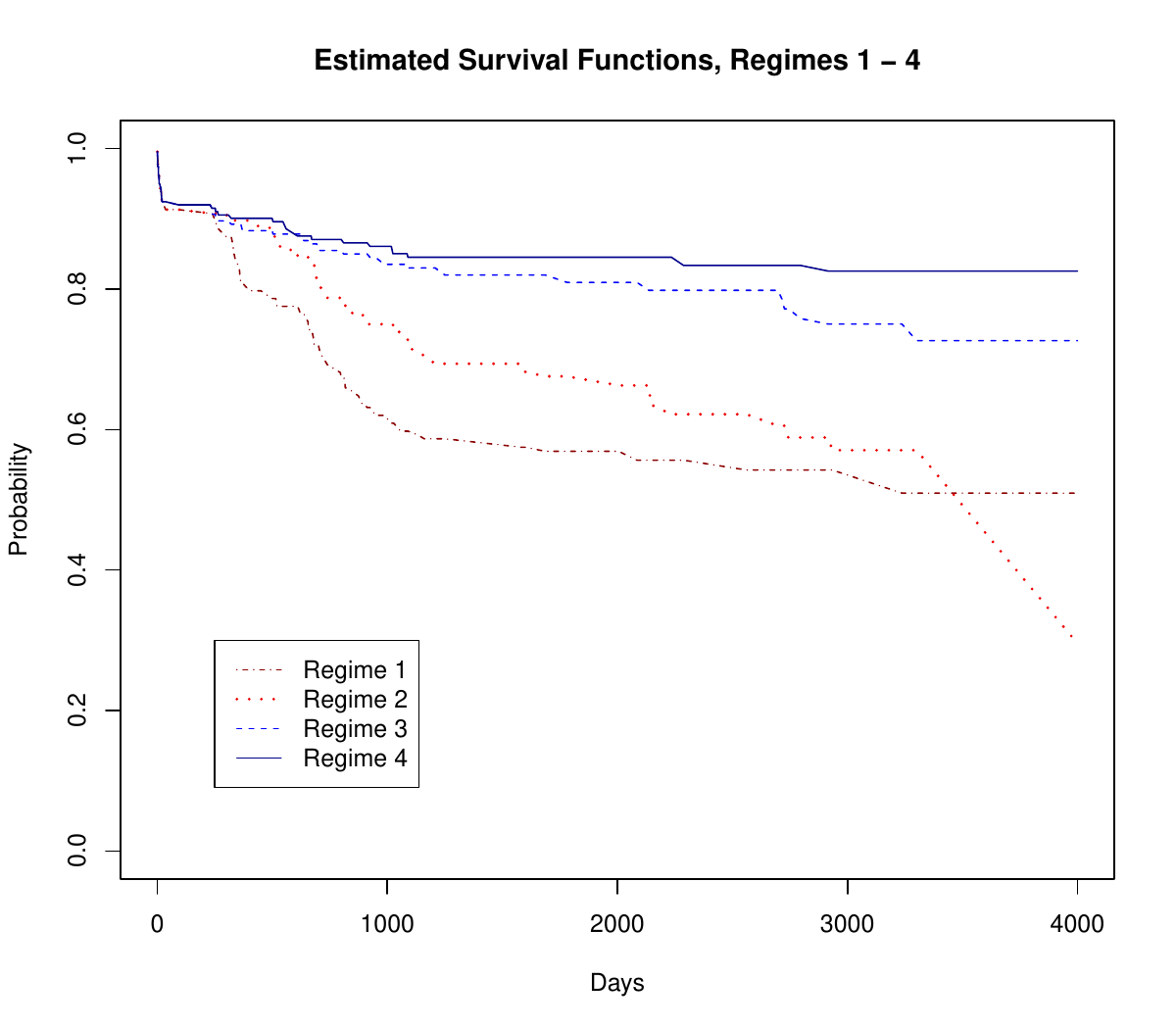}
\caption{\label{f:survcurves}  Estimates of the survival distributions
  $\calS(u,d^j)$ corresponding to embedded Regimes $j=1,\ldots,4$,
 shown with dash-dotted, dotted, dashed, and solid lines, respectively.
As in the text, the regimes are of the form ``Give consolidation
therapy $a$; if subject completes consolidation before the event
occurs, give maintenance therapy $b$,'' where Regimes 1, 2, 3, 4 take 
$(a, b) = (0,0), (0,1), (1,0), (1,1)$, respectively.}
\end{figure}

As in Figure~\ref{f:c9710}, the study involves four embedded regimes
of the form ``Give consolidation therapy $a$; if subject completes
consolidation before the event occurs, give maintenance therapy $b$,''
which we denote as Regimes 1, 2, 3, and 4 as
$(a, b) = (0,0), (0,1), (1,0), (1,1)$.  To visualize the survival
functions $\calS(u,d^j) = P\{ \Ts(d^j) \geq u\}$ corresponding to
Regimes $j = 1,\ldots,4$, we estimated the cumulative hazard functions
$\Lambda(u,d^j) = \int^\infty_0 \lambda(u,d^j) du$ by
$\widehat{\Lambda}(u,d^j) = \int^\infty_0 \big\{\sumin \Omega_i(u,d^j)
dN_i(u)/\sumin \Omega_i(u,d^j) Y_i(u)\big\}$, with estimated
randomization probabilities substituted, obtaining estimates
$\widehat{\calS}(u,d^j) = \exp\{-\widehat{\Lambda}(u,d^j)\}$.
Figure~\ref{f:survcurves} depicts $\widehat{\calS}(u,d^j)$,
$j=1,\ldots,4$, and suggests that administering ATRA + arsenic
trioxide ($a_1=1$) initially, as in Regimes 3 and 4, is more
beneficial than administering ATRA alone as in Regimes 1 and 2,
consistent with the findings of \citet{Powell2010}.  The estimates
suggest further that, of the regimes starting with ATRA + arsenic
trioxide, Regime 4, which gives ATRA+Mtx+MP maintenance to responders,
may yield benefit over Regime 3.

Table~\ref{t:c9710pairs} shows the results of tests of $H_0$ in
(\ref{eq:null}) corresponding to different choices of the set $\calD$
of regimes of interest.  We tested $H_0$ several ways: using (i) the
proposed test statistic $\Zbb(\hatgamma)$ with estimated randomization
probabilities and using the bias correction to $\hatSig_\gamma$,
without incorporation of covariates, denoted $\Zbb_{\textrm{C,nocov}}$
as in Table~\ref{t:results}; (ii) same as (i) but incorporating
selected baseline covariates in $\bX_1$, namely, logarithm of white
blood cell count and ECOG performance status, and intermediate adverse
events in $\bX_2$, nausea and hemorrhage/bleeding (see
Section~\ref{s:webG} of the Appendix), denoted $\Zbb_{\textrm{C,cov}}$; and (iii) the test statistic of
\citet{KidwellLogrank}, denoted $\Zbb_{\textrm{KW}}$ (which uses known
randomization probabilities).  P-values for each test were obtained
from the appropriate $\chi^2$ distribution.  With
$\calD = \{d^1, d^2, d^3, d^4\}$, $D=4$, the set of all embedded
regimes, all tests strongly reject $H_0$, consistent with the evidence
in Figure~\ref{f:survcurves}.  We also took $\calD$ with $D=2$ to
obtain pairwise comparisons; because the components of the test
statistic of \citet{KidwellLogrank} address pairwise comparisons (see
Section~\ref{s:webF} of the Appendix), $\Zbb_{\textrm{KW}}$ in Table~\ref{t:c9710pairs}
represents tests based on (linear combinations of) the relevant
components.  In all cases, the proposed test statistics incorporating
covariates yield the largest value, consistent with the expected
increase in power.  For the key pairwise comparison of ``shared path''
regimes 3 and 4 starting with ATRA + arsenic trioxide, the proposed
test statistics yield evidence of a difference (without adjustment for
multiple comparisons) supporting further investigation of the use of
ATRA+Mtx+MP as maintenance following this consolidation therapy.

\begin{table}[ht]\caption{Test statistics and corresponding p-values
    (in parentheses) for comparisons of embedded regimes as indicated
    in $\calD$ in North American Leukemia Intergroup Study C9710.
    $\Zbb_{\textrm{C,nocov}}$ and $\Zbb_{\textrm{C,cov}}$ are the
    proposed test statistic without and with incorporation of
    covariates, respectively, and using the bias-corrected covariance
    matris; and $\Zbb_{\textrm{KW}}$ is the appropriate test statistic
    based on the approach of \citet{KidwellLogrank}. For all test
    statistics, p-values were obtained from the $\chi^2$ distribution
    with 3 or 1 degree(s) of freedom as appropriate.}
\small \centering
\begin{tabular}{cccc}
  \Hline
$\calD$ & $\Zbb_{\textrm{C,nocov}}$ & $\Zbb_{\textrm{C,cov}}$ &   $\Zbb_{\textrm{KW}}$ \\
\hline
  $\{d^1, d^2, d^3, d^4\}$ & 25.990 ($<0.0001$) & 28.817  ($<0.0001$) & 23.425 ($<0.0001$)\\*[0.05in]
  $\{d^3,d^4\}$ & 3.671 (0.055) & 4.241 (0.040) & 1.344 (0.246) \\*[0.02in]

$\{d^1,d^2\}$ & 1.135 (0.287) & 2.220 (0.138) & 0.832 (0.362)\\*[0.02in]

  $\{d^2,d^4\}$ & 14.638 (0.0001) & 14.605 (0.0001) & 13.277 (0.0003) \\*[0.02in]

  $\{d^1,d^4\}$ & 20.109 ($<0.0001$) & 24.717 ($<0.0001$) & 21.154 ($<0.0001$)\\*[0.02in]
\hline
\end{tabular}
\label{t:c9710pairs}
\end{table}

\section{Discussion}\label{s:discuss}

We have proposed a general statistical framework in which we have
developed a logrank-type test that can be used to compare the survival
distributions of a time-to-event outcome were the patient population
to receive treatment according to each treatment regime within a
specified set of regimes based on the data from a SMART with an
arbitrary number of stages.  The test statistic can be modified to
incorporate covariate information to enhance efficiency, and the test
can also be conducted based on data from a longitudinal observational
study, although we have focused on its use in SMARTs here.  The set of
regimes of interest can be the set of regimes embedded in a SMART, a
subset of these regimes, or an arbitrary set of regimes, where each
regime in the set is ``feasible'' given the available data in the
sense discussed in \citet[Section~6.2.3]{TsiatisBook}.  The test is
valid under assumptions analogous to those underlying the standard
logrank test for comparing single-stage treatments.   Our simulations
demonstrate that the test procedure achieves the nominal level under
null hypotheses and that incorporation of covariates associated with
the outcome leads to increases in power.

\vspace*{0.1in}

\begin{center}
  \textbf{\Large Appendix}
  \end{center}

\vspace*{-0.5in}
  
\setcounter{section}{0}
 \renewcommand{\thesection}{\Alph{section}}

\setcounter{equation}{0}
\renewcommand{\theequation}{A.\arabic{equation}}
  
\section{Representation in Terms of Observed Data}
\label{s:webA}

We wish to show that, under $H_0$ and the identifiability assumptions
in Section~\ref{ss:test}, for $u \geq 0$, an inverse
probability weighted observed data analog to (\ref{eq:scoreLam}), that is,
\begin{equation}
\sumin \sum^D_{j=1} w(u,d^j)\{d\Ns_i(u,d^j)-d\Lambda_0(u)\Ys_i(u,d^j)\} = 0,
\label{eq:scorepot1}
\end{equation}
is given by (\ref{eq:Omega}),
\begin{equation}
\sumin \sum^D_{j=1} \Omega_i(u,d^j)\{ dN_i(u) -d\Lambda_0(u) Y_i(u)\} =0,\,\,\,\,\,
\Omega(u,d) = \frac{C(u,d) I(U \geq u) }{\pi(u,d) K_c\{u | H(u)\}}w(u,d),
\label{eq:scoreobs1}
\end{equation}
and that the ``score equation''  for $\beta_j$ in terms of potential
outcomes,
\begin{equation}
\sumin \int^\infty_0 \, w(u, d^j) \{d\Ns_i(u,d^j) - d\Lambda_0(u)
\Ys_i(u,d^j)\} = 0,
\label{eq:scorepot2}
\end{equation}
has observed data analog
\begin{equation}
\sumin \int^\infty_0 \, \Omega_i(u, d^j) \{dN_i(u) - d\Lambda_0(u)
Y_i(u)\} = 0
\label{eq:scoreobs2}
\end{equation}
for $j = 1,\ldots,D-1$.  Multiplying
(\ref{eq:scorepot1})-(\ref{eq:scoreobs2}) by $n^{-1}$, it suffices to
show that the limit in probability of (\ref{eq:scorepot1}) is the same
as that for (\ref{eq:scoreobs1}), and similarly for
(\ref{eq:scorepot2}) and (\ref{eq:scoreobs2}).  Because the weights
$w(u,d^j)$, $j=1,\ldots,D-1$, are fixed constants for each $u$ and
$d^j$, we disregard them and thus show that, similar to
\citet[Section~8.4.3]{TsiatisBook} and using the definition of
$\Omega(u,d)$ in (\ref{eq:scoreobs1}), for regime $d$,
$$E\left[\frac{C(u,d)I(U \geq u)}{\pi(u,d) K_c\{u | H(u)\} } \{dN(u) -
    d\Lambda_0(u) Y(u)\} \right] = E\{ d\Ns(u,d) - d\Lambda_0(u) \Ys(u,d)\}.$$
Clearly, it suffices to show that 
  \begin{equation}
E\left[\left.\frac{C(u,d)I(U \geq u)}{\pi(u,d) K_c\{u | H(u)\} }\{dN(u) -
    d\Lambda_0(u) Y(u)\} \right| \calW^*\right] = d\Ns(u,d) - d\Lambda_0(u) \Ys(u,d).
\label{eq:potobs}
 \end{equation}
The integrand on the left hand side of (\ref{eq:potobs}) can be
written as
$$\frac{C(u,d)I(U \geq u)}{\pi(u,d) K_c\{u | H(u)\} } I(U \geq u) \{dN(u) -
d\Lambda_0(u) Y(u)\}.$$
By the consistency assumption, this expression can then be written as
\begin{align*}
\frac{C(u,d)I(U \geq u)}{\pi(u,d) K_c\{u | H(u)\} } & \{d\Ns(u,d) -
                                                      d\Lambda_0(u) \Ys(u,d)\} \\
 &= \frac{C(u,d)I(U \geq u)}{\pi(u,d) K_c\{u | H(u)\} }  I\{\Ts(d)
   \geq u\} \{d\Ns(u,d) -  d\Lambda_0(u) \Ys(u,d)\},
  \end{align*}
so that the left hand side of (\ref{eq:potobs}) becomes
\begin{equation}
\{d\Ns(u,d) - d\Lambda_0(u) \Ys(u,d)\} E\left[ \left. \frac{C(u,d)I(U
      \geq u)}{\pi(u,d) K_c\{u | H(u)\} } \right| \calW^*, \Ts(d)
  \geq u \right].
\label{eq:potobs2}
\end{equation}
Thus, from (\ref{eq:potobs2}), we need to show that
\begin{equation}
E \left[ \left. \frac{C(u,d)I(U \geq u)}{\pi(u,d) K_c\{u | H(u)\} } \right| \calW^*, \Ts(d)
  \geq u \right] = 1, \hspace{0.1in} u \geq 0.
\label{eq:prove}
\end{equation}

We argue heuristically that (\ref{eq:prove}) holds using a discrete
time approximation and induction.  Assume that, rather than occurring
at any point in continuous time, decision points can occur only at
equally-spaced time points; e.g., at any whole second, but not at
fractions of a second.  Partition the interval $[0,u]$ into an
equally-spaced grid with grid points $v_m$, $m=0,\ldots,M$, where
$v_0 = 0$ and $v_M = u$, such that any decision points occurring in
the interval $[0,u]$ fall on the grid, and intervals between grid
points are of length $\delta v = u/M$.  Define $A(v_m)$ to be the
treatment option that is received at time $v_m$ from among the
feasible options at $v_m$.  If $v_m$ corresponds to one of the times
$\calT_1,\ldots, \calT_\kappa$ at which decision points are reached, then
$A(v_m)$ is the treatment assigned at $v_m$; if $v_m$ does not
correspond to one these times, then $A(v_m)$ is equal to the sole
option ``do nothing.''  Let $\Hm(v_m)$ to be the history through time
$v_m$ up to but not including $A(v_m)$; i.e., $\Hm(v_m)$ comprises all
prior covariate information, treatments, and times of decision points
up to and including $v_m$ that would be used to determine treatment if
$v_m$ were a decision point.  Thus, with $H(u)$ defined as in the main
paper, $H(v_m) = \{ \Hm(v_m), A(v_m)\}$.  Then, for the regime $d$ of
interest, define $d\{ \Hm(v_m)\}$ to be the ``rule'' that assigns
treatment at time $v_m$ based on the history $\Hm(v_m)$ at time $v_m$;
if $v_m$ corresponds to one of these times, $d\{ \Hm(v_m)\}$ assigns
the treatment indicated by the relevant rule in $d$ given the history;
if not, $d\{\Hm(v_m)\}$ assigns ``do nothing.''  Define the following
quantities:
\begin{eqnarray}
C^M(u,d) & = & \prodmM I\big[ A(v_m) = d\{\Hm(v_m)\} \big] \nonumber\\
\pi^M(u,d) & = & \prodmM P\big[ A(v_m) = d\{\Hm(v_m)\} | \Hm(v_m), U \geq v_m \big] \nonumber\\
K^M_c\{ u | H(u) \} & = & \prodmM P \{ U \geq v_{m+1} | \Hm(v_m),
                          A(v_m), U \geq v_m, \Ts(d) \geq u\} \\
&  = & \prodmM P\{ U \geq v_{m+1} | H(v_m), U \geq v_m, \Ts(d) \geq
       u\}   \label{eq:two4} \\
I^M(u) & = & \prodmM I(U \geq v_{m+1}). \label{eq:two5}
   \end{eqnarray} 
In (\ref{eq:two4}) and (\ref{eq:two5}), $v_{M+1}$ is the next time
point outside of the interval $[0,u]$, i.e., $u + \delta u = u(1 +1/M)$.   Evidently,
$C^M(u,d) = C(u,d)$ and $\pi^M(u,d) = \pi(u,d)$. In 
(\ref{eq:two4}), when $\Ts(d) \geq u$,
\begin{align}
P\{ U \geq v_{m+1} | & U \geq v_m,  H(v_m), \Ts(d) \geq u\} \nonumber\\
&= 1 - P\{ v_m \leq U < v_{m+1} | U \geq v_m,  H(v_m), \Ts(d) \geq
  u\}.
\label{eq:three1}
  \end{align}
Because $v_m \leq u$, $m=0, \ldots, M$ and $\Ts(d) \geq u,$ the only way that $U$ can
fall between $v_m$ and $v_{m+1}$ for $m=0,\ldots, M-1$ is if an
individual is censored, in which case (\ref{eq:three1}) becomes
\begin{equation}
1 - P\{ v_m \leq U < v_{m+1}, \Delta=0 | U \geq v_m,  H(v_m), \Ts(d)
\geq   u\}.
\label{eq:four}
\end{equation}
For $m=M$, the probability in (\ref{eq:three1})  converges to 0 as $M \rightarrow \infty$.
Thus, with the assumption of noninformative censoring, (\ref{eq:four})
is approximately equal to $1 - \lambda_c\{ v_m | H(v_m)\} \delta v$, so that
$$K_c^M\{u | H(u)\} \approx \exp\left[ -\int^u_0 \lambda_c\{ v |
  H(v)\} \, dv\right] = K_c\{ u | H(u)\}$$
for large $M$.  Likewise, if $U$ is continuous, $I^M(u)$ converges
almost surely to $I(U \geq u)$ as $M \rightarrow \infty$.  

Now define the sequence of random variables
$$\calZ^M = \frac{ C^M(u,d) I^M(u)}{\pi^M(u,d) K^M_c\{u | H(u)\}
}.$$  From
the considerations above, under reasonable regularity conditions,
$\calZ^M$ converges almost surely as $M \rightarrow \infty$ to
$$ \frac{ C(u,d) I(U \geq u)}{\pi(u,d) K_c\{u | H(u)\}}.$$
Thus, if we can show that
\begin{equation}
E\left\{ \calZ^M |\calW^*, \Ts(d)  \geq u \right\}  = 1 \,\,\,
\mbox{a.s.},
\label{eq:proveM}
\end{equation}
then under regularity conditions the desired result  (\ref{eq:prove}) holds.

We demonstrate (\ref{eq:proveM}) by induction.  Specifically,
define the partial products 
$$\calZ^M(t) = \prod^t_{m=0} \frac{ I\big[ A(v_m)=d\{\Hm(v_m)\}\big] I(U
  \geq v_{m+1}) }{ P\big[ A(v_m)=d\{\Hm(v_m)\} | \Hm(v_m), U \geq v_m \big]
  P\{  U \geq v_{m+1} |  U \geq v_m, H(v_m), \Ts(d) \geq u\} }$$
for $t = 0, \ldots,M$, so that $\calZ^M(M) = \calZ^M$
and
$$\calZ^M(0) = \frac{ I\{A_1 = d_1(\bH_1)\} I(U \geq v_1)}{P\{A_1 =
  d_1(\bH_1) | \bH_1\} P\{ U \geq v_1 | \bH_1, A_1, \Ts(d) \geq u\}
}.$$ Assume that all probabilities in these expressions are positive.
Using the SRA and the noninformative censoring assumption
\begin{align}
\hspace*{-1in} E&\{ \calZ^M(0) |  \calW^*, \Ts(d) \geq u\}  = E\big[ E\{ \calZ^M(0) |
                                              \bH_1, A_1, \calW^*, \Ts(d) \geq u\} |  \calW^*, \Ts(d) \geq u\big] \nonumber\\
&=E\left( \left. \frac{ I\{A_1 = d_1(\bH_1)\}}{P\{A_1 = d_1(\bH_1) |  \bH_1\}   } \right.\right. \nonumber
  \\
&\hspace*{0.3in} \times \left. \left. E\left[ \left. \frac{I(U \geq v_1)}{P\{  U \geq v_1 | \bH_1,
  A_1, \Ts(d) \geq u\} } \right| \bH_1, A_1, \calW^*, \Ts(d) \geq u
  \right] \right|  \calW^*, \Ts(d) \geq u \right) \nonumber \\
 &= E\left[ \left. \frac{I\{A_1 = d_1(\bH_1) \}}{P\{A_1 = d_1(\bH_1) |
                                                                      \bH_1\}}
   \right| \calW^*, \Ts(d) \geq u\right] \label{eq:Zzero}\\
 &=E\left( \left. E\left[ \left. \frac{I\{A_1 = d_1(\bH_1) \}}{P\{A_1 = d_1(\bH_1) |   \bH_1\}}
   \right| \bH_1, \calW^*, \Ts(d) \geq u\right] \right|\calW^*, \Ts(d)
   \geq u\right) \nonumber\\
 & = 1 \hspace*{0.1in} \mbox{a.s.} \nonumber
\end{align}
Thus, if we can show that
\begin{equation}
E\{ \calZ^M(t) | \calW^*, \Ts(d) \geq u\} = E\{ \calZ^M(t-1) |
\calW^*, \Ts(d) \geq u\} \hspace*{0.1in} \mbox{a.s.},
\label{eq:five1}
  \end{equation}
  (\ref{eq:proveM}) holds by induction.

  To show (\ref{eq:five1}), note that
  $$E\{ \calZ^M(t) | \calW^*, \Ts(d) \geq u\} = E\big[ E\{ \calZ^M(t)
  | \Hm(v_t), U \geq v_t, \calW^*, \Ts(d) \geq u\} | \calW^*, \Ts(d)
  \geq u \big] \hspace*{0.1in} \mbox{a.s.}$$
  The inner expectation is equal almost surely to 
  $$\calZ^M(t) = \calZ^M(t-1)  \left\{ \frac{ E \big( I\big[ A(v_t) = d\{\Hm(v_t)\}
    \big] I(U \geq v_{t+1}) | \Hm(v_t), U \geq v_t, \calW^*, \Ts(d)
    \geq u\big) }{P\big[ A(v_t)=d\{\Hm(v_t)\} | \Hm(v_t), U \geq v_t \big]
  P\{  U \geq v_{t+1} |  U \geq v_t, H(v_t), \Ts(d) \geq u\} } \right\}$$
Then using the SRA and the noninformative censoring assumption, by an
argument analogous to that in (\ref{eq:Zzero}), the term in large
braces is equal to 1 almost surely, and the result follows.

\setcounter{equation}{0}
\renewcommand{\theequation}{B.\arabic{equation}}

\section{More Complex Censoring and
  Modeling/Estimation of  $\omega_k(\bh_k,a_k)$}
\label{s:webB}

\noindent
\textbf{B.1.  Dependence of censoring on history} \\*[-0.1in]

We adopt the assumption of noninformative censoring and further assume
that censoring is independent of $H(u)$, or, if dependent on $H(u)$,
depends on $H(u)$ only through stage 1 (baseline) treatment
assignment.  As we note, these assumptions are analogous to those that
underlie the standard logrank test for a conventional multi-arm
clinical trial.

In some settings, the analyst may believe that censoring, although
noninformative as defined here (so depending only on $H(u)$ and not on
potential outcomes), depends on $H(u)$ in a more complex fashion.  For
example, censoring at $u$ may be associated with baseline covariates,
covariates that evolve over time, or treatments at stage 2 and higher
contained in $H(u)$.  The formulation of the general testing procedure
outlined in Section~\ref{ss:test} allows for this possibility.
Namely, instead of adopting the simplification (\ref{eq:OmegaK}) of
(\ref{eq:Omega}) and taking $w(u,d^j) = K_c(u)$,
$j=1,\ldots,D$, one can maintain the more general expression for
$\Omega(u,d)$ as in (\ref{eq:Omega}) and adopt a model for the
censoring survival function $K_c\{u | H(u)\}$.  The model should be
chosen to reflect beliefs about the censoring process; see
\citet[Section~8.3.3, pages 381--382]{TsiatisBook} for discussion of
considerations for developing and fitting such models.  Specifically,
one can adopt an appropriate proportional hazards model and fit the
model using the ``data'' $\{H_i(U_i), U_i, 1-\Delta_i\}$,
$i = 1,\ldots,n$, where $H(U_i)$ is the history to $U_i$ for subject
$i$.

Two issues must be addressed under these conditions.  One is that
account must be taken of the fact that $K_c\{u | H(u)\}$ has been
modeled and fitted in obtaining the asymptotic covariance matrix of
$\Tbb$; see Section~B.2 below.  In addition, the weight functions
$w(u,d^j)$, $j = 1,\ldots,D$, must be specified.  One choice is to set
these identically equal to 1.  A possibly better alternative is to
choose these in the spirit of stabilizing weights as in, for example,
\citet{Vaccine}; namely, take $w(u, d^j)$ to be the fitted censoring
model evaluated at the averages of components of the history.  

Note that, if one believes that censoring does not depend on $H(u)$ or
depends on $H(u)$ only through stage 1 treatment, one could estimate
$K_c(u)$ (possibly separately by stage 1 treatment as dictated by
$d^j$) using the Kaplan-Meier estimator; however, there is no
advantage over the approach we propose, as the weights can be
chosen to be equal
to the corresponding estimator, in which case $\Omega(u,d^j)$ reduces to (\ref{eq:OmegaK}). \\

\noindent
\textbf{B.2. Modeling and estimation of $\omega_k(\bh_k,a_k)$} \\*[-0.15in]

In Section~\ref{ss:test}, the developments leading
to the test statistic $\Zbb = n^{-1} \Tbb^T \hatSig^{-} \Tbb$ take the
probabilities of treatment assignment
$\omega_k(\bh_k,a_k) = P(A_k=a_k|\bH_k=\bh_k,\kappa \geq k)$,
$k=1,\ldots,K$, to be known functions of $\bh_k$ and $a_k$.  Thus, in
particular, the form of the asymptotic covariance matrix $\bSigma$ of
$\Tbb$ and thus the estimator $\hatSig$ is derived under the condition
that these probabilities are known, so that the test should achieve
the chosen nominal level of significance under $H_0$.  When the
observed data $\calO_i$, $i=1,\ldots,n$, are from a SMART, these
probabilities are the randomization probabilities, which of course are
known by design.  When the data are from an observational study, these
probabilities are unknown; thus, they must be modeled and estimated.
As discussed in detail in Section~\ref{s:webD},
estimating these probabilities by fitting an appropriate parametric
model and substituting the estimators in (\ref{eq:OmegaK}) of the main
paper and thus in $\Tbb$ will lead to a test that is at least as
powerful as that obtained based on known probabilities.

Here, we discuss considerations for modeling and estimating the
probabilities $\omega_k(\bh_k,a_k)$, $k=1,\ldots,K$.  In a SMART, the
randomization probabilities $\omega_k(\bh_k,a_k)$ can be modeled via
an appropriate logistic regression model
$\omega_k(\bh_k,a_k; \bgamma_k)$, say, in terms of a
finite-dimensional parameter $\bgamma_k$.  We consider two examples.\\

\noindent\textit{Example 1: Two-stage SMART.} 
Consider the two-stage SMART in Figure~\ref{f:eight}.  Here,
$\omega_1(\bh_1,a_1) = P(A_1=a_1 | \bH_1=\bh_1, \kappa \geq 1) = P(A_1
= a_1)$ can be represented as
$$\omega_1(\bh_1,a_1; \gamma_1) = \left\{ \frac{\exp(\gamma_1)}{1+\exp(\gamma_1)}\right\}^{I(a_1=1)}
\left\{ \frac{1}{1+\exp(\gamma_1)}\right\}^{I(a_1=0)},$$
which can be fitted by maximum likelihood (ML) using the data
$A_{1i}$, $i=1,\ldots,n$; see below.  
It is straightforward that the ML estimator $\widehat{\gamma}_1$
is such that $\expit(\widehat{\gamma}_1)$ is equal to the sample
proportion of subjects randomized at stage 1 to Trt 1, where $\expit(s) =e^s/(1+e^s)$.   At stage 2, 
letting $r_2$ be the component/function of
$\bh_2$ indicating response status, $\omega_2(\bh_2,a_2) = P(A_2=a_2
| \bH_2=\bh_2, \kappa \geq 2)$ can be represented as
\begin{align*}
 \omega_2&(\bh_2, a_2; \bgamma_2) = \left\{ \frac{\exp(\gamma_{21})}{1+\exp(\gamma_{21})}\right\}^{I(a_1=0, r_2=1,a_2=2)} \left\{ \frac{1}{1+\exp(\gamma_{21})}\right\}^{I(a_1=0, r_2=1,a_2=3)}\\
&\times \left\{ \frac{\exp(\gamma_{22})}{1+\exp(\gamma_{22})}\right\}^{I(a_1=0, r_2=0,a_2=2)} \left\{ \frac{1}{1+\exp(\gamma_{22})}\right\}^{I(a_1=0, r_2=0,a_2=4)}\\
&\times \left\{ \frac{\exp(\gamma_{23})}{1+\exp(\gamma_{23})}\right\}^{I(a_1=1, r_2=1,a_2=2)} \left\{ \frac{1}{1+\exp(\gamma_{23})}\right\}^{I(a_1=1, r_2=1,a_2=5)}\\
&\times \left\{ \frac{\exp(\gamma_{24})}{1+\exp(\gamma_{24})}\right\}^{I(a_1=1, r_2=0,a_2=3)} \left\{ \frac{1}{1+\exp(\gamma_{24})}\right\}^{I(a_1=1, r_2=0,a_2=5)}.
\end{align*}
This model can be fitted by ML to the data $\{\bH_{2i}, A_{2i}\}$ for
all $i$ for whom $\kappa_i = 2$ (see below).  
The ML estimator $\hatgamma_2 = (\widehat{\gamma}_{21},
\widehat{\gamma}_{22}, \widehat{\gamma}_{23}, \widehat{\gamma}_{24})^T$
 is such that, for example, $\expit(\widehat{\gamma}_{21})$ is equal to the sample
proportion of subjects who were randomized at stage 1 to Trt 0,
reached stage 2 without experiencing the event or censoring, responded to
Trt 0, and were randomized to Trt 2.  

The ML estimator for $\bgamma = (\gamma_1,\bgamma_2^T)^T$ is found by
maximizing the loglikelihood for $\bgamma$; the likelihood contribution
for a single subject is given by
\begin{equation}
  \label{eq:like1}
\begin{aligned}
&\frac{\exp\{ \gamma_1I(A_1 = 1)\}}{1+\exp(\gamma_1)}\\
&\times \left( \left[\frac{\exp\{ \gamma_{21}
      I(A_2=2)\}}{1+\exp(\gamma_{21})}\right]^{ I(A_1=0, R_2=1)} 
\left[\frac{\exp\{ \gamma_{22} I(A_2=2)\}}{1+\exp(\gamma_{22})}\right]^{ I(A_1=0, R_2=0)}\right.\\
&\times\left. \left[ \frac{\exp\{ \gamma_{23} I(A_2=2)\}}{1+\exp(\gamma_{23})}\right]^{ I(A_1=1, R_2=1)}
\left[ \frac{\exp\{ \gamma_{24} I(A_2=3)\}}{1+\exp(\gamma_{24})}\right]^{ I(A_1=1, R_2=0)}\right)^{I(\kappa=2)}.
  \end{aligned}
\end{equation}
It is straightforward that the associated score vector is
\begin{equation}
  \bS_\gamma(\bgamma) = \left( \begin{array}{c}
  \{ I(A_1 = 1) - \expit(\gamma_1)\} \\
I(A_1=0, R_2=1, \kappa=2) \{ I(A_2=2) - \expit(\gamma_{21})\} \\
I(A_1=0, R_2=0, \kappa=2) \{ I(A_2=2) - \expit(\gamma_{22})\}\\
I(A_1=1, R_2=1, \kappa=2) \{ I(A_2=2) - \expit(\gamma_{23})\}\\
I(A_1=1, R_2=0, \kappa=2) \{ I(A_2=3) - \expit(\gamma_{24})\} \end{array} \right) 
\label{eq:score1}
\end{equation}

\vspace*{0.2in}

\noindent\textit{Example 2: Observational study with two decision points.}
When the data are from an observational study, the analyst must posit
models $\omega_k(\bh_k,a_k; \bgamma_k)$, $k=1,\ldots,K$, that
incorporate possible dependence of the probabilities of treatment on
past history both to acknowledge the likely dependence of treatment
assignment on history and to account for possible confounding.
Consider a setting with $K=2$ stages where subjects were observed to
receive one of two treatments, coded as 0 and 1, at stage 1, and,
among those who reached stage 2 without having experienced the event
or censoring, responders continued on stage 1 treatment and
nonresponders were observed to receive one of two treatments, also
coded as 0 and 1 (but not necessarily the same as those at stage 1).
Typical logistic regression models are of the form
$$\omega_1(\bh_1,a_1; \gamma_1) = \left\{ \frac{\exp(\bgamma_1^T\tilh_1)}{1+\exp(\bgamma_1^T\tilh_1)}\right\}^{I(a_1=1)}
\left\{ \frac{1}{1+\exp(\bgamma_1^T\tilh_1)}\right\}^{I(a_1=0)},$$
\begin{align*}
\omega_2&(\bh_2, a_2; \bgamma_2) = \left\{ \frac{\exp(\bgamma_{21}^T\tilh_2)}{1+\exp(\bgamma_{21}^T\tilh_2)}\right\}^{I(a_1=0, r_2=0,a_2=1)} \left\{ \frac{1}{1+\exp(\bgamma_{21}^T\tilh_2)}\right\}^{I(a_1=0, r_2=0,a_2=0)}\\
&\times \left\{ \frac{\exp(\bgamma_{22}^T\tilh_2)}{1+\exp(\bgamma_{22}^T\tilh_2)}\right\}^{I(a_1=1, r_2=0,a_2=1)} \left\{ \frac{1}{1+\exp(\bgamma_{22}^T\tilh_2)})\right\}^{I(a_1=1, r_2=0,a_2=0)},
\end{align*}  
where $\tilh_k$ is a vector of functions of elements of $\bh_k$,
ordinarily with a ``1'' in the first position to yield an
``intercept'' term, $k=1, 2$, so could include, for example,
interactions, polynomial terms or splines, and so on.   Analogous to
(\ref{eq:like1}), the contribution of a single subject to the likelihood for $\bgamma =
(\bgamma_1^T,\bgamma_2^T)^T$ is given by
\begin{equation}
  \label{eq:like3}
\begin{aligned}
  & \frac{\exp\{ \bgamma_1^T\tilH_1 I(A_1=1)\}}{1+\exp(\bgamma_1^T\tilH_1)}
\left[ \frac{\exp\{\bgamma_{21}^T\tilH_2I(A_2 =
                1)\}}{1+\exp(\bgamma_{21}^T\tilH_2)}\right]^{I(A_1=0, R_2=0, \kappa=2)}\\
&\hspace*{0.2in}\times \left[ \frac{\exp\{\bgamma_{22}^T\tilH_2I(A_2 =
                1)\}}{1+\exp(\bgamma_{22}^T\tilH_2)}\right]^{I(A_1=1, R_2=0, \kappa=2))},
  \end{aligned}
\end{equation}
and the ML estimator for $\bgamma$ maximizes in $\bgamma$ the
corresponding loglikelihood.  The associated score vector is
\begin{equation}
  \bS_\gamma(\bgamma) = \left( \begin{array}{c}
       \{ I(A_1 = 1) - \expit(\bgamma_1^T\tilH_1)\} \tilH_1 \\                       
I(A_1=0, R_2=0, \kappa=2)\{ I(A_2=1) - \expit(\bgamma_{21}^T\tilH_2)\} \tilH_2\\
(A_1=1, R_2=0, \kappa=2)\{ I(A_2=1) - \expit(\bgamma_{22}^T\tilH_2)\} \tilH_2
\end{array}\right).
 \label{eq:score2}                               
  \end{equation}

\setcounter{equation}{0}
\renewcommand{\theequation}{C.\arabic{equation}}

\section{Representation of $\Tbb$ as a Sum of iid Terms}
\label{s:webC}

We wish to show that $n^{-1/2} \times$ (\ref{eq:scorebetaobs}),
$$n^{-1/2} \Tfrak^j = n^{-1/2}\sumin \int^\infty_0\, \Omega_i(u,d^j)   \{ dN_i(u) -
d\hatLam_0(u)Y_i(u) \},$$ for each $j = 1,\ldots, D-1$, satisfies
(\ref{eq:Tdj}).  We begin by adding and subtracting $d\Lambda_0(u)$ in
the integral, to obtain
\begin{align}
n^{-1/2}  \sumin &\int^\infty_0\, \Omega_i(u,d^j)   \big[ dN_i(u) - \{d\hatLam_0(u) + d\Lambda_0(u) - d\Lambda_0(u)\} Y_i(u) \big] \nonumber\\
   &=n^{-1/2}\sumin \int^\infty_0\, \Omega_i(u,d^j)   \{ dN_i(u) -  d\Lambda_0(u) Y_i(u) \} \nonumber\\
   &\hspace*{0.1in}-n^{-1/2}\sumin \int^\infty_0 \, \Omega_i(u,d^j)
     \{d\hatLam_0(u) - d\Lambda_0(u) \} Y_i(u). \label{eq:Yuterm}
    \end{align} 
Defining 
$d\Nbar_i(u) = \sumjD \Omega_i(u,d^j) dN_i(u)$,
$\Ybar_i(u) = \sumjD \Omega_i(u,d^j) Y_i(u)$, and
$$\hatq(u,d) = \big\{n^{-1} \sumin \Omega_i(u,d) Y_i(u)\big\}/\big\{n^{-1}\sumin
\Ybar_i(u)\big\},$$ so that
$$d\hatLam_0(u) = \big\{ n^{-1}\sumin d\Nbar_i(u)\big\}/\big\{n^{-1}\sumin \Ybar_i(u)\big\},$$ substituting
this expression in (\ref{eq:Yuterm}) and rearranging and using the
definition of  $\hatq(u,d^j) = \big\{n^{-1}\sumin \Omega_i(u,d^j) Y_i(u)\big\}/\big\{n^{-1}\sumin
\Ybar_i(u)\big\}$, (\ref{eq:Yuterm}) becomes
\begin{align}
n^{-1/2}\sumin &\int^\infty_0\, \hatq(u,d^j)\{ d\Nbar_i(u) - d\Lambda_0(u)
         \Ybar_i(u)\} \nonumber \\
       &= n^{-1/2}\sumin \int^\infty_0\, \hatq(u,d^j) \left\{ \sum^D_{j'=1}\Omega_i(u,d^{j'}) dN_i(u)
         - d\Lambda_0(u)  \sum^D_{j'=1}\Omega_i(u,d^{j'}) Y_i(u)
         \right\} \nonumber\\
  &= n^{-1/2}\sumin \sum^D_{j'=1} \int^\infty_0  \Omega_i(u,d^{j'}) \hatq(u,d^j) \{
  dN_i(u) - d\Lambda_0(u)Y_i(u) \}. \label{eq:Yuterm2}
  \end{align}
  Thus, using (\ref{eq:Yuterm2}), we have
  \begin{align}
n^{-1/2}\Tfrak^j &= n^{-1/2}\sumin \left[ \vphantom{\sum^D_{j'=1}}
 \int^\infty_0  \Omega_i(u,d^j) \{ dN_i(u) - d\Lambda_0(u)Y_i(u)\} \right. \nonumber\\
 &\hspace*{0.5in}- \left.\sum^D_{j'=1} \int^\infty_0  \Omega_i(u,d^{j'}) \hatq(u,d^j) \{
   dN_i(u) - d\Lambda_0(u)Y_i(u) \right]. \label{eq:lastterm}
   \end{align}
    For $j=1,\ldots,D-1$, under the consistency assumption, $\hatq(u,d^j)$ converges in probability as
    $n \rightarrow \infty$ to
    $q(u,d^j) = E\big\{ \Omega(u,d^j) \Ys(u,d^j) \big\}/\sum^D_{j'=1}
    E\big\{\Omega(u,d^{j'}) \Ys(u,d^{j'})\big\}$.  Then under
    regularity conditions, $n^{-1/2}$ times (\ref{eq:lastterm}) satisfies
    \begin{align*}
n^{-1/2} \sumin &\left[ \sum^D_{j'=1} \int^\infty_0  \Omega_i(u,d^{j'}) \hatq(u,d^j) \{
   dN_i(u) - d\Lambda_0(u) Y_i(u) \right]\\
&=n^{-1/2} \sumin \left[ \sum^D_{j'=1} \int^\infty_0  \Omega_i(u,d^{j'}) q(u,d^j) \{
   dN_i(u) - d\Lambda_0(u)Y_i(u) \right] + o_p(1),
    \end{align*}
leading to (\ref{eq:Tdj}); namely,
\begin{align*}
 n^{-1/2}   \Tfrak^j &= n^{-1/2} \sumin \left[ \vphantom{\sum^D_{j'=1}}
 \int^\infty_0  \Omega_i(u,d^j) \{ dN_i(u) - d\Lambda_0(u)Y_i(u)
 \} \right.\\
&\hspace*{0.5in}- \left.\sum^D_{j'=1} \int^\infty_0  \Omega_i(u,d^{j'}) q(u,d^j) \{
  dN_i(u) - d\Lambda_0(u)Y_i(u) \right] + o_p(1),\\
& = n^{-1/2} \sumin \Tfrak^j_i + o_p(1),
\end{align*}
which shows that $n^{-1/2} \Tfrak^j$ in (\ref{eq:scorebetaobs}) is
asymptotically equivalent to $n^{-1/2}$ times a sum of iid terms, and
thus so is
$n^{-1/2} \Tbb = n^{-1/2} (\Tfrak^1,\ldots,\Tfrak^{D-1})^T$.
Accordingly, if we define
$\Tbb_i = (\Tfrak^1_i,\ldots,\Tfrak^{D-1}_i)^T$, which are iid for
$i=1,\ldots,n$, so that
$n^{-1/2} \Tbb = n^{-1/2} \sumin \Tbb_i + o_p(1)$, it follows by
standard asymptotic normal theory that the asymptotic covariance
matrix of $n^{-1/2} \Tbb$ can be approximated by
$\bSigma = n^{-1} \sumin (\Tbb_i \Tbb_i^T)$.  Because
$q(u,d^j)$ and $d\Lambda_0(u)$ are not known, it is natural to
substitute $\hatq(u,d^j)$ for $q(u,d^j)$ and $d\hatLam_0(u)$ for
$d\Lambda_0(u)$ to obtain $\widehat{\Tfrak}^j_i$ and
$\widehat{\Tbb}_i = (\widehat{\Tfrak}^1_i,\ldots,
\widehat{\Tfrak}^{D-1}_i)^T$, say, and estimate $\bSigma$ by
$\hatSig = n^{-1} \sumin (\widehat{\Tbb}_i \widehat{\Tbb}_i^T)$.  In fact, it
is straightforward to observe that
$n^{-1/2} \Tbb = n^{-1/2} \widehat{\Tbb} = n^{-1/2} \sumin
\widehat{\Tbb}_i$, so the test statistic can be formed equivalently as 
$\Zbb = n^{-1} \widehat{\Tbb}^T \hatSig^{-} \widehat{\Tbb}$.

\setcounter{equation}{0}
\renewcommand{\theequation}{D.\arabic{equation}}

\section{Improving Efficiency}
\label{s:webD}

\noindent
\textbf{D.1.  Increased efficiency using estimated
  randomization probabilities} \\*[-0.2in]

In Section~\ref{s:webB}, we discuss modeling and
estimation (by ML) of the probabilities $\omega_k(\bh_k,a_k)$,
$k=1,\ldots,K$.  Assuming that suitable models
$\omega_k(\bh_k,a_k; \bgamma_k)$, $k=1,\ldots,K$, have been posited
following the considerations in Section~\ref{s:webB} and that
$\bgamma = (\bgamma_1^T,\ldots,\bgamma_K^T)^T$ is estimated via ML, we
now discuss the implications of substituting estimated probabilities
obtained from the fitted models for the known probabilities in
(\ref{eq:OmegaK}) and thus in $\Tbb$ in forming the test statistic.
Namely, we argue that substituting estimated probabilities for known
probabilities yields a more powerful test as long as the models are
correctly specified, and we demonstrate that estimation of $\bgamma$
must be taken into appropriate account in deriving and estimating the
asymptotic covariance matrix of $\Tbb$ with which to form the test
statistic.  Thus, with data from a SMART, it is advantageous to
estimate the known randomization probabilities and use the test
statistic with this appropriate covariance matrix.  In fact, the
arguments show that, if there are components of the histories $\bH_k$
that are associated with the outcome, it is possible to exploit these
associations to gain further efficiency and thus further increases in
power, analogous to the familiar tactic of covariate adjustment.

Denote $\Tbb$ with elements as in (\ref{eq:scorebetaobs}) with the
models $\omega_k(\bh_k,a_k; \bgamma_k)$, $k=1,\ldots,K$ substituted as
$\Tbb(\bgamma)$, and, as in Section~\ref{s:webB}, let $\hatgamma$ be the ML
estimator for $\bgamma$.  According to Theorem~9.1 of
\citet{TsiatisSemi}, with correctly-specified models, substituting
$\hatgamma$ in these models is equivalent to projecting each of the
$D-1$ components of $\Tbb$, which contain the known probabilities,
onto the space spanned by the score vector associated with the
likelihood under the models.  Because $\Tbb(\hatgamma)$ involves a
projection, the asymptotic covariance matrix of $\Tbb(\hatgamma)$ is
no larger than that of $\Tbb$ with the known probabilities in the
sense of nonnegative definiteness, which implies that estimating
$\bgamma$ results in efficiency gains relative to using the known
probabilities.  Thus, a modified test statistic, $\Zbb(\hatgamma)$,
say, that takes these features into account will yield a more powerful
test than the test statistic $\Zbb$ using the known probabilities.


Letting $\bSigma_\gamma$ be the asymptotic covariance matrix of
$\Tbb(\hatgamma)$, we now present a heuristic argument to motivate our
proposed approach to obtaining an estimator for $\bSigma$,
$\hatSig_\gamma$, to be used to form the test statistic.
 We first provide a representation of the score
vector associated with the ML estimator $\hatgamma$ in generic models
$\omega_k(\bh_k,a_k; \bgamma_k)$, $k=1,\ldots,K$; examples of score
vectors in the specific models considered in Section~\ref{s:webB}  are in
(\ref{eq:score1}) and (\ref{eq:score2}).  Similar to the formulation
in Section~\ref{s:webA}, let $A(u)$ denote the treatment received at time
$u \geq 0$; and let $M$ denote the number of all possible treatment
options that could be administered at $u$, including ``do nothing'' if
$u$ is not the time of a decision point. 
Let $\Hm(u)$ denote the history of
information at time $u$ but not including the treatment received at
time $u$.  Let $\omega\{\ell, u, \Hm(u)\} = P\{ A(u) = \ell | \Hm(u)\}$, the
probability of being assigned treatment $\ell$ at time $u$ given the
history $\Hm(u)$, $\ell = 1,\ldots,M$, where this probability is
positive only for treatments $\ell$ that are feasible at time $u$
given the history $\Hm(u)$.

Consider a model for $\omega\{\ell, u, \Hm(u)\}$ given by
$\omega\{\ell, u,
\Hm(u); \bgamma\}$ indexed by the finite-dimensional parameter
$\bgamma$. Then the contribution to the likelihood for $\bgamma$ for
an individual, under the assumptions of consistency, the SRA, and
positivity, is given by
$$\prod_{u \leq U} \prod^M_{\ell=1} \omega\{\ell, u, \Hm(u); \bgamma\}^{ I\{
  A(u) = \ell\} },$$
so that the contribution to the loglikelihood is
$$\int^U_0 \sum^M_{\ell=1} I\{ A(u) = \ell\} \, \log\big[\omega\{\ell, u, \Hm(u); \bgamma\}\big].$$
Then the score vector is the first partial derivative of this
expression, namely,
$$\bS_\gamma(\bgamma) = \int^U_0 \sum^M_{\ell=1} I\{ A(u) = \ell\} \, 
\frac{\partial/\partial \bgamma \big[ \omega\{\ell, u, \Hm(u); \bgamma\}\big]}{\omega\{\ell, u, \Hm(u); \bgamma\}}.$$
Because $\sum^M_{\ell=1} \omega\{\ell, u, \Hm(u); \bgamma\} = 1$ for
all $\bgamma$, 
 $$\sum^M_{\ell=1} \partial/\partial \bgamma \big[ \omega\{\ell, u,
 \Hm(u); \bgamma\}\big] = 0,$$
 so that
 \begin{equation}
 \bS_\gamma(\bgamma) = \int^U_0 \sum^M_{\ell=1} \big[ I\{ A(u) = \ell\}-
 \omega\{\ell, u, \Hm(u); \bgamma\}\big] \frac{\partial/\partial \bgamma \big[ \omega\{\ell, u,
   \Hm(u); \bgamma\}\big]}{\omega\{\ell, u, \Hm(u); \bgamma\}}.
\label{eq:generalscore}
 \end{equation}

Aligned with \citet[Theorem~9.1]{TsiatisSemi}, consider the space of
``augmentation terms,'' denoted by $\Lambda_2$, defined as all
functions of the observed data, $g(\calO)$, say, such that
$$E\{ g(\calO) \,|\, \calW^*\} = 0.$$
Motivated by the space spanned by score vectors with respect to
$\bgamma$ (\ref{eq:generalscore}), define the subspace
$\Lambda_{\textrm{Aug}}$ of $\Lambda_2$ by the space of functions of
the form
\begin{equation}
\int^U_0 \sum^M_{\ell=1} \big[ I\{ A(u) = \ell\}-
\omega\{\ell, u, \Hm(u); \bgamma_0\}\big] \, b_\ell\{u, \Hm(u)\},
\label{eq:augfunc}
\end{equation}
where $\bgamma_0$ is the true value of $\bgamma$ so that
$\omega\{\ell, u, \Hm(u); \bgamma_0\} = $ the true $\omega\{\ell, u,
\Hm(u) \}$; and 
$b_\ell\{u, \Hm(u)\}$ are arbitrary functions.  That any element
of $\Lambda_{\textrm{Aug}}$ is contained in $\Lambda_2$ follows
because
\begin{align*}
E&\Big( \big[ I\{ A(u) = \ell\} - \omega\{\ell, u, \Hm(u); \bgamma_0\}\big] \,
  b_\ell\{u, \Hm(u)\} | \calW^*\Big)\\
&=E\left\{ \left.E \Big( \big[ I\{ A(u) = \ell\} - \omega\{\ell, u, \Hm(u) ; \bgamma_0\}\big] \,
  b_\ell\{u, \Hm(u)\} | \Hm(u),\calW^*\Big) \right| \calW^*\right\} \\
&= 0,
\end{align*}
using the fact that
$E \Big( \big[ I\{ A(u) = \ell\} | \Hm(u),\calW^*\Big) = \omega\{\ell,
u, \Hm(u); \bgamma_0\}$. Clearly, from (\ref{eq:generalscore}), the space
$\Lambda_{\textrm{Aug}}$ contains the space spanned by
$\bS_\gamma(\bgamma)$.  Thus, by considering test statistics formed by
projecting each of the $D-1$ components of $\Tbb$ onto the space
$\Lambda_{\textrm{Aug}}$, we can obtain tests that are more powerful
than those based on $\Zbb$ using known probabilities.  These
developments suggest the approach presented here for
constructing the test statistic.  Namely, instead of considering the
entire space $\Lambda_{\textrm{Aug}}$, we consider projecting onto a
subspace spanned by a finite number of elements of
$\Lambda_{\textrm{Aug}}$, which are of the form in (\ref{eq:augfunc})
with the elements of $\Lambda_{\textrm{Aug}}$ determined by 
analyst-specified choices for the functions of the history
$b_\ell\{u, \Hm(u)\}$; refer to these functions as basis functions.
As long as this subspace contains all the elements of
$\bS_\gamma(\bgamma)$, so that the chosen basis functions include the
components of $\bS_\gamma(\bgamma)$, the efficiency gain associated with
estimation of $\bgamma$ will be taken into account.

Based on these results, operationally, to obtain $\hatSig_\gamma$ with
which to form a test statistic, for each $j = 1,\ldots,D$, proceed as
follows.  Denote by $\widehat{\Tfrak}^j_i(\hatgamma)$
$\widehat{\Tfrak}^j_i$ (i.e., as in (\ref{eq:Tdj}) with $\hatq(u,d^j)$
and $d\hatLam_0(u)$ substituted) for subject $i$, $i = 1,\ldots,n$,
with the estimated probabilities substituted.  Denote by
$\bS_{\gamma,i}(\hatgamma)$ the score vector associated with the
likelihood for $\bgamma$ for subject $i$ in the chosen models
$\omega_k(\bh_k,a_k; \bgamma_k)$, $k=1,\ldots,K$, with the estimated
probabilities substituted.  Then for each $j$ carry out a linear
regression of the $\widehat{\Tfrak}^j_i(\hatgamma)$ on
$\bS_{\gamma,i}(\hatgamma)$ and form for each subject $i = 1,\ldots,n$
the residuals from this fit, $\widehat{\Tfrak}^{j,R}_i(\hatgamma)$,
say.  Then, defining
$\widehat{\Tbb}^R_i(\hatgamma) = \{
\widehat{\Tfrak}^{1,R}_i(\hatgamma), \ldots,
\widehat{\Tfrak}^{D-1,R}_i(\hatgamma)\}^T$, define
\begin{equation}
\hatSig_\gamma = n^{-1} \sumin \{ \widehat{\Tbb}^R_i(\hatgamma)
\widehat{\Tbb}^R_i(\hatgamma) ^T\}.
\label{eq:hatSigR}
\end{equation}
Letting $\widehat{\Tbb}^R(\hatgamma) = \sumin
\widehat{\Tbb}^R_i(\hatgamma)$, form the test statistic as
\begin{equation}
\Zbb(\hatgamma) = n^{-1} \widehat{\Tbb}^R(\hatgamma)^T \hatSig^{-}_\gamma
\widehat{\Tbb}^R(\hatgamma)
\label{eq:teststat}
\end{equation}


As a demonstration, consider Example 1 in Section~\ref{s:webB}  of a
two-stage SMART.  Here, the randomization probabilities are known, but
the foregoing discussion indicates that a more powerful test can be
obtained by estimating the randomization probabilities using sample
proportions.  In this case, to obtain the test statistic, for each
$j=1,\ldots,8$, perform a linear regression with dependent variable
$\widehat{\Tfrak}^j_i(\hatgamma)$ and design matrix with $i$th row,
from (\ref{eq:score1}),
\begin{equation}
  \label{eq:ex1row}
\begin{aligned}
\big[ \{ I(A_{1i} = 1) &- \expit(\widehat{\gamma}_1)\},
I(A_{1i}=0, R_{2i}=1, \kappa_i=2) \{ I(A_{2i}=2) - \expit(\widehat{\gamma}_{21})\}, \\
&I(A_{1i}=0, R_{2i}=0, \kappa_i=2) \{ I(A_{2i}=2) - \expit(\widehat{\gamma}_{22})\}, \\
&I(A_{1i}=1, R_{2i}=1, \kappa_i=2) \{ I(A_{2i}=2) - \expit(\widehat{\gamma}_{23})\}, \\
&I(A_{1i}=1, R_{2i}=0, \kappa_i=2) \{ I(A_{2i}=3) - \expit(\widehat{\gamma}_{24})\}\big],
\end{aligned}
\end{equation}
and obtain the residuals $\widehat{\Tfrak}^{j,R}_i(\hatgamma)$ from
this fit and then $\hatSig_\gamma$ as in (\ref{eq:hatSigR}) and the
test statistic as in (\ref{eq:teststat}).

Similarly, in Example 2 in Section~\ref{s:webB}  of a two-stage observational
study, where of necessity $\omega_k(\bh_k,a_k)$, $k=1, 2$, must be
modeled and fitted, the test statistic would be found by, for each
$j=1,\ldots,4$, by linear regression with dependent variable
$\widehat{\Tfrak}^j_i(\hatgamma)$ and design matrix with $i$th row,
from (\ref{eq:score2}),
\begin{align*}
 \big[ \{ I(A_{1i} = 1) &- \expit(\hatgamma_1^T\tilH_{1i})\} \tilH_{1i}^T, 
I(A_{1i}=0, R_{2i}=0, \kappa_i=2)\{ I(A_{2i}=1) - \expit(\hatgamma_{21}^T\tilH_{2i})\} \tilH_{2i}^T, \\
&I(A_{1i}=1, R_{2i}=0, \kappa_i=2)\{ I(A_{2i}=1) - \expit(\hatgamma_{22}^T\tilH_{2i})\} \tilH_{2i}^T \big].
  \end{align*}
  Then obtain the residuals $\widehat{\Tfrak}^{j,R}_i(\hatgamma)$ from
  this regression and proceed as above. \\

  \noindent
\textbf{D.2.  Remarks on \citet{Li2014}} \\*[-0.2in]

  It is straightforward to demonstrate that the approach of
  \citet{Li2014} can be viewed as a special case of our general
  testing approach, although formulated explicitly and evaluated in
  their article only for a situation analogous to that of Example 2 in
  Section~\ref{s:webB}.  Thus, the foregoing developments contradict the
  claim of \citet{Li2014} that ``estimation of these (treatment
  selection) probabilities does not have an effect on the asymptotic
  distribution of the weighted logrank statistic.''  In this example
  and in general, the above arguments demonstrate that there is an
  effect of estimation of the probabilities (i.e., estimation of
  $\bgamma$ here).  The proof of Theorem 1 in the Appendix of
  \citet{Li2014} that there is no effect of estimating the
  probabilities hinges on the equality in their equation (5).  If (5)
  is true, it implies that their quantity $A_n$ in their equation (2)
  converges in probability to zero and thus that there is no
  dependence of the large-sample distribution of the quantity on the
  left hand side of their (2) and thus their test statistic on
  estimation of their parameter $\theta$ ($\bgamma$ here).  However,
  we now argue that the equality in (5) does not hold; we provide a
  heuristic sketch of the argument.  Using roughly their notation and
  letting $Z$ denote the data, the expectation on the left hand side
  of their equation (5) is of the form
  \begin{equation}
  E_\theta\left[ \frac{\partial}{\partial \theta}\{ W(Z; \theta) \}
    \left\{ \int^\tau_0 \xi(u) dM(u)  \right\}\right],
  \label{eq:Li1}
  \end{equation}
  where $W(Z; \theta)$ and $dM(u)$ are functions of $Z$ and
  $\xi(u)$ is a deterministic function, and $E_\theta$ indicates
  expectation with respect to the distribution of the data evaluated
  at $\theta$.  The distribution of the data depends on $\theta$, as
  the data involve treatment assignment indicators, so that, letting
  $p(z; \theta)$ denote the density of $Z$, (\ref{eq:Li1}) is
  \begin{equation}
\int \frac{\partial}{\partial \theta}\{ W(z; \theta) \} \left\{ \int^\tau_0 \xi(u) dM(u)\right\} p(z; \theta)\, dz.
\label{eq:Li2}
    \end{equation}
    However, the right hand side of their (5) is, interchanging expectation
    and integration,
    \begin{equation}
      \label{eq:Li3}
      \begin{aligned}
 \frac{\partial}{\partial \theta} &E_\theta\left[ W(Z; \theta) \left\{ \int^\tau_0\xi(u)
   dM(u) \right\}  \right] = \frac{\partial}{\partial \theta} \left[ \int W(z; \theta) \left\{ \int^\tau_0\xi(u) dM(u) \right\} p(z; \theta) \,dz \right] \\
&= \int \frac{\partial}{\partial \theta}\{ W(z; \theta) \} \left\{ \int^\tau_0\xi(u) dM(u)\right\} p(z; \theta)\, dz
+ \int W(z;\theta) \left\{ \int^\tau_0\xi(u) dM(u) \right\} S_\theta(z; \theta) p(z; \theta)\, dz,
\end{aligned}
\end{equation}
where
$S_\theta(z; \theta) = \partial/\partial \theta \{ p(z; \theta)\}/p(z;
\theta)$ is the score associated with ML estimation of $\theta$.
Accordingly, their equation (5) does not hold as shown; rather, from (\ref{eq:Li2})
and (\ref{eq:Li3}),
\begin{align}
&E_\theta\left[ \frac{\partial}{\partial \theta}\{ W(Z; \theta) \}
 \left\{ \int^\tau_0 \xi(u) dM(u) \right\} \right] \nonumber \\ &= \frac{\partial}{\partial \theta} E_\theta\left[ W(Z; \theta) \left\{ \int^\tau_0\xi(u)
   dM(u)  \right\} \right] - E_\theta\left[ W(z;\theta) \left\{ \int^\tau_0\xi(u) dM(u)\right\}
   S_\theta(z; \theta) \right]. \label{eq:Li4}
 \end{align}
 Equation (4) of \citet{Li2014} implies that the first term in
 (\ref{eq:Li4}) is equal to zero; thus, their equation (5) should be
$$E_\theta\left[ \frac{\partial}{\partial \theta}\{ W(Z; \theta) \}
  \left\{ \int^\tau_0\xi(u) dM(u) \right\} \right] = -E_\theta\left[
  W(z;\theta) \left\{ \int^\tau_0\xi(u) dM(u)\right\} S_\theta(z;
  \theta) \right] \neq 0,$$ so that $A_n$ in their  (2) converges in
probability to the right hand side of this expression rather than to
zero.  The implication is that there is in fact an effect of
estimation of their parameter $\theta$ indexing the models for the
treatment probabilities (our $\bgamma$), which depends on the score
vector, analogous to our results.

In fact, because as above estimation of $\bgamma$ leads to an increase
in efficiency and because \citet{Li2014} do not take into account the
effect of estimating $\bgamma$ in the covariance matrix used to form their
test statistic, the covariance matrix they use overstates the uncertainty in
the components of $\Tbb$.  Consequently, under $H_0$, their proposed
test is expected to be conservative under $H_0$.  This behavior is
evident in their Table 1, which shows that, even with sample sizes as
large as $n=10,000$, the empirical type I error rate is lower than the
nominal 0.05 level of significance. \\

\noindent
\textbf{D.3.  Gaining additional efficiency using covariates} \\*[-0.2in]

Returning to our approach, the above arguments and form of the space
$\Lambda_{\textrm{Aug}}$ suggest that, if there are components of the
history at any decision point that are associated with the outcome, it
may be possible to gain further efficiency by exploiting these
associations.  In particular, consider a SMART.  In addition to
including as basis functions the elements of the score vector
associated with estimators of the randomization probabilities as
above, one can also include additional basis functions that are
functions of these components of the history.  For example, in the
SMART in Example 1, in addition to having columns that are determined
by elements of (\ref{eq:score1}) for each subject $i$ as in
(\ref{eq:ex1row}), one could add further columns to the design matrix
by appending to the $i$th row the additional components
\begin{equation}
  \label{eq:basisfunc}
\begin{aligned}\big[ \{ I(A_{1i} = 1) &- \expit(\widehat{\gamma}_1)\} \tilH_{1i}^T,
I(A_{1i}=0, R_{2i}=1, \kappa_i=2) \{ I(A_{2i}=2) -  \expit(\widehat{\gamma}_{21})\} \tilH_{21i,i} ^T, \\
&I(A_{1i}=0, R_{2i}=0, \kappa_i=2) \{ I(A_{2i}=2) - \expit(\widehat{\gamma}_{22})\}\tilH_{22,i} ^T, \\
&I(A_{1i}=1, R_{2i}=1, \kappa_i=2) \{ I(A_{2i}=2) - \expit(\widehat{\gamma}_{23})\}\tilH_{23,i} ^T, \\
&I(A_{1i}=1, R_{2i}=0, \kappa_i=2) \{ I(A_{2i}=3) - \expit(\widehat{\gamma}_{24})\}\tilH_{24,i} ^T\big],
\end{aligned}
\end{equation}
where now $\tilH_1$ and $\tilH_{2\ell}$, $\ell=1,\ldots,4$, are
vectors of basis functions of $\bH_1$ and $\bH_2$, but with no ``1''
in the first position.  For example, if $\bH_1$ contains two baseline
covariates $X_{11},X_{12}$, and $\bH_2$ contains one covariate $X_2$
ascertained between Decisions 1 and 2, one could take
$\tilH_1 = (X_{11},X_{12})^T$ and
$\tilH_{2\ell} = (X_{11},X_{12}, X_2)^T$, $\ell=1,\ldots,4$.  In this
case, the design matrix would be of dimension $(n \times 19)$.  Of
course, the analyst must be judicious in specifying these vectors of
basis functions, as including in them information in the history for
which the association is not strong could lead to finite-sample
``noise'' that diminishes or offsets efficiency gains in finite samples.  To obtain the
test statistic, one would proceed as above and obtain the residuals
$\widehat{\Tfrak}^{j,R}_i(\hatgamma)$ from the regression using this
design matrix, compute $\hatSig_\gamma$ as in (\ref{eq:hatSigR}),
and form the test statistic as in (\ref{eq:teststat}). \\



\noindent
\textbf{D.4.  Remarks on \citet{KidwellLogrank}} \\*[-0.2in]

We provide an argument demonstrating the contention of
Section~\ref{ss:compare} that the components of the vector need not be
martingales with respect to the filtration given by
\citet{KidwellLogrank}.  Because the test of \citet{KidwellLogrank} is
applicable for $K=2$ and the design in Figure~\ref{f:four} without the
control regime, and their formulation does not involve covariates, we
consider this setting with no available covariate information for
simplicity.  In this case, the history up to time $u \geq 0$ is given
by (recall $\calT_1 = 0$ is superfluous, so we eliminate it from the
history for the purpose of this argument)
$$H(u) = \{ A_1, I(\kappa\geq 2, \calT_2 \leq u), 
I(\kappa\geq 2, \calT_2 \leq u) (\calT_2, A_2), I(U < u), (U,\Delta)I(U < u)\}$$
is the filtration of \citet{KidwellLogrank} in the current notation.

Under the null hypothesis $H_0$ in (\ref{eq:null}), the hazard rates
$\lambda(u, d^j)$, $j = 1,\ldots,D = 4$ are all equal to the common
hazard rate $\lambda_0(u)$.  In computing the covariance matrix of the
numerator of their test statistic under $H_0$, \citet{KidwellLogrank}
invoke the theory of counting process martingales.  Specifically, the
authors assume that the increment
$dM(u) = dN(u) - d\Lambda_0(u) Y(u)$, where
$N(u) = I(U\leq u, \Delta=1)$ and $Y(u) = I(U \geq u)$, is a $H(u^-)$
martingale increment; i.e.,
$$E\{ dM(u)\,| \, H(u^-)\} = 0.$$
Consequently, for each regime $j$,
\begin{align*}
E\{ \Omega(u, d^j) dM(u) \,|\, H(u^-)\} = \Omega(u, d^j)  E\{
  dM(u) \,|\, H(u^-)\} = 0.
\end{align*}
However, in general, the intensity process for the counting process
$N(u)$ is
\begin{align*}
E\{ dN(u) \,|\, H(u^-)\}  &=  \big[ d\Lambda_{NR}(u, A_1) \{ 1 -
                                  I(\kappa\geq 2, \calT_2 \leq u) \} \\
  & + d\Lambda_R(u, A_1, \calT_2,
  A_2) I(\kappa\geq 2, \calT_2 \leq u) \} \big] Y(t).
\end{align*}
Here, $d\Lambda_{NR}(u, A_1)$ is the hazard of experiencing the event
at time $u$ for an individual who is at risk at time $u$ and has not
yet had a response, which may be a function of $u$ and $A_1$; and
$d\Lambda_R(u, A_1, \calT_2, A_2)$ is the hazard of experiencing the
event at time $u$ for an individual who is at risk at time $u$, has
achieved a response at time $\calT_2 < u$, and received treatments
$A_1$ and $A_2$ at Decisions 1 and 2, which may be a function of $u$,
$A_1$, $\calT_2$, and $A_2$.

In the data generative scenarios of \citet{KidwellLogrank} (our
Scenarios 1 and 2, for example) and our Scenarios 3-5, under the null
hypothesis, $d\Lambda_{NR}(u, A_1) = d\Lambda_{NR}(u)$, so does not
depend on $A_1$; similarly, $d\Lambda_R(u, A_1, \calT_2, A_2) =
d\Lambda_R(u)$, so does not depend on $A_1, \calT_2, A_2$.  However,
$ d\Lambda_{NR}(u)$ need not be equal to  $d\Lambda_{R}(u)$.  Thus,
under these generative scenarios,
\begin{align*}
E\{ dN(u) \,|\, H(u^-)\}  &=  \big[ d\Lambda_{NR}(u) \{ 1 -
                                  I(\kappa\geq 2, \calT_2 \leq u) \} \\
  & + d\Lambda_R(u) I(\kappa\geq 2, \calT_2 \leq u) \} \big] Y(u),
\end{align*}
in which case
\begin{equation}
  \label{eq:dMuLamR}
\begin{aligned}
E\{ dM(u)\,| \, H(u^-)\} &= \big[ d\Lambda_{NR}(u) \{ 1 - I(\kappa\geq 2, \calT_2 \leq u) \} \\
  & + d\Lambda_R(u) I(\kappa\geq 2, \calT_2 \leq u) -d\Lambda_0(u)\} \big] Y(u).
\end{aligned}
\end{equation}
(\ref{eq:dMuLamR}) is not equal to zero unless $d\Lambda_{NR}(u) =
  d\Lambda_{R}(u),$ in which case
  $$d\Lambda_{NR}(u) = d\Lambda_{R}(u)=d\Lambda_{0}(u),$$
and (\ref{eq:dMuLamR}) equals zero.
  From Section~\ref{s:webA}, $E\{ \Omega(u, d^j) dM(u) \} =0$, which is the
  basis of our approach; however, as the above demonstrates,
  $E\{ \Omega(u, d^j) dM(u)\,|\, H(u^-)\}$ does not necessarily
  equal to zero unless $d\Lambda_{NR}(u) = d\Lambda_{R}(u)$.

  The fact that $dM(u)$ is not necessarily a $H(u^-)$ martingale
  increment does not affect the unbiasedness of the vector of paired
  comparisons on which the test statistic of \citet{KidwellLogrank} is
  based, but may impact the relevance of the authors' derivation of
  the covariance matrix of this vector, which depends on the assumed
  martingale structure.  As we discuss further in Section~\ref{s:webF}, this
  feature may underlie the simulation results reported in
  Section~\ref{s:sims}.

  \setcounter{equation}{0}
\renewcommand{\theequation}{E.\arabic{equation}}

\section{Second Order Correction to $\hatSig$}
\label{s:webE}

As discussed in Section~\ref{ss:improving}, in small
samples, the proposed test may be anticonservative.  This feature is a
consequence of the fact that the estimator $\hatSig$ for the
asymptotic covariance matrix $\bSigma$ of $\Tbb$ understates the
uncertainty in the components of $\Tbb$ in finite samples when $n$ is
not sufficiently large.  Such behavior is not uncommon with methods
based on first-order semiparametric theory.

Accordingly, to improve the finite-sample performance of the test so
that it achieves the nominal level of significance, we propose a
second-order correction to the estimator for $\bSigma$ in the spirit of
similar corrections in other contexts
\citep[e.g.][]{Kauer,Schaubel,WangTurner}.  Because we have observed
anticonservatism of the test in simulations of SMARTs with small
sample sizes in which the randomization probabilities were taken to be
known or estimated, for simplicity, we present the argument leading to
the corrected estimator for $\bSigma$ in the case that the
randomization probabilities are known.  The same correction tactic can
be used with estimated randomization probabilities and with inclusion
of additional covariates to gain efficiency, as demonstrated in Section~\ref{s:webD}.

We present a heuristic argument leading to the proposed correction.
Recall that
$$d\hatLam_0(u) = \frac{\sumin \sumjD \Omega(u,d^j)dN_i(u)}{\sumin \Ybar_i(u)},$$
$$\hatq(u,d^j) = \frac{\sumin \Omega_i(u,d^j) Y_i(u)}{\sumin\Ybar_i(u)}$$
where $\Ybar_i(u) = \sumjD \Omega_i(u,d^j) Y_i(u)$.  As before, let $q(u,d^j)$ be
the limit in probability of $\hatq(u,d^j)$.  For brevity, define
$$dM_i(u,\Lambda_0) = dN_i(u) - d\Lambda_0(u) Y_i(u).$$
Then, from (\ref{eq:Tdj}), write $\Tfrak^j_i $ as
\begin{equation}
\Psi_i(d^j; q,\Lambda_0) = \int^\infty_0 \left\{ \Omega_i(u,d^j) -
  q(u,d^j) \sum^D_{j'=1} \Omega_i(u,d^{j'}) \right\} dM_i(u,\Lambda_0),
\label{eq:one1}
\end{equation}
and let
$$\bPsi_i(q,\Lambda_0) = \{ \Psi_i(d^1; q,\Lambda_0), \ldots,
\Psi_i(d^{D-1}; q,\Lambda_0)\}^T,$$ which is the same as $\Tbb_i$, so
that the asymptotic covariance matrix of
$n^{-1/2} \Tbb = n^{-1/2}\sumin \Tbb_i =n^{-1/2}\sumin
\bPsi_i(q,\Lambda_0)$ can be approximated to first order by
\begin{equation}
  n^{-1} \bSigma = n^{-1}\sumin \{ \bPsi_i(q,\Lambda_0)  \bPsi_i(q,\Lambda_0)^T\}.
\label{eq:nSig}
  \end{equation}
  We propose estimating (\ref{eq:nSig}) by
  \begin{equation}
    n^{-1}\sumin \{ \bPsi_i(\hatq,\hatLam_0)  \bPsi_i(\hatq,\hatLam_0)^T\}.
 \label{eq:nSighat}
\end{equation}
To identify the source of the finite-sample bias and to correct it,
write
$$\bPsi_i(\hatq,\hatLam_0) = \bPsi_i(q,\Lambda_0) - \{ \bPsi_i(q,\Lambda_0)-\bPsi_i(q,\hatLam_0)\}
- \{ \bPsi_i(q,\hatLam_0) - \bPsi_i(\hatq,\hatLam_0)\},$$
so that the estimator (\ref{eq:nSighat}) can be written
as
\begin{align}
n^{-1} \sumin \{ \bPsi_i(\hatq,\hatLam_0)  &\bPsi_i(\hatq,\hatLam_0)^T\}  =
                               n^{-1}    \sumin \{\bPsi_i(q,\Lambda_0)  \bPsi_i(q,\Lambda_0)^T\} \nonumber \\
&- n^{-1}\sumin \left[ \bPsi_i(q,\Lambda_0) \{ \bPsi_i(q,\Lambda_0)-\bPsi_i(q,\hatLam_0)\}^T\right]\label{eq:if4}\\
&- n^{-1}\sumin \left[ \bPsi_i(q,\Lambda_0)\{\bPsi_i(q,\Lambda_0)-\bPsi_i(q,\hatLam_0)\}^T \right]^T\nonumber \\
&- n^{-1}\sumin\left[\bPsi_i(q,\Lambda_0) \{ \bPsi_i(q,\hatLam_0)-\bPsi_i(\hatq,\hatLam_0)\}^T\right]\label{eq:if5}\\
&- n^{-1}\sumin\left[ \bPsi_i(q,\Lambda_0)\{\bPsi_i(q,\hatLam_0)-\bPsi_i(\hatq,\hatLam_0)\}^T \right]^T\nonumber \\
&+ \mbox{smaller order terms}, \label{eq:smaller}
  \end{align}
  where we disregard the terms in (\ref{eq:smaller}) as they involve
  sample averages of the product of the differences
  $\{ \bPsi_i(q,\Lambda_0)-\bPsi_i(q,\hatLam_0)\}$ and
  $\{ \bPsi_i(q,\hatLam_0) - \bPsi_i(\hatq,\hatLam_0)\}$.

  First consider (\ref{eq:if4}) and define for brevity
  $$A_i(u,q,d^j) = \left\{ \Omega_i(u,d^j) -
    q(u,d^j) \sum^D_{j'=1} \Omega_i(u,d^{j'}) \right\},$$
and consider a typical term in the matrix in a summand of (\ref{eq:if4}).  I.e.,
for regimes $d^j$ and $d^{j'}$, consider
$$\Psi_i(d^j; q,\Lambda_0) \{ \Psi_i(d^{j'}; q, \Lambda_0) -
\Psi_i(d^{j'}; q, \hatLam_0) \},$$ which by (\ref{eq:one1}) is equal
to
\begin{equation}
\int^\infty_0 A_i(u,q,d^j) dM_i(u,\Lambda_0) \, \int^\infty_0
A_i(u,q,d^{j'}) \{ d\hatLam_0(u) - d\Lambda_0(u)\} Y_i(u).
\label{eq:five11}
 \end{equation} 
 But
 \begin{equation}
\{ d\hatLam_0(u) - d\Lambda_0(u)\} = \frac{ n^{-1} \sumln \sumjpD \Omega_\ell(u,d^{j'}) dM_\ell(u,\Lambda_0)}{ n^{-1} \sumln \Ybar_\ell(u)}
\label{eq:five2}
\end{equation}
Note that the denominator of (\ref{eq:five2}),
$n^{-1} \sumln \Ybar_\ell(u)$,
will converge in probability as $n \rightarrow \infty$, so we can
regard this quantity as fixed in subsequent arguments, as presumably
the difference between it and its limit in probability will be a
second order effect.  

Thus, interchanging sums, (\ref{eq:five11}) is equal to
\begin{equation}
n^{-1} \sumln \frac{ \left\{ \int^\infty_0 A_i(u,d^j) dM_i(u,\Lambda_0) \, \int^\infty_0
A_i(u,q,d^{j'}) \sumjpD \Omega_\ell(u,d^{j'}) dM_\ell(u,\Lambda_0)\right\}
Y_i(u)}{ n^{-1} \sumln \Ybar_\ell(u)}.
\label{eq:six1}
\end{equation}  
The expectation of a summand in (\ref{eq:six1}) when $\ell \neq i$ is
zero; thus, the expectation of (\ref{eq:six1}) is the same as the
expectation of 
$$\left\{\int^\infty_0 A_i(u,q,d^j) dM_i(u,\Lambda_0) \right\}
\frac{ \left\{ \int^\infty_0 A_i(u,q,d^j) \sumjpD \Omega_i(u,d^{j'})
    dM_i(u,\Lambda_0) \right\} }{n^{-1} \sumln \Ybar_\ell(u)},$$
where we have used the fact that $dM_i(u,\Lambda_0) Y_i(u) =
dM_i(u,\Lambda_0)$.

Based on these developments, letting
$\bA_i(u,q) =\{ A_i(u,q,d^1),\ldots,A_i(u,q,d^{D-1})\}^T$, we propose
estimating (\ref{eq:if4}) by
$$n^{-2} \sumin \left\{\int^\infty_0 \bA_i(u) dM_i(u,\Lambda_0) \right\}
\frac{ \left\{ \int^\infty_0 \bA_i(u,q)^T \sumjpD \Omega_i(u,d^{j'})
    dM_i(u,\Lambda_0) \right\} }{n^{-1} \sumln \Ybar_\ell(u)},$$
where in practice we replace $q(u,d^j)$ and $\Lambda_0(u)$ by the
estimators $\hatq(u,d^)$ and $\hatLam_0(u)$.   Defining further
$$\bG_i(q,\Lambda_0) = \frac{ \left\{ \int^\infty_0 \bA_i(u,q) \sumjpD \Omega_i(u,d^{j'})
    dM_i(u,\Lambda_0) \right\} }{n^{-1} \sumln \Ybar_\ell(u)},$$
and noting that
$$\int^\infty_0 \bA_i(u,q) dM_i(u,\Lambda_0) = \bPsi_i(q,\Lambda_0),$$
we estimate (\ref{eq:if4}) by
$$n^{-2} \sumin  \bPsi_i(q,\Lambda_0) \bG_i(q,\Lambda_0)^T,$$
where in practice we replace $q(u,d^j)$ and $\Lambda_0(u)$ by the
estimators $\hatq(u,d^j)$ and $\hatLam_0(u)$.  

A similar argument can be used to derive an estimator for
(\ref{eq:if5}). First note that (\ref{eq:if5}) is equal to
$$n^{-1}\sumin\left[\bPsi_i(q,\Lambda_0) \{
  \bPsi_i(q,\Lambda_0)-\bPsi_i(\hatq,\Lambda_0)\}^T\right]+
\mbox{smaller order terms}.$$
Thus, consider a typical term in the matrix in a summand; i.e.,
for regimes $d^j$ and $d^{j'}$, 
$$\Psi_i(d^j; q,\Lambda_0) \{ \Psi_i(d^{j'}; q, \Lambda_0) -
\Psi_i(d^{j'}; \hatq, \Lambda_0) \},$$
which is equal to
\begin{equation}
\left\{ \int^\infty_0 A_i(u,q,d^j) dM_i(u,\Lambda_0)\right\} \int^\infty_0
  \{\hatq(u,d^{j'})-q(u,d^{j'})\}
  \sumjpD \Omega_i(u, d^{j'}) dM_i(u,\Lambda_0).  
\label{eq:eight2}
  \end{equation}
It is straightforward to show that $\{\hatq(u,d^{j'})-q(u,d^{j'})\}$
is equal to
$$n ^{-1} \sumln \frac{ \Omega_\ell(u,d^{j'}) Y_\ell(u) - q(u,d^{j'})
  \sum^D_{j''=1} \Omega_\ell(u,d^{j''}) Y_\ell(u) }{n^{-1} \sumln \Ybar_\ell(u)}.$$
Thus, substituting, (\ref{eq:eight2}) equals
\begin{equation}
  \label{eq:nine1}
\begin{aligned}
n ^{-1} \sumln &\left\{ \int^\infty_0 A_i(u,q,d^j) dM_i(u,\Lambda_0)\right\}
\frac{ \left\{ \Omega_\ell(u,d^{j'}) Y_\ell(u) - q(u,d^{j'})
  \sum^D_{j''=1} \Omega_\ell(u,d^{j''}) Y_\ell(u) \right\}}{n^{-1} \sumln \Ybar_\ell(u)}\\
&\times \sum^D_{j''=1}  \Omega_i(u, d^{j''}) dM_i(u,\Lambda_0).
\end{aligned}
  \end{equation}
As above, when $\ell \neq i$, the expectation of a summand in
(\ref{eq:nine1}) is zero.  Thus, using $dM_i(u,\Lambda_0) Y_i(u) = dM_i(u,\Lambda_0)$,
we replace (\ref{eq:nine1}) by 
$$\left\{ \int^\infty_0 A_i(u,q,d^j) dM_i(u,\Lambda_0)\right\}
\frac{ \Omega_i(u,d^{j'}) - q(u,d^{j'})
  \sum^D_{j''=1} \Omega_i(u,d^{j''}) }{n^{-1} \sumln \Ybar_\ell(u)}
\sum^D_{j''=1}  \Omega_i(u, d^{j''}) dM_i(u,\Lambda_0).$$
Then, using the definitions of $\bA_i(u,q)$ and $\bG_i(q,\Lambda_0)$ above, 
we estimate (\ref{eq:if5}) by
$$n^{-2} \sumin  \bPsi_i(q,\Lambda_0) \bG_i(q,\Lambda_0)^T,$$
where in practice we replace $q(u,d^j)$ and $\Lambda_0(u)$ by the
estimators $\hatq(u,d^)$ and $\hatLam_0(u)$.  

Collecting these results, we propose the bias-corrected estimator for
$\bSigma$ given by
  \begin{align}
n^{-1} \sumin \{ \bPsi_i(\hatq,\hatLam_0) & \bPsi_i(\hatq,\hatLam_0)^T\}
+ n^{-2} \sumin \left\{ 2 \bPsi_i(\hatq,\hatLam_0) \bG_i(\hatq,\hatLam_0)^T
  + 2 \bG_i(\hatq,\hatLam_0) \bPsi_i(\hatq,\hatLam_0)^T\right\}
                                     \nonumber \\
&= \hatSig + n^{-1} \left[ n^{-1} \sumin \left\{ 2 \widehat{\Tbb}_i\bG_i(\hatq,\hatLam_0)^T
  + 2 \bG_i(\hatq,\hatLam_0) \widehat{\Tbb}_i^T\right\} \right], \label{eq:Sigest}
  \end{align}
  where $\hatSig$ is our prposed estimator and
  $\widehat{\Tbb}_i = \bPsi_i(\hatq,\hatLam_0)$.   The
second term in (\ref{eq:Sigest}) effects a bias correction that will
be nonnegligible when $n$ is not large, so that (\ref{eq:Sigest})
provides a more faithful representation of the uncertainty in
estimation of $\bSigma$.  

If in a SMART the randomization probabilities are estimated or modeled
as discussed in Section~\ref{s:webD}, with the possible addition of basis
functions of the history to enhance efficiency/power, to obtain a
bias-corrected covariance matrix estimator, form all of the quantities
above involving $\Omega(u,d^j)$, $j=1,\ldots,D$, by substituting the
fitted models $\omega_k(\bh_k,a_k; \hatgamma)$, $k=1,\ldots,K$, and 
replace
$\hatSig$ and $\widehat{\Tbb}_i$ in (\ref{eq:Sigest}) by
$\hatSig_\gamma$ and $\widehat{\Tbb}^R_i(\hatgamma)$.

\setcounter{equation}{0}
\renewcommand{\theequation}{F.\arabic{equation}}

\section{Simulation Details}
\label{s:webF}

\noindent
\textbf{F.1.  Implications of martingale property}\\*[-0.2in]

We examine the implications of Section D.4, namely, that the
martingale property need not hold, for the data generative Scenarios
1-3 in the simulation in Section~\ref{s:sims}, which
may offer insight into the nature of the simulation results.

First consider Scenarios 1(a) and 2(a) under the null hypothesis; as
noted in Section~\ref{s:sims}, Scenario 1(a) is the
first null scenario of \citet{KidwellLogrank}.  Let $\pi_R = P(R=1)$;
$\pi_R = 0.4$ in Scenarios 1(a) and 2(a).  Because $H_0$ holds, we
suppress the subscripts indicating treatment; thus, for $R=0$, let
$T^{NR}$, the potential event time, be exponential($\lambda_1$).  If
$R=1$, let the potential time to response, $T^R$, be
exponential($\lambda_2$) and the potential time to event, $T^{RE}$, be
exponential($\lambda_3$).  Then it is straightforward to show that the
hazard of experiencing the event prior to response at time $u$ is
given by
\begin{align}
\frac{(1-\pi_R) \lambda_1 \exp(-\lambda_1 u)}{(1-\pi_R) \exp(-\lambda_1
  u) + \pi_R  \exp(-\lambda_2 u)}.
  \label{eq:kidwell12}
  \end{align}
The hazard of experiencing the event after response ($R=1$) is equal
to $\lambda_3$ from above.  According to Section D.4, the
martingale property will hold if (\ref{eq:kidwell12}) is equal to
$\lambda_3$.  Moreover, it can be shown that
\begin{align}
\lambda_0(u) = \frac{(1-\pi_R) \lambda_1 \exp(-\lambda_1 u) + \pi_R
  \lambda_2 \lambda_3 \{\exp(-\lambda_3u) -
  \exp(-\lambda_2u)\}/(\lambda_2-\lambda_3)}
  { (1-\pi_R) \exp(-\lambda_1 u) + \pi_R \{\lambda_2\exp(-\lambda_3 u)
  - \lambda_3\exp(-\lambda_2 u)\}/(\lambda_2-\lambda_3)}.
\label{eq:kidwelllam0}
\end{align}

\begin{figure}
\centering
\includegraphics[width=12cm]{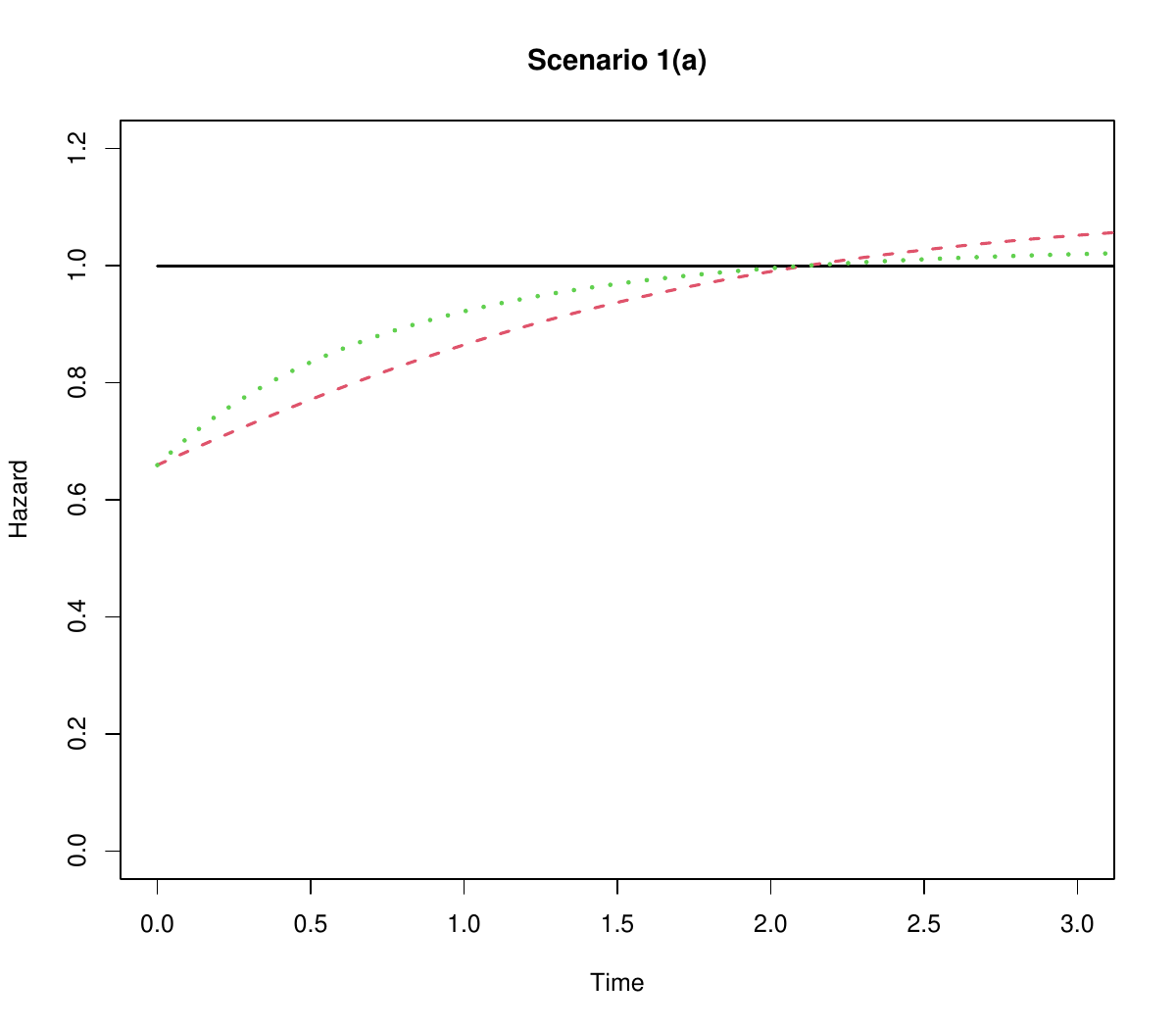}
\caption{\label{f:scenario1a}  Hazards under Scenario 1(a).  The solid
  line is $\lambda_3$, the dashed line is (\ref{eq:kidwell12}), and
  the dotted line is (\ref{eq:kidwelllam0}).}
\end{figure}

\begin{figure}
\centering
\includegraphics[width=12cm]{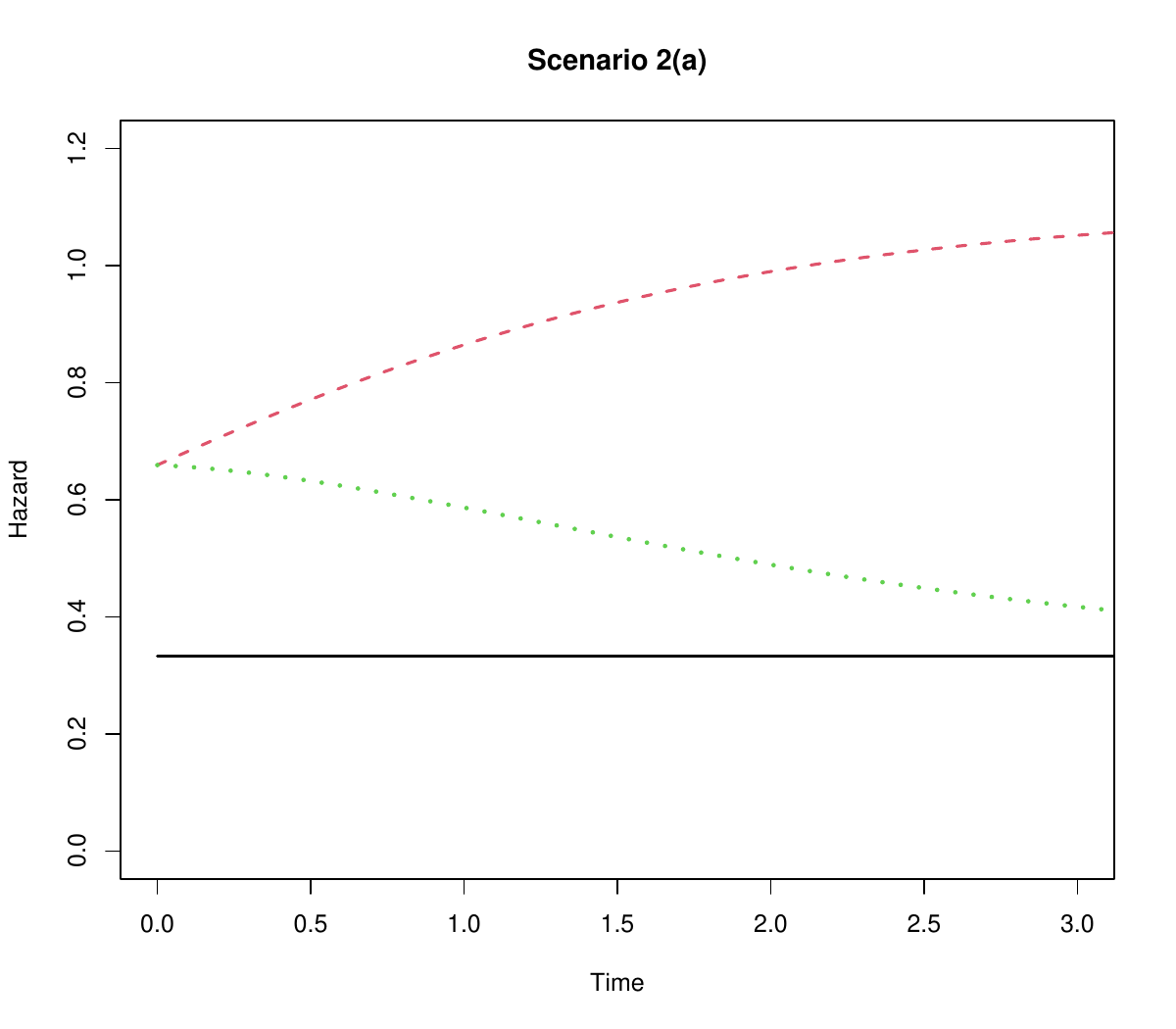}
\caption{\label{f:scenario2a}  Hazards under Scenario 2(a).  The solid
  line is $\lambda_3$, the dashed line is (\ref{eq:kidwell12}), and
  the dotted line is (\ref{eq:kidwelllam0}).}
\end{figure}

In Scenario 1(a), $\lambda_1 = 1/0.91 = 1.099$,
$\lambda_2 = 1/0.5 = 2$, and $\lambda_3 = 1$.  In Figure~\ref{f:scenario1a}, for these
values and $\pi_R=0.4$ we plot (\ref{eq:kidwell12}) (dashed line),
$\lambda_3$ (solid line), and (\ref{eq:kidwelllam0}) (dotted line).
The hazard in (\ref{eq:kidwell12}) differs from $\lambda_3$ over most
of the time range, but the greatest disparity, for small $u$, is not
substantial.  We conjecture that this relatively mild departure from
the martingale property is reflected in the fact that the test of
\cite{KidwellLogrank} achieves the nominal level.  In contrast, Figure~\ref{f:scenario2a}
shows the same plot for Scenario 2(a), for which
$\lambda_1 = 1/0.91 = 1.099$, $\lambda_2 = 1/0.5 = 2$, and
$\lambda_3 = 1/3$.  Here, the disparity between $\lambda_3$ (solid)
and (\ref{eq:kidwell12}) (dashed) is substantial, and we we conjecture
that the conservatism of the test of \cite{KidwellLogrank} in this
case, which persists as the sample size increases, is a consequence of
this feature.

Consider now Scenario 3 under the null hypothesis ($\zeta=0$) in the
simplified case in which there are no covariate effects, so that
$\psi=0$.  Under this generative scenario, the hazard of experiencing
the event prior to response at time $u$ is given by
$\lambda_1 = \exp(\alpha_{1D})$ and the hazard of experiencing the
event after response ($R=1$) is given by
$\lambda_3 = \exp(\alpha_{2AL})$.  Moreover, the hazard for the
potential time to Decision 2 $\lambda_2 = \exp(\alpha_{1SS})$, and
\begin{align}
  \lambda_0(u) = \frac{(\lambda_1+\lambda_2)
  (\lambda_1-\lambda_3)\exp\{ - (\lambda_1+\lambda_2)u\} + \lambda_2
  \lambda_3 \exp(-\lambda_3 u)}
  {(\lambda_1-\lambda_3)\exp\{ - (\lambda_1+\lambda_2)u\}  + \lambda_2
  \exp(-\lambda_3 u)  }.
\label{eq:beckylam0}
\end{align}
According to Section D.4, the martingale property holds in this
setting if $\lambda_1 = \lambda_3$.  In Scenario 3(a),
$\alpha_{1D} = \alpha_{2AL} = -5.5$; thus, without covariate effects
($\psi=0$), the martingale property holds exactly, and likely holds
approximately when covariate effects are included as in
Section~\ref{s:sims} ($\psi=1.5$).  As seen in Table~\ref{t:results},
the test of \citet{KidwellLogrank} achieves the nominal level, as
expected under these conditions.  In Scenario 3(b),
$\alpha_{1D} = -4.5$ and $\alpha_{2AL} = -5.5$, so that the martingale
property does not hold exactly or approximately; we conjecture that
the conservatism of the test of \cite{KidwellLogrank} in this case,
which persists as the sample size increases, is a consequence of this
feature.  In Scenario 3(c), $\alpha_{1D} = -5.5$ and
$\alpha_{2AL} = -3.5$, so that the martingale property does not hold
exactly or approximately.  Again, we conjecture that the
anti-conservatism of the test of
\cite{KidwellLogrank} reflects this feature.\\

\noindent
\textbf{F.2.  Details for Scenarios 4 and 5}\\*[-0.2in]

Scenario 4, which is based on the SMART design in Figure~\ref{f:four},
with an additional control regime, is similar to Scenario 3, with the
following modifications to incorporate the control regime.  Code the
control as $a_1 = 2$, so that $\calA_1 = \{0, 1, 2\}$.  We generated
$A_1$ as multinomial with probabilities 1/3, 1/3, 1/3 for
$a_1 = 0, 1, 2$, respectively.  For given $\alpha_{1D} = -5.5$,
$\alpha_{1SS} = -4.2$, $\alpha_{2AL} = -5.5$, $\psi = 1.5$, and
$\zeta$, we took
$\btheta_{1D} = (\alpha_{1D}, 0.5 \psi,0.5 \psi, -0.26 \zeta, 0.15
\zeta)^T$ and
$\btheta_{1SS} = (\alpha_{1SS}, 0.5 \psi,0.5 \psi, 0.24 \zeta -0.13
\zeta)^T$, generated potential event time $T_D$ as
exponential($\lambda_{1D}$),
$\lambda_{1D} = \exp\{ \theta_{1D,1} + \theta_{1D,2} X_{11} +
\theta_{1D,3} X_{12}+ \theta_{1D,4} I(A_1=1) + \theta_{1D,5}
I(A_1=2)\}$ and potential time to Decision 2 $T_{SS}$ as
exponential($\lambda_{1SS}$),
$\lambda_{1SS} = \exp\{\theta_{1SS,1} + \theta_{1SS,2} X_{11} +
\theta_{1SS,3} X_{12} + \theta_{1SS,4} A_1\}$, and took
$S = \min(T_D,T_{SS})$ and $R = I(T_{SS} < T_D)$.
If $R=1$, with
$\btheta_{X_2} = (0.2, 0.5\psi, 0.4 \psi, 0.12 \zeta, 0.1 \zeta)^T$
and
$\btheta_{2AL} = (\alpha_{2AL}, 0.5 \psi, -0.52 \psi, 0.6 \psi, -0.1
\zeta, 0.15 \zeta, -0.11 \zeta)^T$, we generated $X_2$ as
Bernoulli($p_{X_2}$),
$p_{X_2} = \expit\{\theta_{X_2,1} + \theta_{X_2,2} X_{11} +
\theta_{X_2,3} X_{12} + \theta_{X_2,4} I(A_1=1) + \theta_{X_2,5}
I(A_1=2)\}$, $A_2$ as Bernoulli(0.5), and ``added life'' post-response
$T_{AL}$ as exponential($\lambda_{2,AL}$),
$\lambda_{2AL} = \exp\{\theta_{2AL,1} + \theta_{2AL,2} X_{11} +
\theta_{2AL,3} X_{12} + \theta_{2AL,4} X_2 + \theta_{2AL,5} I(A_1=1) +
\theta_{2AL,6}I(A_1=2) + \theta_{2AL,7}A_2I(A_1<2) \}$.  All other
features of the scenario are as in Section~\ref{s:sims}.

  For Scenario 5, data generation is as in Scenario 3 at the first
  stage of the design, with $\alpha_{1D} = -5.5$ and
  $\alpha_{1SS} = -3.5$, $\psi = 1.5$, and $\btheta_{1D}$ and
  $\btheta_{1SS}$ as in Scenario 3 but with potential event time $T_D$
  as exponential($\lambda_{1D}$),
  $\lambda_{1D} = \exp(\theta_{1D,1} + \theta_{1D,2} X_{11} +
  \theta_{1D,3} X_{12}+ \theta_{1D,4} A_1)$ and potential time to
  Decision 2 $T_{SS}$ as exponential($\lambda_{1SS}$),
  $\lambda_{1SS} = \exp(\theta_{1SS,1} + \theta_{1SS,2} X_{11} +
  \theta_{1SS,3} X_{12} + \theta_{1SS,4} A_1)$.  We then took
  $S = \min(T_D,T_{SS})$ as before and additionally
  $\Gamma_{SS} = I(T_{SS} < T_D)$.  For subjects who would reach
  Decision 2, with $\Gamma_{SS} = 1$, we generated response status $R$
  as Bernoulli($p_R$), where
  $p_R = \expit(0.3 + 0.15 X_{11} +0.15 X_{12} + 0.2 \zeta A_1)$, and
  $X_2$ as in Scenario 3.  For these subjects with $\Gamma_{SS} = 1$,
  we generated $A_2$ as Bernoulli(0.5) within their observed value of $(A_1, R)$.
Then with $\btheta_{2AL} = (\alpha_{2AL}, 0.5 \psi, -0.52 \psi, 0.6 \psi, -0.1
\zeta, -0.11 \zeta, -0.3 \zeta)^T$, we generated ``added life''
  post-response $T_{AL}$ as exponential($\lambda_{2,AL}$),
  $\lambda_{2AL} = \exp\{\theta_{2AL,1} + \theta_{2AL,2} X_{11} +
  \theta_{2AL,3} X_{12} + \theta_{2AL,4} X_2 + \theta_{2AL,5} A_1
  + \theta_{2AL,6} A_2 + \theta_{2AL,7}R \}$.  We then took 
$T = (1-\Gamma_{SS}) T_D +  \Gamma_{SS} (T_{SS} + T_{AL})$ and, with $C$ 
uniform($0, c_{\textrm{max}})$,  $U = \min(T,C)$, $\Delta=I(T \leq
C)$.  If  $\Gamma_{SS}=1$ but $C < T_{SS}$ or $\Gamma_{SS} = 0$, take
$R$ to be undefined.   For subjects for whom $\Gamma_{SS} = 1$ and $C
\geq T_{SS}$, $\calT_2 = T_{SS}$ and $\kappa = 2$; if $\Gamma_{SS} = 1$
and $C < T_{SS}$ or $\Gamma_{SS} = 0$, $\kappa = 1$. \\

\noindent
\textbf{F.3.  Simulation studies of tests of pairwise comparisons of regimes}\\*[-0.2in]

The test statistic of \citet{KidwellLogrank} involves a $(3 \times 1)$
vector, where each component addresses a pairwise comparison of the
hazards for two of the four regimes in a SMART analogous to that in
Figure~\ref{f:four} without the additional control
regime.  Specifically, as in the figure, represent these regimes as
being of the form ``Give treatment $a$ inititally; if the event does
not occur before response status is ascertained and nonresponse,
continue, otherwise, if response, give treatment $b$,'' the four
regimes correspond to $(a,b) = (0,0), (0,1), (1,0), (1, 1)$.  Denoting
these as Regimes 1, 2, 3, and 4 in this order, the three components
involve the comparisons of 3 versus 4, 2 versus 4, and 1 versus 4;
that is, Regime 4 is the reference regime against which the other
three regimes are compared.  The comparison of Regimes 3 and 4
involves ``shared path'' regimes starting with the same stage 1
treatment, while the other two comparisons start with different stage
1 treatments.  The test statistic of \citet{KidwellLogrank} is formed
as a quadratic form involving this $(3 \times 1)$ vector and an
approximation to its sampling covariance matrix using martingale
theory, which, as discussed in Section~\ref{ss:compare}, may be
violated.  It is thus possible to construct from these elements three
test statistics as a component of the vector divided by its standard
error obtained from this sampling covariance matrix, where these test
statistics address the comparisons of Regimes 3 versus 4, 2 versus 4,
and 1 versus 4, respectively.

The above configuration is the situation in simulation Scenarios 1 - 3
in Section~\ref{s:sims}.  Study C9710,, depicted in
Figure~\ref{f:c9710}, is analogous, involving four embedded regimes of
the form ``Give consolidation therapy $a$; if subject completes
consolidation (responder) before the event occurs, give maintenance
therapy $b$,'' which we denote as Regimes 1, 2, 3, and 4 as
$(a, b) = (0,0), (0,1), (1,0), (1,1)$.

To evaluate the performance of the proposed methods and the test of
\citet{KidwellLogrank} for making pairwise comparisons, we considered
Scenario 1(a), which as in Section~\ref{s:webF}  is
expected to be robust to departure from the martingale property;
Scenario 2(a), which involves a strong departure from the martingale
property; Scenario 3(a), for which the martingale property holds
roughly except for covariate associations; and Scenario 3(c), which
reflects a departure from the martingale property and includes
covariate associations.  For each scenario, we used the test
statistics obtained from the overall test statistic of
\citet{KidwellLogrank} as decribed above and the various versions of
the proposed test statistic denoted as $\Zbb_{\textrm{U,nocov}}$,
$\Zbb_{\textrm{U,nocov}}$, $\Zbb_{\textrm{U,cov}}$, and
$\Zbb_{\textrm{C,cov}}$, where $\calD$ was taken to
be the subset of embedded regimes involved in the indicated pairwise
comparison (so $D=2$).  

Table~\ref{t:pair} shows the results.  Under the null hypothesis,
under Scenarios 1(a) and 3(a), the proposed methods with using the
bias-corrected covariance matrix and the test statistics of
\citet{KidwellLogrank} mostly achieve the nominal level, reflecting
for the latter that the martingale property is not significantly
violated.  However, under Scenarios 2(a) and 3(c), although the
proposed methods continue to achieve the nominal level, that of
\citet{KidwellLogrank} for testing the ``shared path'' comparison (3
versus 4) in Scenario 2(a) is anti-conservative, suggesting that the
anti-conservative performance of the overall test shown in
Table~\ref{t:results} stems from this comparison.  Likewise, under
Scenario 3(c), although the test of \citet{KidwellLogrank} achieves
the nominal level for the ``shared path'' comparison (3 versus 4), the
non-shared path comparisons (2 versus 4, 1 versus 4) exhibit
anti-conservatism, suggesting that the performance of the overall test
shown in Table~\ref{t:results} reflects this behavior.  We also show
power under the same alternatives in Section~\ref{s:sims} for
Scenarios 3(a) and 3(c), which involve covariates associated with the
event time; the comparison of Regimes 2 versus 4 is omitted under
Scenario 3(c) because the hazards for these regimes are very similar,
so that power is extremely low using all test statistics.  As for the
proposed tests of the overall null hypothesis in (\ref{eq:null}), the
proposed tests of pairwise comparisons show enhanced power when
covariates are incorporated.

\begin{table}[h]
  \caption{Simulation results for comparisons of pairs of embedded
    regimes under data generative scenarios described in
    Section~\ref{s:sims} under both the null hypothesis $H_0$ in
    (\ref{eq:null}) and alternatives based on 5000 Monte Carlo (MC)
    data sets, with all tests conducted with level of significance
    0.05.  Entries are as in Table~\ref{t:results} 
    except that $\Zbb_{\textrm{KW}}$ denotes the test based on the
    relevant component of the test statistic proposed by
    \citet{KidwellLogrank}.  For each scenario, ``3v4'' denotes the
    ``shared path'' comparison of regimes 3 and 4 (first component of
    the test of \citet{KidwellLogrank} and ``xv4'' denotes the
    ``non-shared path'' comparison of regimes x and 4 (second/third
    component of the test of \citet{KidwellLogrank}).}
  \label{t:pair}
  \centering
  \begin{footnotesize}
    \begin{tabular}{clcccccc} \Hline
\textbf{Scenario} &  $n$ & $\zeta$ &$\Zbb_{\textrm{U,nocov}}$
      & $\Zbb_{\textrm{C,nocov}}$  & $\Zbb_{\textrm{U,cov}}$  &  $\Zbb_{\textrm{C,cov}}$
      & $\Zbb_{\textrm{KW}}$\\
      \hline 
  &  \multicolumn{7}{c}{\textbf{Null Scenarios}} \\*[0.06in]   

1(a) 3v4 & 250 & -- & 0.066 & 0.048 & 0.064 & 0.048 & 0.056\\ %
              & 500 & -- & 0.058 & 0.047 & 0.056 & 0.046 & 0.050\\ %
              & 1000 & -- & 0.057 & 0.051 & 0.057 & 0.053 & 0.052 \\*[0.06in]   %

1(a) 2v4 & 250 & -- & 0.068 & 0.057 & 0.067 & 0.056 & 0.060\\ %
              & 500 & -- & 0.054 & 0.047 & 0.056 & 0.051 & 0.051\\
              & 1000 & -- & 0.052 & 0.050 & 0.052 & 0.050 & 0.052 \\*[0.06in]   

1(a) 1v4 & 250 & -- & 0.062 & 0.054 & 0.065 & 0.054 & 0.057\\ %
              & 500 & -- & 0.057 & 0.051 & 0.056 & 0.050 & 0.056\\
              & 1000 & -- & 0.055 & 0.051 & 0.053 & 0.050 & 0.052 \\*[0.06in]   
      
 2(a) 3v4 & 250 & -- & 0.068  &  0.048 & 0.068 & 0.051 & 0.090 \\ %
               & 500 & -- & 0.060  &  0.051 & 0.060 & 0.050 & 0.085 \\  %
               & 1000 & -- & 0.053 & 0.047 & 0.051 & 0.046 & 0.080\\*[0.06in]  %
    
2(a) 2v4  & 250 & -- & 0.061  &  0.047 & 0.067 & 0.054 & 0.055 \\ %
               & 500 & -- & 0.056  &  0.048 & 0.057 & 0.052 & 0.053 \\ %
               & 1000 & -- & 0.052 & 0.048 & 0.053 & 0.051 & 0.049\\*[0.06in]      

2(a) 1v4  & 250 & -- & 0.058  &  0.049 & 0.062 & 0.051 & 0.054 \\ %
               & 500 & -- & 0.058  &  0.050 & 0.061 & 0.053 & 0.053 \\ %
               & 1000 & -- & 0.053 & 0.049 & 0.057 & 0.053 & 0.053\\*[0.06in]      
      
  3(a), 3v4 & 250 & 0.00 & 0.063 & 0.050 & 0.060 & 0.049 & 0.053\\
                 & 500 & 0.00 & 0.050 & 0.045 & 0.051 & 0.045 & 0.047\\
                 & 1000 & 0.00 & 0.052 & 0.049 & 0.056 & 0.043 & 0.046\\*[0.06in]     
      
  3(a), 2v4 & 250 & 0.00 & 0.058 & 0.049 & 0.056 & 0.044 & 0.056\\
                 & 500 & 0.00 & 0.055 & 0.052 & 0.053 & 0.049 & 0.055\\
                 & 1000 & 0.00 & 0.051 & 0.046 & 0.054 & 0.052 & 0.054\\*[0.06in]     

  3(a), 1v4 & 250 & 0.00 & 0.056 & 0.047 & 0.057 & 0.049 & 0.053\\
                 & 500 & 0.00 & 0.053 & 0.049 & 0.057 & 0.051 & 0.053\\
                 & 1000 & 0.00 & 0.054 & 0.051 & 0.051 & 0.047 & 0.056\\*[0.06in]     
      
3(c), 3v4 & 250 & 0.00 & 0.059 & 0.046 & 0.060 & 0.0.046 & 0.053 \\
               & 500 &  0.00 & 0.054 & 0.048 & 0.050 & 0.045 & 0.055 \\
               & 1000 & 0.00 & 0.051 & 0.047 & 0.056 & 0.050 & 0.057 \\*[0.06in]

3(c), 2v4 & 250 & 0.00 & 0.055 & 0.048 & 0.058 & 0.049 & 0.073 \\
               & 500 &  0.00 & 0.050 & 0.046 & 0.051 & 0.045 & 0.073 \\
               & 1000 & 0.00 & 0.052 & 0.049 & 0.051 & 0.049 & 0.080 \\*[0.06in]

3(c), 1v4 & 250 & 0.00 & 0.057 & 0.050 & 0.058 & 0.050 & 0.070 \\
               & 500 &  0.00 & 0.055 & 0.049 & 0.056 & 0.051 & 0.075 \\
               & 1000 & 0.00 & 0.047 & 0.045 & 0.048 & 0.046 & 0.074 \\*[0.06in]

                  & \multicolumn{7}{c}{\textbf{Alternative Scenarios}} \\*[0.06in]


3(a), 3v4  & 1000 & 1.75 & 0.195 & 0.188 & 0.242 & 0.234 & 0.147\\
3(a), 2v4  & 1000 & 1.75 & 0.581 & 0.570 & 0.678 & 0.671 & 0.563\\
3(a), 1v4  & 1000 & 1.75 & 0.845 & 0.840 & 0.915 & 0.911 & 0.848\\*[0.06in]    
      
3(c), 3v4  & 1000 & 3.50 & 0.625 & 0.614 & 0.766 & 0.755 & 0.476 \\
3(c), 1v4  & 1000 & 3.50 & 0.340 & 0.334 & 0.444 & 0.435 & 0.343 \\*[0.06in]                                           

    \hline
    \end{tabular}
    \end{footnotesize}
\end{table}

\noindent
\textbf{F.4.  Simulation study of comparison of more complex regimes}\\*[-0.2in]

To demonstrate that the proposed methods can be used to compare the
regimes in an arbitrary set of regimes that are ``feasible'' given the
available data in the sense discussed in
\citet[Section~6.2.3]{TsiatisBook}, consider Scenario 3(a) of 
Section~\ref{s:sims}, where $\calA_1 = \{0, 1\}$ and
$\calA_2 = \{0, 1\}$.  Recall that by construction the covariates are
associated with the event time outcome.  We take the set of regimes of interest to
be $\calD = \{d^1, d^2,d^3\}$ ($D=3$), where the regimes are defined as
follows:
\begin{itemize}
\item Regime 1, $d^1$, with rules $d^1_1(\bh_1) = I(X_{12}
  \geq 0.3)$, $d^1_2(\bh_2) = I(X_{12} \geq 0.4, X_2=1, R=1)$

  \item Regime 2, $d^2$, with rules $d^2_1(\bh_1) = I(X_{12}
  \leq 0.5)$, $d^2_2(\bh_2) = I(X_{12} \geq 0.6, X_2=1, R=1)$

 \item Regime 3, $d^3$, with rules $d^3_1(\bh_1) = I(X_{12}
  \geq 0.7)$, $d^3_2(\bh_2) = I(X_{12} \geq 0.8, X_2=0, R=1)$.
  \end{itemize}
The choice of these regimes is completely arbitrary and for
demonstration only.  
  
  Table~\ref{t:fancy} shows the results under the null hypothsis of no
  difference in hazard rate among these three regimes and demonstrates
  that the proposed methods yield a test achieving the nominal
  level of significance.  

\begin{table}[h]
  \caption{Simulation results for comparisons of a set of three
    arbitrary regimes under the null hypothesis $H_0$ in
    (\ref{eq:null}) based on 5000 Monte Carlo (MC) data sets, with all
    tests conducted with level of significance 0.05.  Entries are as
    in Table~\ref{t:results}.} 
  \label{t:fancy}
  \centering
  \begin{footnotesize}
    \begin{tabular}{clccccc} \Hline
\textbf{Scenario} &  $n$ & $\zeta$ &$\Zbb_{\textrm{U,nocov}}$
      & $\Zbb_{\textrm{C,nocov}}$  & $\Zbb_{\textrm{U,cov}}$  &  $\Zbb_{\textrm{C,cov}}$\\
      \hline 
  &  \multicolumn{6}{c}{\textbf{Null Scenario}} \\*[0.06in]   

      3(a)  & 250 & 0.00 & 0.056 & 0.045 & 0.062 & 0.049 \\ %
              & 500 & 0.00 & 0.054 & 0.049 & 0.053 & 0.048 \\ %
              & 1000 & 0.00 & 0.051 & 0.048 & 0.053 & 0.050 \\*[0.06in]   %
    \hline
    \end{tabular}
    \end{footnotesize}
\end{table}

\noindent
\textbf{F.5.  Simulation study of three decision SMART}\\*[-0.2in]

All of the simulations reported in Section~\ref{s:sims} and so far in
this section involve SMARTs with two decision points, which is
arguably by far the most common situation in practice.  To demonstrate
that the proposed methods can be used in the more complex setting of a
SMART with three decision points, we consider an extension of Scenario
3 to involve an additional decision point.  The generative scenario is
as follows and is such that the SMART embeds eight regimes in which
subjects who reach the second and third decision points are are
re-randomized.

We first generated $\bX_1 = (X_{11},X_{12})^T$, where
$X_{11} \sim \calN(0,1)$ and $X_{12} \sim$ uniform$(0,1)$, and $A_1$
as Bernoulli(0.5).  For given $\alpha_{1D}$, $\alpha_{1SS}$,
$\psi$, and $\zeta$, we took
$\btheta_{1D} = (\alpha_{1D}, 0.5 \psi,0.5 \psi, -0.26 \zeta)^T$ and
$\btheta_{1SS} = (\alpha_{1SS}, 0.5 \psi,0.5 \psi, 0.24 \zeta)^T$,
generated potential event time $T_{D1}$ as
exponential($\lambda_{1D}$),
$\lambda_{1D} = \exp\{\theta_{1D,1} + \theta_{1D,2} X_{11} +
\theta_{1D,3} (X_{12}-0.5) + \theta_{1D,4} (A_1-0.5)\}$ and potential
time to Decision 2 $T_{SS}$ as
exponential($\lambda_{1SS}$),
$\lambda_{1SS} = \exp\{\theta_{1SS,1} + \theta_{1SS,2} X_{11} +
\theta_{1SS,3} (X_{12}-0.5) + \theta_{1SS,4} (A_1-0.5)\}$, and took
$S_2 = \min(T_{D1},T_{SS})$, $R_2 = I(T_{SS} < T_{D1})$.  If $R_2=1$, 
with $\btheta_{X_2} = (0.2, 0.5\psi, 0.4 \psi, 0.12 \zeta)^T$, we generated $X_2$ as
Bernoulli($p_{X_2}$),  $p_{X_2} = \expit(\theta_{X_2,1} + \theta_{X_2,2} X_{11} + \theta_{X_2,3}
X_{12} + \theta_{X_2,4} A_1)$, and $A_2$ as Bernoulli(0.5).  
For given $\alpha_{2D}$ and $\alpha_{2TS}$, we took
$\btheta_{2D} = (\alpha_{2D}, 0.5 \psi,-0.52 \psi, 0.6 \psi,-0.1 \zeta,-0.11 \zeta)^T$ and
$\btheta_{2TS} = (\alpha_{2TS}, 0.5 \psi,-0.52 \psi, 0.6 \psi,-0.1 \zeta,-0.11 \zeta)^T$,
and generated potential event time $T_{D2}$ as 
exponential($\lambda_{2D}$),
$\lambda_{2D} = \exp\{\theta_{2D,1} + \theta_{2D,2} X_{11} +
\theta_{2D,3} (X_{12}-0.5) + \theta_{2D,4} (X_2-p_{X_2}) +
\theta_{2D,5}(A_1-0.5) + \theta_{2D,6} (A_2 - 0.5)\}$ and potential
time to Decision 3 $T_{TS}$ as
exponential($\lambda_{2TS}$),
$\lambda_{2TS} = \exp\{\theta_{2TS,1} + \theta_{2TS,2} X_{11} +
\theta_{2TS,3} (X_{12}-0.5) + \theta_{2TS,4} (X_2-p_{X_2}) +
\theta_{2TS,5} (A_1-0.5) + \theta_{2TS,6} (A_2-0.5)\}$, and took
$S_3 = \min(T_{SS}+T_{D2},T_{SS}+T_{TS})$, $R_3 = I(T_{TS} < T_{D2})$.
With $\btheta_{X_3} = (0.2, 0.5\psi, 0.4 \psi, 0.12 \zeta, -0.15 \zeta)^T$,
if $R_3=1$, we generated $X_3$ as
  Bernoulli($p_{X_3}$),  $p_{X_3} = \expit(\theta_{X_3,1} + \theta_{X_3,2} X_{11} + \theta_{X_3,3}
X_{12} + \theta_{X_3,4} X_2 + \theta_{X_3,5} A_1 + \theta_{X_3,6}A_2)$, and $A_3$ as Bernoulli(0.5).
Finally, for given $\alpha_{3AL}$, with 
$\btheta_{3AL} = (\alpha_{3AL}, 0.5 \psi,-0.52 \psi, 0.6 \psi,0.1
\psi,-0.1 \zeta,-0.11 \zeta,0.2 \zeta)^T$,
we generated ``added life''  $T_{AL}$ as exponential($\lambda_{3AL}$), 
$\lambda_{3AL} = \exp\{\theta_{3AL,1} + \theta_{3AL,2} X_{11} +
\theta_{3AL,3} (X_{12}-0.5) + \theta_{3AL,4} (X_2 - p_{X2}) +
\theta_{3AL,5} (X_3 - p_{X3}) + \theta_{1AL,6} (A_1-0.5) +
\theta_{1AL,7}(A_2-0.5) +\theta_{1AL,8}(A_3-0.5)\}$.
We then took 
$T = (1-R_2) T_{D1} + R_2(1-R_3) (T_{SS} + T_{D2}) + R_2R_3 (T_{SS} +
T_{TS} + T_{AL})$, and, with $C$ 
uniform($0, c_{\textrm{max}})$,
$U = \min(T,C)$, $\Delta=I(T \leq
C)$; if $R_2=1$ and $C < S_2$, redefine $R_2=0$, and if $R_3=1$ and
$C< S_3$, redefine $R_3=0$.   For $R_2=1$, $\calT_2
= T_{SS}$ and, if $R_3=0$, $\kappa=2$.  If $R_2=1$ and $R_3=1$, $\calT_3
= T_{SS}+T_{TS}$, $\kappa=3$; else, $\kappa=1$.

Under this scenario, the eight regimes are of the form ``Give
treatment $a$ inititally; if the event does not occur before treatment
with $a$ concludes, continue to Decision 2 and give treatment $b$; if
the event does not occur before treatment with $b$ concludes, continue
to Decision 3 and give treatment $c$,''  The eight regimes, which we
denote in order as regimes 1, 2, $\ldots$, 8, correspond
to
$(a,b,c) = (0,0,0), (0,1,0), (1,0,0), (1, 1,0), (0,0,1), (0,1,1),
(1,0,1), (1, 1,1)$.

For the simulations reported in Table~\ref{t:three}, we took
$\zeta=0.0$, so focusing on performance under the null hypothesis;
$\alpha_{1D}=-4.5$, $\alpha_{1SS} = -3.2$, $\alpha_{2D} = -4.0$,
$\alpha_{2TS} = -2.7$, and $\alpha_{3AL} = -3.0$; and
$c_{\textrm{max}}) = 350$, corresponding to about 15\% censoring.  The
comparison of regimes 1 and 8 corresponds to comparison of
``non-shared path'' regimes; that of regimes 1 and 2 corresponds to
comparison of ``shared path'' regimes that share the same first stage
treatment; and that of regimes 1 and 5 corresponds to comparison of
``shared path'' regimes that share the same first and second stage
treatments.  The results show that, for comparison of all eight
embedded regimes, all tests are generally anti-conservative until
$n=4000$, where $\Zbb_{\textrm{C,nocov}}$ and $\Zbb_{\textrm{C,cov}}$
achieve the nominal level (within Monte Carlo error), suggesting that
large sample sizes may be needed to achieve the nominal level for an
overall test.  This result may not be too surprising, as with three
decision points the numbers of subjects with experience consistent
with any regime may be small when $n$ is not large.  Comparison of
pairs of ``non-shared path'' and ``shared path'' regimes can be carried
out reliably with a much smaller overall sample size, as the nominal
level is achieved for the tests based on $\Zbb_{\textrm{C,nocov}}$ and
$\Zbb_{\textrm{C,cov}}$.  Not surprisingly, comparison of four ``shared path''
regimes requires a larger sample size to achieve the nominal level,
but considerably smaller than that required for the overall comparison
of all eight regimes.

\begin{table}[h]
  \caption{Simulation results for comparisons of sets of regimes in
    the three decision SMART scenario under the null hypothesis $H_0$
    in (\ref{eq:null}) based on 5000 Monte Carlo (MC) data sets, with
    all tests conducted with level of significance 0.05.  The first
    column indicates the set of regimes $\calD$ and thus the
    comparison of interest; ``All'' indicates that $\calD$ corresponds
    to all eight embedded regimes.  Otherwise, entries are
    as in Table~\ref{t:results}.}
  \label{t:three}
  \centering
  \begin{footnotesize}
    \begin{tabular}{clccccc} \Hline
\textbf{Comparison} &  $n$ & $\zeta$ &$\Zbb_{\textrm{U,nocov}}$
      & $\Zbb_{\textrm{C,nocov}}$  & $\Zbb_{\textrm{U,cov}}$  &  $\Zbb_{\textrm{C,cov}}$\\
      \hline 
  &  \multicolumn{6}{c}{\textbf{Null Scenario}} \\*[0.06in]   

   All   & 1000 & 0.00 & 0.072 & 0.067 & 0.071 & 0.063 \\ %
          & 1500 & 0.00 & 0.058 & 0.056 & 0.067 & 0.061 \\ %
          & 2000 & 0.00 & 0.059 & 0.057 & 0.059 & 0.055 \\ %
          & 4000 & 0.00 & 0.057 & 0.056 & 0.054 & 0.053   \\*[0.06in]   %

  1,8   & 500 & 0.00 & 0.060 & 0.053 & 0.059 & 0.049 \\ %
          & 1000 & 0.00 & 0.053 & 0.049 & 0.051 & 0.047 \\ %
          & 1500 & 0.00 & 0.051 & 0.048 & 0.049 & 0.044   \\*[0.06in]   %

  1,2   & 500 & 0.00 & 0.062 & 0.052 & 0.066 & 0.054 \\ %
          & 1000 & 0.00 & 0.055 & 0.049 & 0.057 & 0.048 \\ %
          & 1500 & 0.00 & 0.047 & 0.043 & 0.054 & 0.048   \\*[0.06in]   %

  1,5   & 500 & 0.00 & 0.067 & 0.054 & 0.065 & 0.045 \\ %
          & 1000 & 0.00 & 0.056 & 0.049 & 0.059 & 0.048 \\ %
          & 1500 & 0.00 & 0.058 & 0.051 & 0.054 & 0.047   \\*[0.06in]   %

1,2,5,6  & 1000 & 0.00 & 0.067 & 0.062 & 0.066 & 0.057 \\ %
             & 1500 & 0.00 & 0.058 & 0.054 & 0.054 & 0.048   \\*[0.06in]   %

    \hline
    \end{tabular}
    \end{footnotesize}
\end{table}

\setcounter{equation}{0}
\renewcommand{\theequation}{G.\arabic{equation}}

\section{Data Analysis Details}
\label{s:webG}

\noindent
\textbf{G.1.  Data definitions and data cleaning} \\*[-0.2in]

The data from North American Leukemia Intergroup Study C9710 provided
to us by the Alliance for Clinical Trials in Oncology comprise
information on 538 subjects; this data set is the same as that
considered by \citet{Hager}.  Ten (10) baseline covariates (components
of $\bX_1$) were ideally collected on each subject at the time of
registration into the trial: age, gender, race, ethnicity, Eastern
Cooperative Oncology Group (ECOG) performance status, risk group,
white blood cell count, platelet count, serum creatinine, and
hemoglobin.  Table~\ref{t:baseline} shows these covariates.  For
responders, $\bX_2$ includes variables summarizing adverse events and
toxicities occurring during induction or consolidation therapy.  We
consider only adverse events experienced by at least 5\% of patients.
Table~\ref{t:adverse} shows the adverse events, which are coded with
an eight-digit Medical Dictionary for Regulatory Activities (medDRA)
code.  In the data, the adverse events were recorded as occurring
during induction therapy (coded by ``I''), during consolidation
therapy (coded by ``C''), and during either induction or consolidation
therapy (coded by ``IC'').  Thus, the binary variable ``IC\_90004060''
represents whether or not hemorrhage or bleeding with grade 3 or 4
thrombocytopenia occurred during either induction or consolidation
therapy.  There are 46 such variables in $\bX_2$; three of the
variables are trinary; all others are binary.

\begin{table}[ht]\caption{Baseline covariates available from the North
    American Leukemia Intergroup Study C9710. Race is redefined as
    indicated in the text.}
	\small
	\centering
	\begin{tabular}{l  p{10cm} }
		\Hline
		Variable & Meaning\\
		\hline
		Age & Age at registration\\ 
		Gender & 1=Male, 2=Female \\ 
		Race & 1=White, 2=Hispanic American, 3=Black/African American, 9=Other\\ 
		Ethnicity & 1=Hispanic, 2=Non-Hispanic, 9=Unknown \\
		WBC & White blood cell count at registration ($10^3/\mu$L)\\ 
		Platelet & Platelet count at registration ($10^3/\mu$L) \\
		ECOG Performance Status& 0 =
                                         fully active, 1= restricted
                                         in strenuous activity, 2 =
                                         restricted in work activity,
                                         3 = capable of limited self
                                         care, 4 = completely
                                         disabled\\ 
		Risk Group & 1=Low (WBC $\leq$ 10 and Platelet $>$
                             40), 2=Intermediate (WBC $\leq$ 10 and
                             Platelet $\leq$ 40), 3=High (WBC $>$ 10) \\
		Creatinine & Serum creatinine (mg/dL) \\ 
		Hemoglobin & Hemoglobin (g/dL)\\
		\hline
	\end{tabular}
	\label{t:baseline}
\end{table}

\begin{table}[ht]\caption{medDRA codes and definitions of the adverse events experienced by at least 5\% of patients in North American Leukemia Intergroup Study C9710.}
	\small
	\centering
	\begin{tabular}{ c p{5cm} p{7.75cm} }
		\Hline
		medDRA & Adverse Event Category & Adverse Event\\
		\hline                     
		10002646 & Gastrointestinal & Anorexia\\
		10012457 & D\MakeLowercase{ERMATOLOGY/SKIN} & Rash/desquamation\\
		10012745 & Gastrointestinal & Diarrhea \\
		10013442 & C\MakeLowercase{OAGULATION} & Disseminated intravascular coagulation\\
		10013972 & P\MakeLowercase{ULMONARY} & Dyspnea (shortness of breath)\\
		10016288 & Infection & Febrile neutropenia, fever of
                                       unknown origin without
                                       clinically or microbiologically
                                       documented infection (ANC
                                       $<$1.0 x 10e9/L, fever $\geq$ 8.5 degrees C) \\
		10018876 & Blood/bone marrow & Hemoglobin \\
		10019218 & Pain & Headache\\
		10020637 & Metabolic/laboratory & Glucose serum-high (hyperglycemia)\\
		10020947 & Metabolic/laboratory & Calcium serum-low (hypocalcemia)\\
		10021015 & Metabolic/laboratory & Potassium serum-low (hypokalemia)\\
		10021143 & P\MakeLowercase{ULMONARY/UPPER RESPIRATORY} & Hypoxia \\
		10021842 & I\MakeLowercase{NFECTION/FEBRILE NEUTROPENIA} & Infection without neutropenia\\
		10024285 & Blood/bone marrow & Leukocytes (total WBC)\\
		10025327 & Blood/bone marrow & Lymphopenia\\
		10028813 & Gastrointestinal & Nausea \\
		10029363 & Blood/bone marrow & Neutrophils/granulocytes (ANC/AGC)\\
		10033359 & Blood/bone marrow & Transfusion: packed red blood cells\\
		10035528 & Blood/bone marrow & Platelets\\
		10035543 & Blood/bone marrow & Transfusion: platelets\\
		10043607 & Vascular & Thrombosis/thrombus/embolism\\
		90004060 & H\MakeLowercase{EMORRHAGE} & Hemorrhage/bleeding with grade 3 or 4 thrombocytopenia\\
		90004070 & I\MakeLowercase{NFECTION/FEBRILE NEUTROPENIA} & Infection (documented clinically or microbiologically) with grade 3 or 4 neutropenia (ANC $<$1.0 x 10e9/L) \\                
		\hline 
	\end{tabular}
	\label{t:adverse}
\end{table}

For the data analysis reported in Section~\ref{s:example}, we
considered a subset of these data, similar to \citet{Hager}.  First,
the following subjects were removed from the data set: 57 subjects
deemed ineligible according to the age criterion in
\citet{Powell2010}, 7 subjects with more than two covariates missing,
and 6 subjects for whom the date of initiation of induction therapy is
missing.  Fifteen (15) subjects are missing the start date of
maintenance therapy; the actual event or censoring time for these
subjects was artificially censored at the start date of consolidation
therapy.  One (1) such subject is also missing the start date of
consolidation therapy; this subject was also removed from the data set
(\citet{Hager} did not remove this subject).  After these deletions,
$n=467$ subjects were available for analysis.

Of these subjects, the first 50 to enroll in the study had different
maintenance treatment options; of the 47 of these subjects not already
removed, all reached stage 2, and the actual event or censoring time
for these subjects was artificially censored at the start date of
maintenance therapy.  For four (4) other subjects, the start date for
maintenance therapy was recorded as being before or equal to the date
of registration in the study; the actual event/censoring time for
these subjects was artificially censored at the date of start of
consolidation therapy.

Taking into account all of these conventions, among the $n=467$
subjects, 310 reached stage 2 and were randomized to maintenance
therapy options.  

As in \citet{Hager}, several of the baseline covariates were redefined
as follows:
\begin{itemize}
\item The number of race groups is reduced to only include ``White,''
  ``Hispanic American,'' ``Black/African American,'' and ``Other.''
  Patients previously classified as ``Asian,'' ``Native Hawaiian or
  Pacific Islander,'' ``American Indian or Alaska Native,'' ``Indian
  Subcontinent,'' or ``Multiple Races Reported'' are classified as
  ``Other.''
\item 1 subject has a platelet count of 5300, which is much higher
  than those for all other subjects, who had values in the range
  1--232. This observation was divided by 100 for consistency with the
  range given in \citet{Powell2010}.

\item 4 subjects have hemoglobin levels of 87, 92, 81, and 80, which
  are much higher than those for the rest of the subjects, who had
  values in the range of 4.3--14.6.  It is believed that these values were
  entered in units of g/L instead of g/dL, so we divided these hemoglobin
  observations by 10. 

\item 7 subjects have creatinine levels of 76, 95, 87, 92, 90, 52, and
  53, which are much higher than those for the rest of the subjects,
  who had values in a range of 0.1--10.4. It is believed these were
  entered in units of $\mu$mol/L instead of mg/dL. These creatinine
  observations are divided by 88.4 to correct the
  units. 
\end{itemize} 
The data after these adjustments are consistent with the information
presented in \citet{Powell2010}.

Of the $n=467$ subjects, 446 have complete data on all baseline
covariates, and the remainder have missing values for no more than 2
covariates.  The missing values were imputed as follows:
\begin{itemize}

\item Race is set to ``Other'' for the 8 subjects for whom it is missing. 
	
\item 5 subjects are missing creatinine. A linear model to estimate
  creatinine using all other baseline covariates was fitted to the
  data from the complete cases and used to impute the missing values.

\item 1 subject is missing white blood cell count (WBC). A linear
  model to estimate WBC using all baseline covariates except for WBC
  and Risk group was fitted to the data from the complete cases and
  used to impute the missing WBC value.

\item 4 subjects are missing ECOG performance status. A multinomial
  logistic regression model for ECOG performance status using all
  other baseline covariates was fitted to the data from the complete
  cases. The missing ECOG performance status values were imputed using
  this model.

\item 4 subjects are missing hemoglobin. A linear model to estimate
  hemoglobin using all other baseline covariates was fitted to the
  data from the complete cases and used to impute the missing values.
\end{itemize}

\noindent
\textbf{G.2.  Data analysis} \\*[-0.2in]

At stage 1, let $A_1 = 0$ if a subject was randomized to ATRA
consolidation therapy and $A_1 = 1$ if randomized to ATRA+Arsenic
Trioxide.  Here, a subject is a ``nonresponder'' if the subject
experienced the event or was censored before completing consolidation
therapy, with $\kappa=1$.  A subject is a ``responder'' if the subject
completed consolidation without experiencing the event or censoring,
$\kappa=2$; for such subjects, let $A_2 = 0$ if a subject was
randomized at stage 2 to ATRA maintenance therapy and $A_2 = 1$ if
randomized to ATRA+Mtx+MP.  For consistency with the examples in
Sections~\ref{s:webB} and \ref{s:webD}, define $R_2 = I(\kappa=2)$, although this
definition is redundant, as all subjects for whom $\kappa=2$ are ``responders.''

For the proposed methods, as in Section~\ref{s:webD}, we estimated the
randomization probabilities by positing and fitting logistic
regression models
$$\omega_1(\bh_1,a_1; \gamma_1) = \left\{ \frac{\exp(\gamma_1)}{1+\exp(\gamma_1)}\right\}^{I(a_1=1)}
\left\{ \frac{1}{1+\exp(\gamma_1)}\right\}^{I(a_1=0)}$$
at stage 1 and
\begin{align*}
 \omega_2&(\bh_2, a_2; \bgamma_2) = \left\{ \frac{\exp(\gamma_{21})}{1+\exp(\gamma_{21})}\right\}^{I(a_1=0, r_2=1,a_2=1)} \left\{ \frac{1}{1+\exp(\gamma_{21})}\right\}^{I(a_1=0, r_2=1,a_2=0)}\\
&\times \left\{ \frac{\exp(\gamma_{22})}{1+\exp(\gamma_{22})}\right\}^{I(a_1=1, r_2=1,a_2=1)} \left\{ \frac{1}{1+\exp(\gamma_{22})}\right\}^{I(a_1=1, r_2=1,a_2=0)}
\end{align*}
at stage 2.  Letting $\hatgamma$ be the ML estimator for
$\bgamma = (\gamma_1,\gamma_{21},\gamma_{22})^T$, as in
Section~\ref{s:webD}, to obtain our test statistic without
incorporating covariates to gain efficiency, for each regime
$j = 1,\ldots,4$, we performed a linear regression with dependent
variable $\widehat{\Tfrak}^j_i(\hatgamma)$ and design matrix with
$i$th row
\begin{equation}
  \label{eq:designmat1}
\begin{aligned}
\big[ \{ I(A_{1i} = 1) &- \expit(\widehat{\gamma}_1)\},
I(A_{1i}=0, R_{2i}=1, \kappa_i=2) \{ I(A_{2i}=1) - \expit(\widehat{\gamma}_{21})\}, \\
&I(A_{1i}=1, R_{2i}=1, \kappa_i=2) \{ I(A_{2i}=1) -  \expit(\widehat{\gamma}_{22})\}
\end{aligned}
\end{equation}
 to obtain the residuals $\widehat{\Tfrak}^{j,R}_i(\hatgamma)$ and thus
 $\hatSig_\gamma$.

 To attempt to enhance power by incorporating covariates, we
 informally examined associations between the covariates and outcome.
 To gain a sense of the extent to which components of the baseline
 covariates $\bX_1$ in Table~\ref{t:baseline} are potentially
 associated with outcome, we fit separate proportional hazards models
 to the data $(U,\Delta)$ for the $n=467$ subjects, where in each
 model we included each component of $\bX_1$ as the sole covariate.
 Because the distribution of WBC is extremely skewed, we considered
 the logarithm of WBC as a covariate, and all discrete covariates were
 treated as categorical.  From these fits, we identified
 $(X_{11},X_{12},X_{13}) = \{\log(\textrm{WBC}), I(\textrm{ECOG
   Performance Status} > 1), I(\textrm{Risk Group} > 2)\}$ as potentially
 associated with the EFS outcome.  As $X_{11}$ and $X_{13}$ are highly
 associated owing to the definition of the latter, we opted to
 consider $X_{11}$ and $X_{12}$ as covariates for the purpose of
 gaining efficiency as discussed in Section~\ref{s:webD}.  Thus, analogous
 to (\ref{eq:basisfunc}), we defined the vector of basis functions
 $\tilH_1 = (X_{11},X_{12})^T$.

To gain an informal sense of the extent to which adverse event
variables in $\bX_2$ are associated with outcome, we considered only
the 310 subjects who were observed to reach the second decision point
($\kappa = 2$), and fit separate proportional hazards models to the
data $(U,\Delta)$ including each component of $\bX_2$ as the sole
covariate, treating each covariate as categorical.  From these fits,
we identified the binary variables $X_{21} = $ I\_10028813, Nausea,
and $X_{22} = $ IC\_90004060, hemorrhage
or bleeding with grade 3 or 4 neutropenia, as potentially associated
with the EFS outcome.  Thus, analogous to (\ref{eq:basisfunc}), we
defined vectors of basis functions $\tilH_{21} = \tilH_{22} =
(X_{11},X_{12},X_{21},X_{22})^T$,  appended to the $i$th row of the design
matrix in (\ref{eq:designmat1}) the additional components
\begin{align*}
\big[ \{ I(A_{1i} = 1) &- \expit(\widehat{\gamma}_1)\} \tilH_1^T,
I(A_{1i}=0, R_{2i}=1, \kappa_i=2) \{ I(A_{2i}=1) -
                         \expit(\widehat{\gamma}_{21})\} \tilH_{21}^T, \\
&I(A_{1i}=1, R_{2i}=1, \kappa_i=2) \{ I(A_{2i}=1) -  \expit(\widehat{\gamma}_{22})\}\tilH_{22}^T
\end{align*}
and carried out the regression with dependent variable
$\widehat{\Tfrak}^j_i(\hatgamma)$ and obtained the residuals
$\widehat{\Tfrak}^{j,R}_i(\hatgamma)$ and thus $\hatSig_\gamma$ and
the test statistic $\Zbb(\hatgamma)$.

\backmatter

\section*{Acknowledgments}

The authors thank the Alliance for Clinical Trials in Oncology for
providing the Study C9710 data.   This research was supported by
NIH grant R01CA280970.

\bibliographystyle{biom}\bibliography{logrank}

\begin{thebibliography}{}

\bibitem[\protect\citeauthoryear{Almirall, Nahum-Shani, Sherwood, and
  Murphy}{Almirall et~al.}{2014}]{almirall2014introduction}
Almirall, D., Nahum-Shani, I., Sherwood, N.~E., and Murphy, S.~A. (2014).
\newblock Introduction to {SMART} designs for the development of adaptive
  interventions: with application to weight loss research.
\newblock {\em Translational Behavioral Medicine} {\bf 4,} 260--274.

\bibitem[\protect\citeauthoryear{Bigirumurame, Uwimpuhwe, and
  Wason}{Bigirumurame et~al.}{2022}]{bigirumurame2022sequential}
Bigirumurame, T., Uwimpuhwe, G., and Wason, J. (2022).
\newblock Sequential multiple assignment randomized trial studies should report
  all key components: a systematic review.
\newblock {\em Journal of Clinical Epidemiology} {\bf 142,} 152--160.

\bibitem[\protect\citeauthoryear{Boos}{Boos}{1992}]{Boos}
Boos, D.~D. (1992).
\newblock On generalized score tests.
\newblock {\em The American Statistician} {\bf 46,} 327--333.

\bibitem[\protect\citeauthoryear{Feng and Wahed}{Feng and
  Wahed}{2008}]{FengWahed}
Feng, W. and Wahed, A.~S. (2008).
\newblock Supremum weighted log-rank test and sample size for comparing
  two-stage adaptive treatment strategies.
\newblock {\em Biometrika} {\bf 95,} 695--707.

\bibitem[\protect\citeauthoryear{Guo and Tsiatis}{Guo and
  Tsiatis}{2005}]{GuoTsiatis}
Guo, X. and Tsiatis, A.~A. (2005).
\newblock A weighted risk set estiamtor for survival distributions in two-stage
  randomization designs with censored survival data.
\newblock {\em The International Journal of Biostatistics} {\bf 1,} 1.

\bibitem[\protect\citeauthoryear{Hager, Tsiatis, and Davidian}{Hager
  et~al.}{2018}]{Hager}
Hager, R., Tsiatis, A.~A., and Davidian, M. (2018).
\newblock Optimal two-stage dynamic treatment regimes from a classification
  perspective with censored survival data.
\newblock {\em Biometrics} {\bf 74,} 1180--1192.

\bibitem[\protect\citeauthoryear{Kauermann and Carroll}{Kauermann and
  Carroll}{2001}]{Kauer}
Kauermann, G. and Carroll, R.~J. (2001).
\newblock A note on the efficiency sandwich matrix estiamtion.
\newblock {\em Journal of the American Statistical Association} {\bf 96,}
  1387--1396.

\bibitem[\protect\citeauthoryear{Kidwell}{Kidwell}{2014}]{KidwellCancer}
Kidwell, K.~M. (2014).
\newblock {SMART} designs in cancer research: {L}earning from the past, current
  limitations and looking toward the future.
\newblock {\em Clinical Trials} {\bf 11,} 445--456.

\bibitem[\protect\citeauthoryear{Kidwell and Wahed}{Kidwell and
  Wahed}{2013}]{KidwellLogrank}
Kidwell, K.~M. and Wahed, A.~S. (2013).
\newblock Weighted log-rank statistic to compare shared-path adaptive treatment
  strategies.
\newblock {\em Biostatistics} {\bf 14,} 299--312.

\bibitem[\protect\citeauthoryear{Lavori and Dawson}{Lavori and
  Dawson}{2004}]{Lavori}
Lavori, P.~W. and Dawson, R. (2004).
\newblock Dynamic treatment regimes: {P}ractical design considerations.
\newblock {\em Clinical Trials} {\bf 1,} 9--20.

\bibitem[\protect\citeauthoryear{Li and Murphy}{Li and Murphy}{2011}]{LiMurphy}
Li, Z. and Murphy, S.~A. (2011).
\newblock Sample size formulae for two-stage randomized trials with survival
  outcomes.
\newblock {\em Biometrika} {\bf 98,} 503--518.

\bibitem[\protect\citeauthoryear{Li, Valenstein, Pfeiffer, and Ganoczy}{Li
  et~al.}{2014}]{Li2014}
Li, Z., Valenstein, M., Pfeiffer, P., and Ganoczy, D. (2014).
\newblock A global logrank test for adpative treatment strategies based on
  observational studies.
\newblock {\em Statistics in Medicine} {\bf 33,} 760--771.

\bibitem[\protect\citeauthoryear{Lorenzoni, Petracci, Scarpi, Baldi, Gregori,
  and Nanni}{Lorenzoni et~al.}{2023}]{lorenzoni2023use}
Lorenzoni, G., Petracci, E., Scarpi, E., Baldi, I., Gregori, D., and Nanni, O.
  (2023).
\newblock Use of sequential multiple assignment randomized trials (smarts) in
  oncology: systematic review of published studies.
\newblock {\em British Journal of Cancer} {\bf 128,} 1177--1188.

\bibitem[\protect\citeauthoryear{Lunceford, Davidian, and Tsiatis}{Lunceford
  et~al.}{2002}]{Lunceford}
Lunceford, J.~K., Davidian, M., and Tsiatis, A.~A. (2002).
\newblock Estimation of survival distributions of treatment policies in
  two-stage randomization designs in clinical trials.
\newblock {\em Biometrics} {\bf 58,} 48--57.

\bibitem[\protect\citeauthoryear{Murphy}{Murphy}{2005}]{Murphy2005}
Murphy, S.~A. (2005).
\newblock An experimental design for the development of adaptive treatment
  strategies.
\newblock {\em Statistics in Medicine} {\bf 24,} 1455--1481.

\bibitem[\protect\citeauthoryear{Nahum-Shani and Almirall}{Nahum-Shani and
  Almirall}{2019}]{nahum2019introduction}
Nahum-Shani, I. and Almirall, D. (2019).
\newblock An introduction to adaptive interventions and {SMART} designs in
  education. ncser 2020-001.
\newblock {\em National Center for Special Education Research} .

\bibitem[\protect\citeauthoryear{Orellana, Rotnitzky, and Robins}{Orellana
  et~al.}{2010}]{Orellana}
Orellana, L., Rotnitzky, A., and Robins, J.~M. (2010).
\newblock Dynamic regime marginal structural mean models for estimation of
  optimal dynamic treatment regimes, part i: main content.
\newblock {\em The International Journal of Biostatistics} {\bf 6,} 8.

\bibitem[\protect\citeauthoryear{Powell, Moser, Stock, Gallegher, Willman,
  Stone, Rowe, Coutre, Feusner, Gregory, Couban, Appelbaum, Tallman, and
  Larson}{Powell et~al.}{2010}]{Powell2010}
Powell, B.~L., Moser, B., Stock, W., Gallegher, R.~E., Willman, C.~L., Stone,
  R.~M., Rowe, J.~M., Coutre, S., Feusner, J.~H., Gregory, Couban, S.,
  Appelbaum, F.~R., Tallman, M.~S., and Larson, R.~A. (2010).
\newblock Arsenic trioxide improves event-free and overall survival for adults
  with acute promyelocytic leukemia: {N}orth {A}merican {L}eukemia {I}ntergroup
  {S}tudy {C}9710.
\newblock {\em Blood} {\bf 116,} 3751--3757.

\bibitem[\protect\citeauthoryear{Schaubel}{Schaubel}{2005}]{Schaubel}
Schaubel, D.~E. (2005).
\newblock Variance estimation for clustered recurrent event data with a small
  number of clusters.
\newblock {\em Statistics in Medicine} {\bf 24,} 3037--3051.

\bibitem[\protect\citeauthoryear{Tsiatis}{Tsiatis}{2006}]{TsiatisSemi}
Tsiatis, A.~A. (2006).
\newblock {\em Semiparametric Theory and Missing Data}.
\newblock Springer, New York.

\bibitem[\protect\citeauthoryear{Tsiatis and Davidian}{Tsiatis and
  Davidian}{2022}]{Vaccine}
Tsiatis, A.~A. and Davidian, M. (2022).
\newblock Estimating vaccine efficacy over time after a randomized study is
  unblinded.
\newblock {\em Biometrics} {\bf 78,} 825--838.

\bibitem[\protect\citeauthoryear{Tsiatis, Davidian, Holloway, and
  Laber}{Tsiatis et~al.}{2020}]{TsiatisBook}
Tsiatis, A.~A., Davidian, M., Holloway, S.~T., and Laber, E.~B. (2020).
\newblock {\em Dynamic Treatment Regimes: {S}tatistical Methods for Precision
  Medicine}.
\newblock Chapman and Hall/CRC Press, Boca Raton, FL.

\bibitem[\protect\citeauthoryear{Wang, Turner, and Li}{Wang
  et~al.}{2023}]{WangTurner}
Wang, X., Turner, E.~L., and Li, F. (2023).
\newblock Improving sandwich variance estimation for marginal cox analysis of
  cluster randomized trials.
\newblock {\em Biometrical Journal} {\bf 65,} 2200113.

\bibitem[\protect\citeauthoryear{Wu, Wang, and Wahed}{Wu
  et~al.}{2023}]{WahedChi}
Wu, L., Wang, J., and Wahed, A.~S. (2023).
\newblock Interim monitoring in sequential multiple assignment randomized
  trials.
\newblock {\em Biometrics} {\bf 79,} 368--380.

\bibitem[\protect\citeauthoryear{Yang, Tsiatis, and Blazing}{Yang
  et~al.}{2018}]{Yangetal}
Yang, S., Tsiatis, A.~A., and Blazing, M. (2018).
\newblock Modeling survival distribution as a function of time to treatment
  discontinuation: A dynamic treatment regime approach.
\newblock {\em Biometrics} {\bf 74,} 900--909.

\end{thebibliography}
 
\end{document}